\documentclass[12pt]{article}

\usepackage{graphicx}
\usepackage{amsmath}
\usepackage{amssymb}
\usepackage{color}

\numberwithin{equation}{section}

\newcommand{\splus}{
\setbox1=\hbox{$+$}
\setbox2=\hbox to \wd1 {}
\box2
}

\newcommand{\spacemaker}[1]{
\setbox1=\hbox{$#1$}
\setbox2=\hbox to \wd1 {}
\box2
}

\newcommand{\fduu}[3]{
\setbox1=\hbox{$f$}
\setbox2=\vbox to \ht1 {}
f_{#1}\box2^{#2#3}
}

\newcommand{\fddu}[3]{
\setbox1=\hbox{$f$}
\setbox2=\vbox to \ht1 {}
f_{#1#2}\box2^{#3}
}

\newcommand{\llra}{ {\bf \longleftrightarrow} }

\newcommand{\coef}[6]{
\setbox1=\hbox{$A$}
\setbox2=\vbox to \ht1 {}
\setbox3=\vbox to \ht1 {}
\setbox4=\hbox{${}^{#1}_{#2}$}
\setbox5=\hbox{${}^{#3}_{#4}$}
\setbox6=\hbox{${}^{#5}_{#6}$}
\box1^{\hbox to \wd4 {\scriptsize\hfil$#1$\hfil}}_{\hbox to \wd4 {\scriptsize\hfil$#2$\hfil}} 
\box2^{\hbox to \wd5 {\scriptsize\hfil$#3$\hfil}}_{\hbox to \wd5 {\scriptsize\hfil$#4$\hfil}} 
\box3^{\hbox to \wd6 {\scriptsize\hfil$#5$\hfil}}_{\hbox to \wd6 {\scriptsize\hfil$#6$\hfil}} 
}

\newcommand{\ambig}[4]{
\setbox1=\hbox{$c$}
\setbox2=\vbox to \ht1 {}
\setbox4=\hbox{${}^{#1}_{#3}$}
\setbox5=\hbox{${}^{#2}_{#4}$}
\box1^{\hbox to \wd4 {\scriptsize\hfil$#1$\hfil}}_{\hbox to \wd4 {\scriptsize\hfil$#3$\hfil}} 
\box2^{\hbox to \wd5 {\scriptsize\hfil$#2$\hfil}}_{\hbox to \wd5 {\scriptsize\hfil$#4$\hfil}} 
}

\def\thickhrulefill{\leavevmode \leaders \hrule height 1pt\hfill  }

\newcommand{\updown}[4]{
\vbox{%
\hbox{\hbox to #3pt{\thickhrulefill}#1\hbox to #4pt{\thickhrulefill}}%
\hbox{\hbox to #4pt{\thickhrulefill}#2\hbox to #3pt{\thickhrulefill}}}%
}

\newcommand{\upupdown}[4]{
\setbox1=\hbox{\footnotesize\hspace{7pt}#1\hspace{#4pt}#2\hspace{7pt}}
\setbox2=\hbox to \wd1{\footnotesize{\thickhrulefill}#1{\thickhrulefill}#2{\thickhrulefill}}
\setbox3=\hbox to \wd1{\footnotesize{\thickhrulefill}#3{\thickhrulefill}}
\vbox{\box2 \box3}
}

\newcommand{\downdownup}[4]{
\setbox1=\hbox{\hspace{7pt}$#1$\hspace{#4pt}$#2$\hspace{7pt}}
\setbox2=\hbox to \wd1{\thickhrulefill$#1$\thickhrulefill$#2$\thickhrulefill}
\setbox3=\hbox to \wd1{\thickhrulefill$#3$\thickhrulefill}
\vbox{\box3 \box2}
}

\begin{document}


\thispagestyle{empty}
\begin{flushright}\footnotesize
\texttt{CALT-68-2666}\\
\texttt{NI-07085}\\
\end{flushright}

\renewcommand{\thefootnote}{\fnsymbol{footnote}}
\setcounter{footnote}{0}
\vspace{1.5cm}

\begin{center}
{\Large\textbf{
Algebra of transfer-matrices and Yang-Baxter equations \\
on the string worldsheet in $AdS_5 \times S^5$}\par}

\vspace{2.1cm}

\textrm{Andrei Mikhailov\footnote{On leave from
 Institute for Theoretical and 
Experimental Physics, 
117259, Bol. Cheremushkinskaya, 25, 
Moscow, Russia} and Sakura Sch\"afer-Nameki}
\vspace{1.2cm}

\textit{California Institute of Technology\\
1200 E California Blvd., Pasadena, CA 91125, USA } \\
\texttt{andrei@theory.caltech.edu, ss299@theory.caltech.edu} \\
\vspace{0.5cm}
\textit{and}\\
\vspace{0.5cm}
\textit{
Isaac Newton Institute for Mathematical Sciences\\
20 Clarkson Road, Cambridge, CB3 0EH, UK  
}


\par\vspace{1.2cm}

\textbf{Abstract}\vspace{5mm}
\end{center}

\noindent 
Integrability of the string worldsheet theory in $AdS_5\times S^5$ is related
to the existence of a flat connection depending on the spectral parameter. 
The transfer matrix is the open-ended Wilson line of this flat connection.
We study the product of transfer matrices in the near-flat space 
expansion of the $AdS_5\times S^5$ string theory in the pure spinor 
formalism. The natural operations on Wilson lines with insertions are
described in terms of $r$- and $s$-matrices satisfying a generalized
classical Yang-Baxter equation. The formalism is especially transparent
for infinite or closed Wilson lines with simple gauge invariant insertions.

\vspace*{\fill}

\setcounter{page}{1}
\renewcommand{\thefootnote}{\arabic{footnote}}
\setcounter{footnote}{0}

\newpage

\tableofcontents


\section{Introduction}

Integrability of superstring theory in $AdS_5\times S^5$ has been a vital input 
for recent progress in understanding  the 
AdS/CFT correspondence. However quantum integrability of the string worldsheet
sigma-model is far from having been established.
The notion of quantum integrability is well developed for relativistic
massive quantum field theories, which describe scattering of particles in two
space-time dimensions.   But the string worldsheet theory is a 
very special type
of a quantum field theory,  and certainly not a relativistic massive theory.
It may not be the most natural way to think of the string worldsheet theory as describing
a system of particles. It may be better to think of it as describing certain
operators, or rather equivalence classes of operators. What does integrability
mean in this case?
Progress in this direction could be key to understanding the exact quantum
spectrum, which goes beyond  the infinite volume spectrum that is obtained from the 
asymptotic Bethe ansatz \cite{Beisert:2006ib, Beisert:2006ez}.  

The transfer matrix usually plays an important role in integrable models,
in particular in conformal ones \cite{Bazhanov:1998dq}. 
The renormalization group usually acts nontrivially on the 
transfer matrix \cite{Bachas:2004sy, Alekseev:2007in}. But the string worldsheet
theory is special.  The transfer matrix
on the string worldsheet is BRST-invariant, and there is a conjecture that 
it is not renormalized. This was demonstrated in a  one-loop calculation in \cite{Mikhailov:2007mr}. 

In this paper we will revisit the problem of calculating the Poisson brackets
of the worldsheet transfer matrices 
\cite{Das:2004hy,Das:2005hp,Bianchi:2006im,Dorey:2006mx,Mikhailov:2006uc,Kluson:2007ua}. 
The transfer matrix is a monodromy of a certain flat connection on the worldsheet,
which exists because of classical integrability. One can think of it as a kind of
Wilson line: given an open contour $C$, we calculate $T[C]=P\exp -\int_C J$.
Instead of calculating the Poisson bracket we consider the product of two transfer
matrices for two different contours, and considering the limit when
one contour is on top of another: 

\vspace{7pt}

\hbox to \linewidth{\hfill  \includegraphics[width=1.5in]{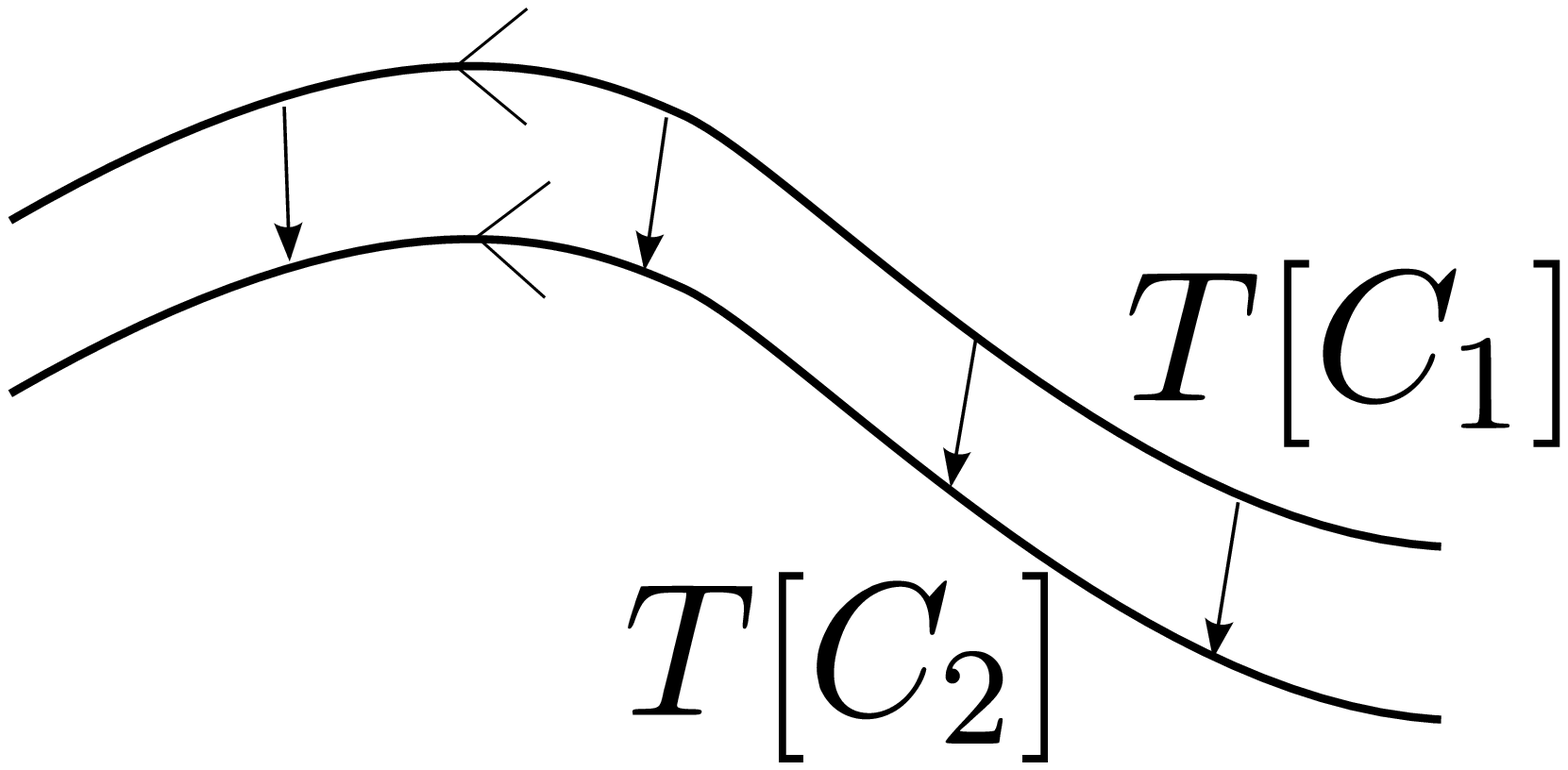}  \hfill}

\noindent
At first order of perturbation theory studying this limit is equivalent
to calculating the Poisson brackets -- we will explain this point in detail. We find that the typical object appearing
in this calculation is a  dynamical (=field-dependent) R-matrix suggested
by J.-M.~Maillet \cite{Maillet:1985ec,Maillet:1985ek,Maillet:1985fn}. 
The Maillet approach was discussed recently for the superstring in $AdS_5 \times S^5$
in \cite{Kluson:2006wq,Dorey:2006mx,Kluson:2007ua}.  

The transfer matrix is a parallel-transport type of object. Given two points $x$ and $y$ 
on the string worldsheet, we can consider the tangent spaces to the target at these
two points, $T_x(AdS_5\times S^5)$ and $T_y(AdS_5\times S^5)$. The transfer matrix
allows us to transport various vectors, tensors and spinors between
$T_x(AdS_5\times S^5)$ and $T_y(AdS_5\times S^5)$. This allows to construct 
operators on the worldsheet by inserting the tangent space objects (for example $\partial_+x$)
at the endpoints of the Wilson line:

\vspace{7pt}

\hbox to \linewidth{\hfill  \includegraphics[width=1.5in]{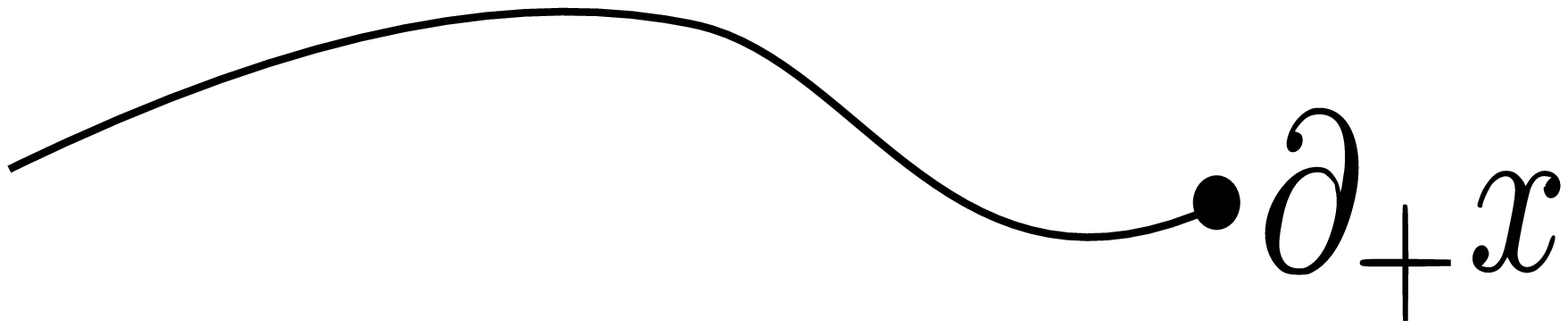}  \hfill}

\noindent
or inside the Wilson line:

\vspace{7pt}

\hbox to \linewidth{\hfill  \includegraphics[width=1.5in]{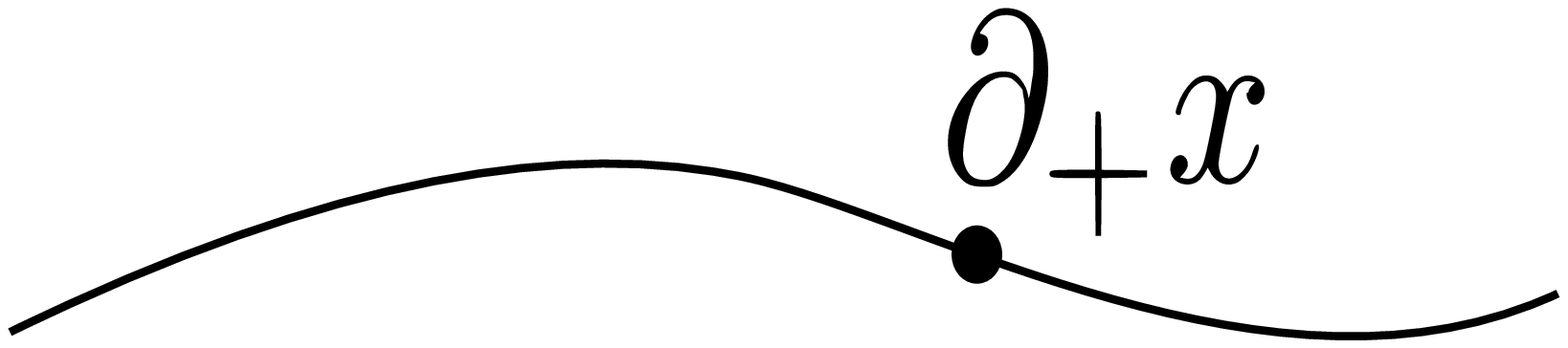}  \hfill}

\noindent
We study the products of the simplest objects of this type at the first order of 
perturbation theory.
The results are summarised in Section \ref{sec:TwoTransferMatrices}. 
The subsequent sections contain derivations, the main points are in
Sections \ref{sec:StructureOfDelta} and \ref{sec:BoundaryEffects}.
In Section \ref{sec:GCYBE} we discuss the consistency conditions 
(generalized Yang-Baxter equations).

\newpage

\section{Summary of results}
\label{sec:TwoTransferMatrices}
This section contains a summary of our results, and in the subsequent sections 
we will describe the derivation.

\subsection{Definitions}
\subsubsection{The definition of the transfer matrix}

Two dimensional integrable systems are characterized by the existence of 
certain currents $J^a$, which have the property that the transfer matrix
\begin{equation}\label{DefTransferMatrix}
T[C]=P\exp\left( -\int_C J^a e_a \right) \,,
\end{equation}
is independent of the choice of the contour. In this definition
$e_a$ are generators of some algebra. The algebra usually has many
different representations, so the transfer matrix is labelled 
by a representation. We will write $T_{\rho}[C]$ where the generators
$e_a$ act in the representation $\rho$.

For the string in $AdS_5\times S^5$ the algebra is the twisted loop
algebra $L\mathfrak{psu(2,2|4)}$ and the coupling of the currents to the
generators is the following:
\begin{eqnarray}
&&
J_+=
(J_{0+}^{[\mu\nu]}-N_{0+}^{[\mu\nu]})e^0_{[\mu\nu]}+
J_{3+}^{\alpha}e^{-1}_{\alpha}+ 
J_{2+}^{\mu}e^{-2}_{\mu} +
J_{1+}^{\dot{\alpha}}e^{-3}_{\dot{\alpha}} +
N_{0+}^{[\mu\nu]}e^{-4}_{[\mu\nu]} 
\label{APlusZ}\\
&&        
J_-=
(J_{0-}^{[\mu\nu]}-N_{0-}^{[\mu\nu]})e^0_{[\mu\nu]}+
J_{1-}^{\alpha}e^{1}_{\dot{\alpha}}+ 
J_{2-}^{\mu}e^{2}_{\mu} +
J_{3-}^{\dot{\alpha}}e^{3}_{\alpha} +
N_{0-}^{[\mu\nu]}e^{4}_{[\mu\nu]} \,.
\label{AMinusZ} 
\end{eqnarray}
Here $e^m_a$ are the generators of the twisted loop algebra. We will use the evaluation
representation of the loop algebra. In the evaluation representation $e_a^m$ are related
to the generators of some representation of the finite-dimensional 
algebra $\mathfrak{psu(2,2|4)}$ in the following
way:
\begin{equation} \label{LoopAlgebraGenerators}
e^{-3}_{\alpha}=z^{-3}t^1_{\alpha}, \;\;
e^{-2}_{\mu}=z^{-2}t^2_{\mu}, \;\;
e^1_{\alpha}=zt^1_{\alpha} \;\; \mbox{\it etc.}
\end{equation}
where $z$ is a complex number, which is called ``spectral parameter''.
Further details on the conventions can be found in
Section \ref{sec:Notations} and in  \cite{Mikhailov:2007mr}.

\subsubsection{Setup: expansion around flat space and expansion in powers of fields}
\label{sec:NearFlatSpace}

The gauge group $\mathfrak{g}_{\bar{0}}\subset \mathfrak{psu(2,2|4)}$
acts on the currents in the following way:
\begin{eqnarray}
&& \delta_{\xi_{\bar{0}}} J_{\bar{1}} = [\xi_{\bar{0}},J_{\bar{1}}]\;,\;\;
   \delta_{\xi_{\bar{0}}} J_{\bar{2}} = [\xi_{\bar{0}},J_{\bar{2}}]\;,\;\;
   \delta_{\xi_{\bar{0}}} J_{\bar{3}} = [\xi_{\bar{0}},J_{\bar{3}}]\;,\;\;
\nonumber
\\
&& \delta_{\xi_{\bar{0}}} J_{\bar{0}} = -d\xi_{\bar{0}} + [\xi_{\bar{0}},J_{\bar{0}}]
\;,\;\;\;\;\mbox{where }
\xi_{\bar{0}}\in \mathfrak{g}_{\bar{0}} \,.
\label{GaugeInvarianceOnCurrents}
\end{eqnarray}
In terms of the coordinates of the coset space:
\begin{equation}
J=-dgg^{-1}\;,\;\; g\in PSU(2,2|4) \,.
\end{equation}
The gauge invariance (\ref{GaugeInvarianceOnCurrents}) acts on $g$ as follows:
\begin{equation}
g\mapsto hg\;,\;\; h=e^{\xi}\;,\;\; \xi\in\mathfrak{g}_{\bar{0}} \,.
\end{equation}
There are two versions of the transfer matrix. 
One is $T$ given by Eq. (\ref{DefTransferMatrix})
and the other is $g^{-1}Tg$. Notice that $g^{-1}Tg$ is gauge invariant,
while $T$ is not. We should think of $T[C]$ as a map from the (supersymmetric)
tangent space  $T(AdS_5\times S^5)$ at the starting point of $C$ to
$T(AdS_5\times S^5)$ at the endpoint of $C$.

The choice of a point in $AdS_5\times S^5$ leads to the special gauge, which we
will use in this paper:
\begin{equation}\label{SpecialGauge}
g=e^{R^{-1}(\vartheta_L+\vartheta_R)}e^{R^{-1}x} \,.
\end{equation}
Here $R$ is the radius of AdS space, and it is introduced in (\ref{SpecialGauge})
for convenience. The action has a piece quadratic in $x,\vartheta$ and interactions
which we can expand in powers of $x,\vartheta$. There are also pure spinor ghosts
$\lambda,w$. All the operators can be
expanded\footnote{ 
The expansion in powers of elementary fields is especially transparent in the
classical theory where it can be explained in the spirit of \cite{Rosly:1996vr}.
We write 
\[
x  =  \sum_{a=1}^N \epsilon_a e^{i k_a \overline{w} + i \overline{k}_a w} + 
	 +\sum_{ab} G_{ab}(k_a,k_b) \epsilon_a\epsilon_b  
	e^{i (k_a+k_b)\overline{w} + i (\overline{k}_a+\overline{k}_b)w} + \ldots  \,.
\]
where $\epsilon_a$, $a=1,2,\ldots, N$ are nilpotents: $\epsilon_a^2=0$ for every $a$.
The nilpotency of $\epsilon_a$ implies that the powers of $x$ higher than $x^N$ 
automatically drop out.} in powers of $x,\vartheta,\lambda,w$. 
We will refer to this expansion as 
``expansion in powers of elementary fields'', or
``expansion in powers of $x$''. Every power of elementary field carries a factor $R^{-1}$.
The overall power
of $R^{-1}$ is equal to twice the number of propagators plus the number
of uncontracted elementary fields. A propagator is a contraction of two
elementary fields.

The currents are invariant under the global symmetries, up to gauge
transformations. For example  the global shift
\begin{equation}\label{GlobalShift}
	S_{g_0} x = x + \xi + {1\over 3R^2} [x, [x, \xi]] +\ldots
\end{equation}
results in the gauge transformation of the currents with the parameter
\begin{equation}
 	h(\vartheta,x; e^{\xi}) = 
	\exp\left( -{1\over 2R^2}[x,\xi] +\ldots\right) \,.
\label{CompensatingGaugeTransformation}
\end{equation}
To have the action invariant we should also transform the pure spinors with
the same parameter:
\begin{equation}
\delta_{\xi}\lambda = -\left[ {1\over 2R^2}[x,\xi]\; , \; \lambda \right],\;\;\;
\delta_{\xi}w_+ = -\left[ {1\over 2R^2}[x,\xi]\; , \; w_+ \right]
\end{equation}
and same rules for $\hat{w}_-,\hat{\lambda}$.


\subsection{Fusion and exchange of transfer matrices}
\subsubsection{The product of two transfer matrices}
\label{sec:TheProductOfTwoT}

Consider the transfer matrix in the tensor
product of two representations $\rho_1\otimes \rho_2$. 
There are two ways of defining this object. One way is to take the usual 
definition of the Wilson line
\begin{equation}
P\exp\left( -\int J^a(z) e_a\right) \,,
\end{equation}
and use for $e_a$ the usual definition of the tensor product of generators
of a Lie superalgebra:
\begin{equation}
\rho_1(e_a)\otimes {\bf 1} + (-)^{F\bar{a}}\otimes \rho_2(e_a) \,,
\end{equation}
where $\bar{a}$ is $0$ if $e_a$ is an even element of the superalgebra,
and $1$ if $e_a$ is an odd element of the superalgebra.

Another possibility is to consider two Wilson lines
$T_{\rho_1}$ and $T_{\rho_2}$ and put them on
top of each other. In other words, consider the product
$T_{\rho_2}T_{\rho_1}$.
In the classical theory these two definitions of the ``composite''
Wilson line are equivalent, because of this identity:
\begin{equation}
e^{\alpha}\otimes e^{\beta} = e^{\alpha\otimes 1 + 1\otimes \beta}
\,.
\end{equation}
But at the first order in $\hbar$ there is a difference.
The difference is related to the  singularities in the operator product of 
two currents.

Consider the example when the product of the
currents has the following form:
\begin{equation}
J^a_+(w) J^b_+(0) = {1\over w} 
A^{ab}_{c} J^c_+ 
+\ldots \,,
\end{equation}
where dots denote regular terms. 
Take two contours $C_1$ and $C_2$ and calculate the product
\begin{equation}\label{T1T2}
T_{\rho_2}[C_2] \; T_{\rho_1}[C_1] \,,
\end{equation}
where the indices $\rho_1$ and $\rho_2$ indicate that we are
calculating the monodromies in the representations $\rho_1$ and
$\rho_2$ respectively.
For example, suppose that the contour $C_1$ is the line
$\tau=0$ (and $\sigma$ runs from $-\infty$ to $+\infty$),
and the contour $C_2$ is at $\tau=y$ 
(and $\sigma\in [-\infty,+\infty]$).
Suppose that we bring the contour of $\rho_2$ on top of the contour
of $\rho_1$, in other words $y\to 0$. Let us expand
both $T_{\rho_2}[C_2]$ and $T_{\rho_1}[C_1]$ in powers
of $R^{-2}$, and think of them as series of multiple integrals
of $J$. Consider for example a term in which one $\int J$ comes
from $T_{\rho_2}[C_2]$ and another $\int J$ comes from
$T_{\rho_1}[C_1]$. We get:
\begin{eqnarray}
\int\int d\sigma_1 d\sigma_2 \; J_+^a(y,\sigma_2) (e_a\otimes 1) \;
J_+^b(0,\sigma_1) (1\otimes e_b) =
\nonumber \\
\int\int d\sigma_1 d\sigma_2 \; {1\over \sigma_2-\sigma_1+iy} 
A^{ab}_{c} J^c_+ \;
(e_a\otimes 1) (1\otimes e_b)  \,.
\end{eqnarray}
The pole ${1\over \sigma_2-\sigma_1+iy}$ leads to the difference 
between $\lim_{y\to 0}T_{\rho_2}[C+y]T_{\rho_1}[C]$ and
$T_{\rho_2\otimes \rho_1}[C]$.
Indeed, the natural definition of the double integral
when $y=0$ would be that when $\sigma_1$ collides with $\sigma_2$
we take a principle value:
\begin{equation}\label{FiniteIntegral}
\mbox{V.P.}\int\int d\sigma_1 d\sigma_2 \;  
J_+^a(0,\sigma_2) (e_a\otimes 1) \;
J_+^b(0,\sigma_1) (1\otimes e_b) \,.
\end{equation}
Here V.P. means that we treat the integral as the principal value
when $\sigma_1$ collides with $\sigma_2$.
Modulo the linear divergences, which we neglect, the integral
(\ref{FiniteIntegral}) is finite. This is because $e_a\otimes 1$
commutes with $1\otimes e_b$. But  such a VP
integral is different from what we would get in the limit $y\to 0$, by
a finite piece. Indeed:
\begin{eqnarray}
\int dw J_+^a(w+i\epsilon) J_+^b(0)
& = &
\mbox{V.P.} \int dw J_+^a(w) J_+^b(0) 
+ \label{VP1}\\
&&
+\pi i A^{ab}_{c} J_+^c(0) \label{VP2}\,.
\end{eqnarray}
The second row is the difference between the VP prescription
and the $\lim\limits_{y\to 0}$ prescription. 
The additional piece 
$\pi i  A^{ab}_{c} J_+^c(0)$
could also be interpreted as the deformation of the generator
to which $J_+^c$ couples in the definition of the transfer matrix:
\begin{equation}\label{Coproduct}
J_+^c(e_c\otimes 1 + (-)^{F\bar{c}}\otimes e_c) \mapsto
J_+^c\left(e_c\otimes 1 + (-)^{F\bar{c}}\otimes e_c + 
\pi i A^{ab}_{c} e_a(-)^{F\bar{b}}\otimes e_b\right) \,.
\end{equation}
We have two different definitions of the transfer matrix in the
tensor product of two representations. Is it true that these two 
definitions actually give the same object?
There are several logical possibilities:
  \begin{enumerate}
  
  \item There are several ways to define the transfer matrix, and they all give
  essentially different Wilson line-like operators.
  
  \item We should interpret Eq. (\ref{Coproduct}) as defining the deformed
  coproduct on the algebra of generators. The algebra of generators is 
  in our case a twisted loop algebra of $\mathfrak{psu(2,2|4)}$. There are
  at least three possibilities:
  
    \begin{enumerate}
    
    \item	The proper definition of the transfer matrix actually requires
    	the deformation of the algebra of generators $e^a$, and the deformed
    	algebra has deformed coproduct.
    
    \item	The algebra of generators is the usual loop algebra, but it
    	has a nonstandard coproduct; 
    	$\lim_{y\to 0}T_{\rho_2}[C+y]T_{\rho_1}[C]$ is different
    	from $T_{\rho_1\otimes \rho_2}[C]$, the difference being the use of
	a nonstandard coproduct. We are not aware of a mathematical theorem which forbids
    	such a nontrivial coproduct.
    
    \item	\label{en:Conjectured}
		The coproduct defined by Eq. (\ref{Coproduct}) is equivalent to
		the standard one, in a sense that it is obtained from the standard
		coproduct by a conjugation:
\begin{eqnarray}
\Delta^0(e^c) & = & e_c\otimes 1 + (-)^{F\bar{c}}\otimes e_c
\\[5pt]
\Delta(e^c) & = & e_c\otimes 1 + (-)^{F\bar{c}}\otimes e_c + 
\pi i A^{ab}_{c} e_a\otimes (-)^{F\bar{c}}e_b =
\nonumber
\\
& = & e^{{\pi i\over 2}r} ( e_c\otimes 1 + (-)^{F\bar{c}}\otimes e_c ) 
e^{-{\pi i\over 2}r} \,.
\label{Xdefinition}
\end{eqnarray}
    
    \end{enumerate}

\end{enumerate}
We will argue that what actually happens (at the tree level) 
is a generalization of \ref{en:Conjectured}.  
The deformation (\ref{Xdefinition}) is almost enough to account
for the difference between $\lim_{y\to 0}T_{\rho_2}[C+y]T_{\rho_1}[C]$ and 
$T_{\rho_1\otimes \rho_2}[C]$, but in addition to (\ref{Xdefinition}) one has to 
do a field-dependent generalized gauge transformation\footnote{Generalized gauge
transformation is 
$J\mapsto f(d+J)f^{-1}$. If $f\in \mbox{exp}\; \mathfrak{g}_{\bar{0}}$ 
then this is a usual (or ``proper'' gauge transformation as defined 
in Section \ref{sec:NearFlatSpace}. If we relax this condition we get
the ``generalized gauge transformation'' see 
Section \ref{sec:FixingTotalDerivatives}.}. The correct statement 
is:

\vspace{8pt}
\centerline{\em for a contour $C$ going from the point $A$ to the point $B$}
\begin{equation}\label{IntroRHat}
\lim_{y\to 0}T_{\rho_2}[C+y]T_{\rho_1}[C] =
e^{{\pi i\over 2}\hat{r}(A)}
T_{\rho_1\otimes \rho_2}[C]e^{-{\pi i\over 2}\hat{r}(B)}
\end{equation}
where $\hat{r}$ is field dependent (``dynamical'').
In fact $\hat{r}$ is of the order $\hbar$. This paper is all about the tree level.
Therefore all we are saying is:
\begin{equation}\label{TreeLevelOnly}
\lim_{y\to 0}T_{\rho_2}[C+y]T_{\rho_1}[C] =
T_{\rho_1\otimes \rho_2}[C]+
{\pi i\over 2} \left(\; \hat{r}(A)\;T_{\rho_1\otimes \rho_2}[C] -
T_{\rho_1\otimes \rho_2}[C] \; \hat{r}(B)\; \right)
+\ldots
\end{equation}
where dots stand for loop effects.
The hat over the letter $r$ shows that this is a field-dependent object.
We will also use a field-independent $r$-matrix which will be denoted $r$ without
a hat; $r$ is the leading term in the near-flat-space expansion of $\hat{r}$,
which is the expansion in powers of elementary fields explained in
Section \ref{sec:NearFlatSpace}:
\begin{eqnarray}
\hat{r}\;\;\;=& r
&
-{\pi i\over 2} \left(\; ((z_1^{-2}-z_1^2) t^2)\otimes [t^2,x]
                      -[t^2,x]\otimes ((z_2^{-2}-z_2^2)t^2) \;\right)
-
\nonumber \\
&&
-{\pi i\over 2} \left(\; ((z_1^{-3}-z_1) t^1)\otimes \{t^3,\vartheta_L\}
                      -\{t^3,\vartheta_L\}\otimes ((z_2^{-3}-z_2) t^1) \;\right)
-
\nonumber \\
&&
-{\pi i\over 2} \left(\; ((z_1^{-1}-z_1^3) t^3)\otimes \{t^1,\vartheta_R\}
                      -\{t^1,\vartheta_R\}\otimes ((z_2^{-1}-z_2^3)t^3) \;\right)+
\nonumber \\
& +\ldots &\label{XDependentTerms}
\end{eqnarray}

Here $r$ is given by Eq. (\ref{IntroClassicalR}) and dots stand
for the terms of quadratic and higher orders in $x$ and $\vartheta$.
The pure spinor ghosts do not enter into the expression for $\hat{r}$, only the
matter fields $x$ and $\vartheta$.

\begin{figure}[th]
\centerline{\includegraphics[width=10cm]{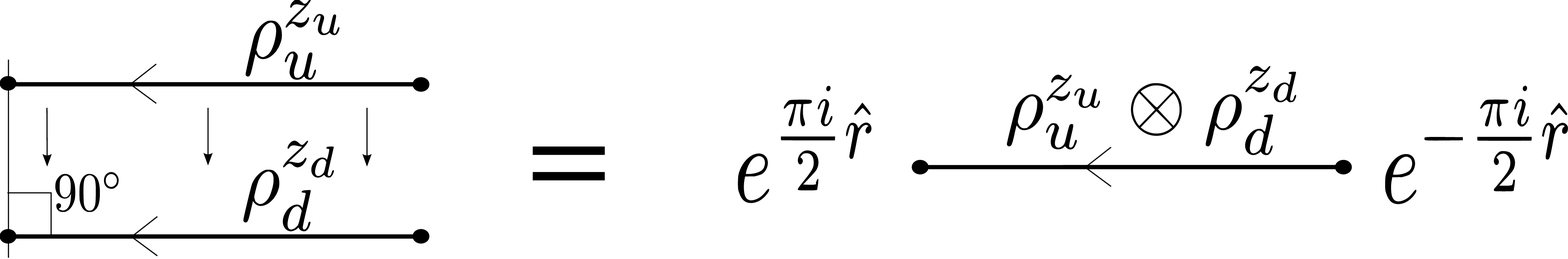}}
\caption{\label{fig:Fusion}\small
Fusion of transfer matrices.}
\vspace{0.4cm}
\centerline{\includegraphics[width=10cm]{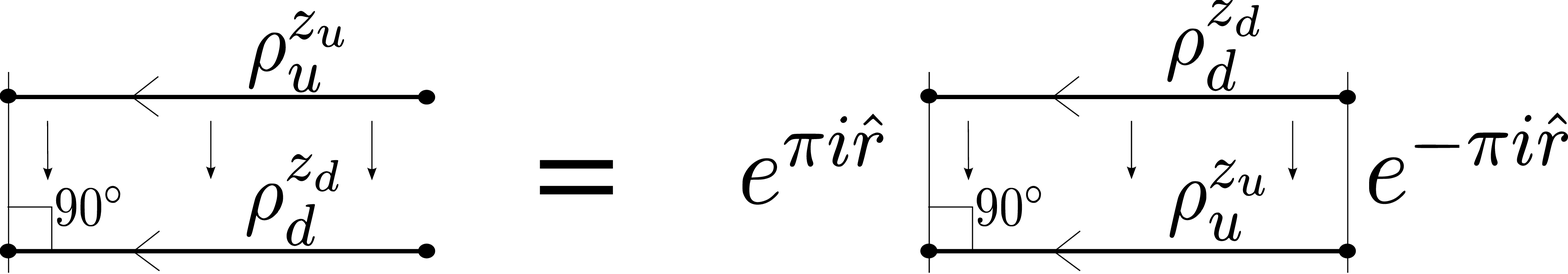}}
\caption{\label{fig:UpDownUp}\small
Exchange of transfer matrices.}
\end{figure}

The special thing
about the constant term $r$ is that it is a rational function of the 
spectral parameter with the first order pole at $z_u=z_d$. 
The coefficients of the $x,\vartheta$-dependent terms are all polynomials
in $z_u,z_d,z_u^{-1},z_d^{-1}$. 
 The field dependence of the $\hat{r}$ matrix in this example is related
to the fact that
the pair of Wilson lines with ``loose ends'' is not a gauge invariant 
object.\footnote{We use the special gauge (\ref{SpecialGauge}), therefore
in our formalism the lack of gauge invariance translates into the lack
of translational invariance.}

Eq. (\ref{IntroRHat}) is schematically illustrated in
Figure \ref{fig:Fusion}.
A consequence of (\ref{IntroRHat}) is the equivalence relation for the
exchange of the order of two transfer matrices, see Figure \ref{fig:UpDownUp}:
\begin{equation}
\lim_{C_{u}\searrow C_{d}} T_{C_{u}}({\rho_{u}^{z_{u}}})
T_{C_{d}}({\rho_{d}^{z_{d}}})
=
\exp(\pi i\;\hat{r}) \left[\lim_{C_{u}\nearrow C_{d}} T_{C_{u}}({\rho_{u}^{z_{u}}})
T_{C_{d}}({\rho_{d}^{z_{d}}})\right] \exp(-\pi i\;\hat{r}) \,.
\end{equation}


\subsubsection{Relation to Poisson brackets}
At the tree level the calculation of the fusion of transfer matrices is equivalent to the
calculation of the Poisson brackets. This follows from the definition of the Poisson bracket:
\begin{equation}
\{ T_{\rho_1} , T_{\rho_2} \} = \lim_{\hbar\to 0}{1\over i\hbar}
\left(
\lim_{y\to 0+}
T_{\rho_1}[C+y]T_{\rho_2}[C]-
\lim_{y\to 0+}
T_{\rho_2}[C+y]T_{\rho_1}[C]
\right)
\end{equation}
and the equation:
\begin{equation}\label{AverageIsVP}
\lim_{y\to 0+}
T_{\rho_1}[C+y]T_{\rho_2}[C]+
\lim_{y\to 0+}
T_{\rho_2}[C+y]T_{\rho_1}[C]=2T_{\rho_1\otimes \rho_2}[C]+O(\hbar^2)
\end{equation}
which holds to the first order in $\hbar$. These two equations and Eq. (\ref{TreeLevelOnly}) 
imply 
\begin{equation}
\{T_{\rho_1}[C],T_{\rho_2}[C]\}
=
\pi  \left(\; \hat{r}(A)\;T_{\rho_1\otimes \rho_2}[C] -
T_{\rho_1\otimes \rho_2}[C] \; \hat{r}(B)\; \right)
\end{equation}
and therefore the calculation of $\hat{r}$ is actually equivalent to the calculation
of the Poisson brackets. 

To derive (\ref{AverageIsVP}) we expand the product $T[C+y]T[C]$ as normal ordered
product plus contractions. At the tree level only one contraction is needed;
schematically we get 
\[ J(w)J(0)=:J(w)J(0):+F(w,\bar{w}) \]
where $F(w,\bar{w})$ is
$1\over w$ or $1\over w^2$ or $1\over \bar{w}$ or $1\over \bar{w}^2$ times some expression
regular at $w\to 0$; see Section \ref{sec:ShortDistance}. 
Then eq. (\ref{AverageIsVP})  follows from the relation
\begin{equation}
 \hbox{lim}_{\epsilon \rightarrow 0^+}\left(
{ 1 \over (w + i\epsilon)^n } + 
{ 1 \over (w - i\epsilon)^n}\right)
= 2\mbox{V.P.} {1\over w^n}
\end{equation}
applied to the singular part of $F(w,\bar{w})$. 

The "standard" calculation of the Poisson bracket of two transfer matrices involves the equal time
Poisson brackets of the currents $\{J(\sigma),J(\sigma')\}$. This is proportional
to $\delta(\sigma-\sigma')$ or $\partial_{\sigma}\delta(\sigma-\sigma')$. This is equivalent
to what we are doing because:
\begin{equation}
\hbox{lim}_{\epsilon \rightarrow 0^+}\left(
{ 1 \over (w + i\epsilon)^n } - 
{ 1 \over (w - i\epsilon)^n}\right)
= {2\pi i(-1)^{n}\over (n-1)!}\;\partial_{\sigma}\delta(\sigma-\sigma')
\end{equation}
We conclude that the difference between our approach based on the notion of "fusion"
and the "standard" approach to calculating the Poisson brackets is a matter of notations.
(But we believe that our notations are more appropriate for calculating beyond the tree level.)


\subsubsection{$r$- and $s$-matrices and generalized classical YBE}

The open ended contours like the ones shown in 
Figures \ref{fig:Fusion} and \ref{fig:UpDownUp} are strictly speaking not
gauge invariant. In our approach we fix the gauge (\ref{SpecialGauge})
and therefore it is meaningful to consider these operators as operators
in the gauge fixed theory. Nevertheless we feel that these are probably not the
most natural objects to study, at least from the point of view of the 
differential geometry of the worldsheet.

\begin{figure}[th]
\centerline{\includegraphics[width=1.9in]{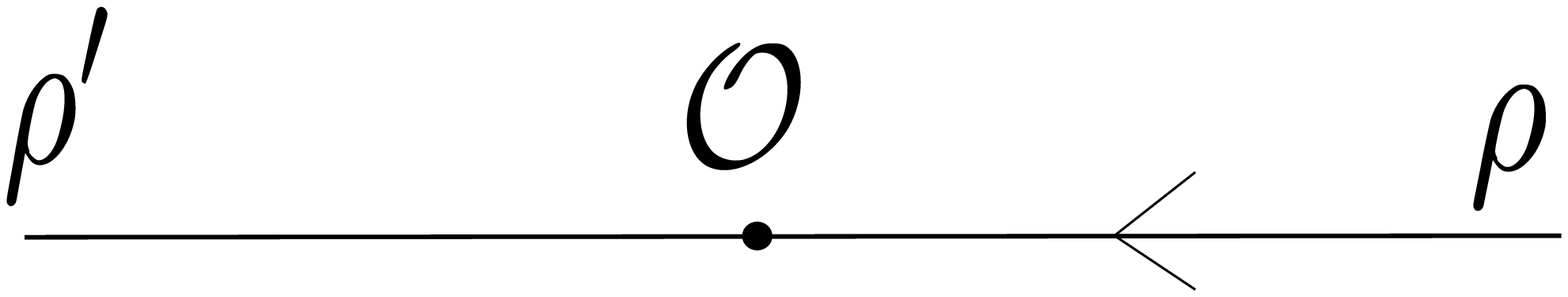}}
\caption{\small An infinite Wilson line with an operator insertion
\label{fig:InfiniteWilsonLineWithInsertion}}
\end{figure}
\noindent

The natural objects to consider are {\em infinite} (or periodic) Wilson lines with
various operator {\em insertions}, see Figure \ref{fig:InfiniteWilsonLineWithInsertion}.
How to describe the algebra formed by such operators?
What is the relation between \includegraphics[width=0.7in]{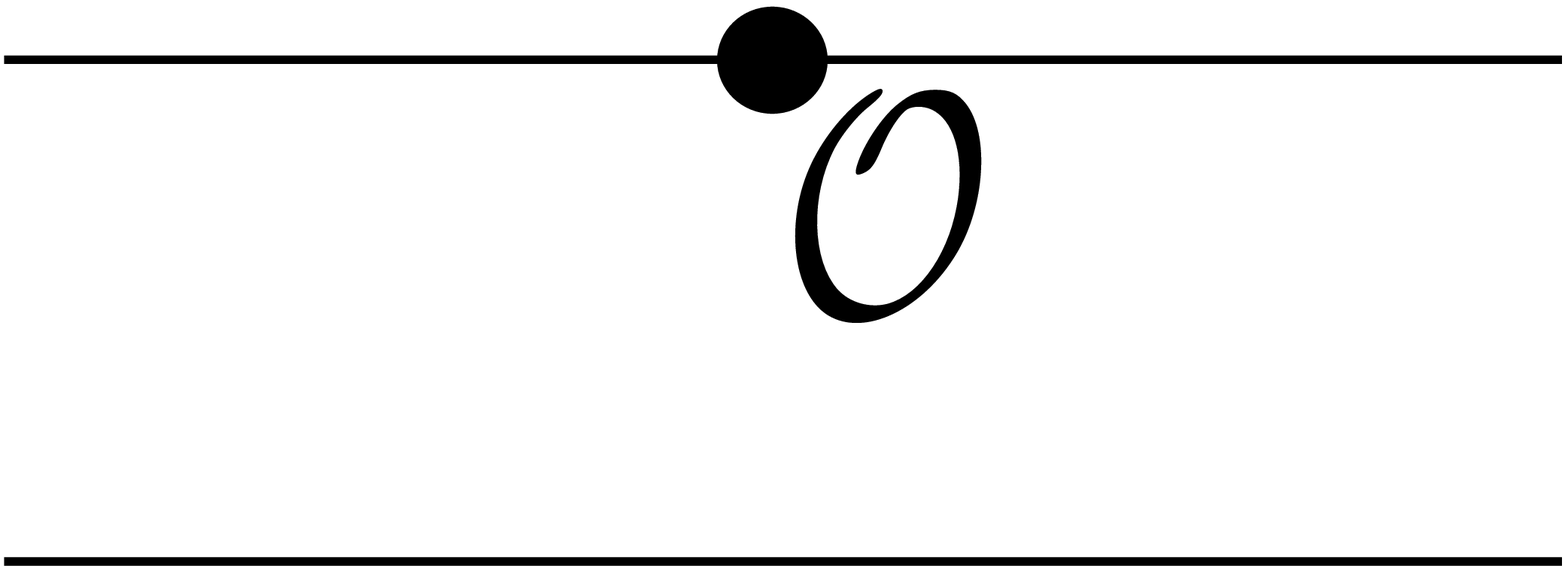}
and \includegraphics[width=0.7in]{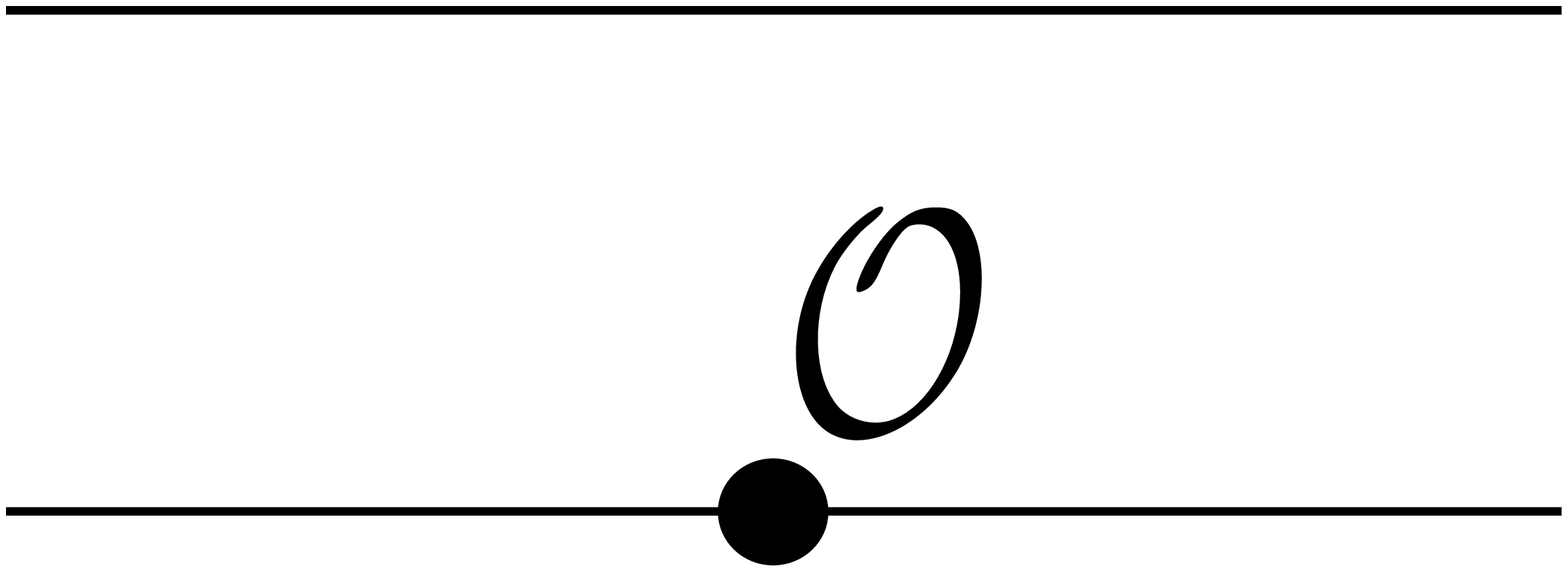}\hspace{4pt}? \hspace{5pt}
We will find that the description of this algebra involves matrices $r$ 
and $s$ which have the following form:
\begin{eqnarray}
r & = & 
 {\Phi (z_1, z_2)\over z_1^4 - z_2^4}
(z_1z_2^3 t^1 \otimes t^3 + z_1^3z_2 t^3 \otimes t^1 + 
z_1^2z_2^2 t^2 \otimes t^2) 
+2{\Psi(z_1, z_2) \over z_1^4 - z_2^4} t^0 \otimes t^0 \,,
\label{IntroClassicalR}
\\
s & = & 
 (z_1^{-1} z_2^{-3} - z_1^3 z_2)  t^3\otimes t^1 
+(z_1^{-2} z_2^{-2} - z_1^2 z_2^2)t^2\otimes t^2
+(z_1^{-3} z_2^{-1} - z_1   z_2^3)t^1\otimes t^3 \,,
\label{IntroClassicalS}
\end{eqnarray}
where
\begin{eqnarray}
\Phi(z_1, z_2) & = & (z_1^2 -z_1^{-2})^2 + (z_2^2 -z_2^{-2})^2
\nonumber
\\
\Psi(z_1, z_2) & = & 1+ z_1^4 z_2^4 - z_1^4- z_2^4  \,.
\nonumber
\end{eqnarray}
The notations used in (\ref{IntroClassicalR}), (\ref{IntroClassicalS}) are explained in
Section \ref{sec:Notations}. 
In section \ref{sec:GCYBE} we will study the consistency conditions 
for $r$ and $s$, which generalize the standard classical Yang-Baxter algebra. 
At the tree level 
we will get a generalization of the  classical Yang-Baxter equations:
\begin{equation}
[(r_{12} + s_{12}), (r_{13} + s_{13})] + [(r_{12} + s_{12}), (r_{23} + s_{23})] 
+ [(r_{13} + s_{13}), (r_{23} - s_{23})] = t_{123} \,,
\end{equation}
where the RHS is essentially a gauge transformation; the explicit expression
for $t$ is (\ref{RHS}). 
Note that neither $r$ nor $s$ 
satisfy the standard classical YBE on their own, and even
  the combination $r\pm s$ 
satisfies an analogue of the cYBE only when acting on gauge invariant quantities.
Therefore we have a {\em generalization} of the classical Yang-Baxter equations with 
the {\em gauge invariance} built in. 

\subsection{Infinite Wilson lines with insertions}
\label{sec:WhatToExpect}

To explain how $r$ and $s$ enter in the description of the algebra of
transfer matrices, we have to introduce some notations.

\subsubsection{General definitions}
Consider a Wilson line with an operator insertion, shown in Fig. 
\ref{fig:InfiniteWilsonLineWithInsertion}.
For this object to be gauge invariant, we want ${\cal O}$ 
to transform under the gauge transformations 
in the representation $\rho'\otimes \rho^*$ of the gauge group
$\mathfrak{g}_{\bar{0}}\subset \mathfrak{psu(2,2|4)}$. 
We will introduce the notation ${\bf H}(\rho_1\otimes \rho_2)$ for
the space of operators transforming in the representation $\rho_1\otimes\rho_2$
of $\mathfrak{g}_0$. With this notation\footnote{If $\rho'$ is a trivial 
(zero-dimensional) representation,
then the Wilson line terminates:
\includegraphics[width=0.9in]{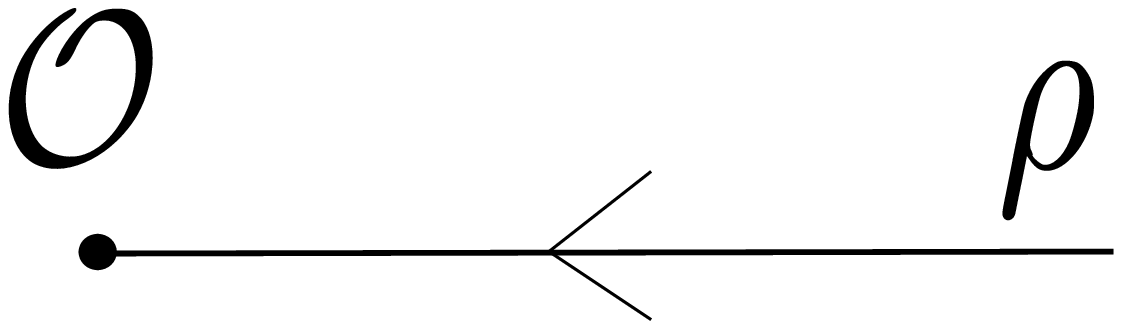}. In this case 
${\cal O}\in {\bf H}(\rho^*)$.}:
\begin{equation}
{\cal O}\in {\bf H}(\rho'\otimes \rho^*) \,.
\end{equation}
Here $\rho^*$ means the representation dual to $\rho$.

For example, we can take $\rho$  the evaluation representation of the loop
algebra corresponding to the adjoint of $\mathfrak{psu(2,2|4)}$, with some
spectral parameter $z$, and take ${\cal O}=J_{2+}$:
\begin{equation}
J_{2+}\in {\bf H}(\mbox{ad}^z\otimes (\mbox{ad}^z)^*) \,.
\end{equation}
In other words, consider:
\begin{equation}
P\exp\left(-\int_0^{+\infty} \mbox{ad}(J(z))\right)\; 
\mbox{ad}(J_{\bar{2}+}) \;
P\exp\left(-\int_{-\infty}^0 \mbox{ad}(J(z))\right) \,.
\end{equation}
This is gauge invariant because 
$\mbox{ad}\subset \mbox{ad}\otimes \mbox{ad}^*$ as a representation 
of $\mathfrak{psu}(2,2|4)$ and therefore also as a representation of 
$\mathfrak{g}_{\bar{0}}$. Of course, we could also pick 
${\cal O}=\mbox{ad}(J_{\bar{1}+})$ or 
$\mbox{ad}(J_{\bar{3}+})$.
These operators have engineering dimension $(1,0)$. Geometrically they correspond
to $\partial_+x$ or $\partial_+\vartheta$.

\begin{figure}[th]
\centerline{\includegraphics[width=3.3in]{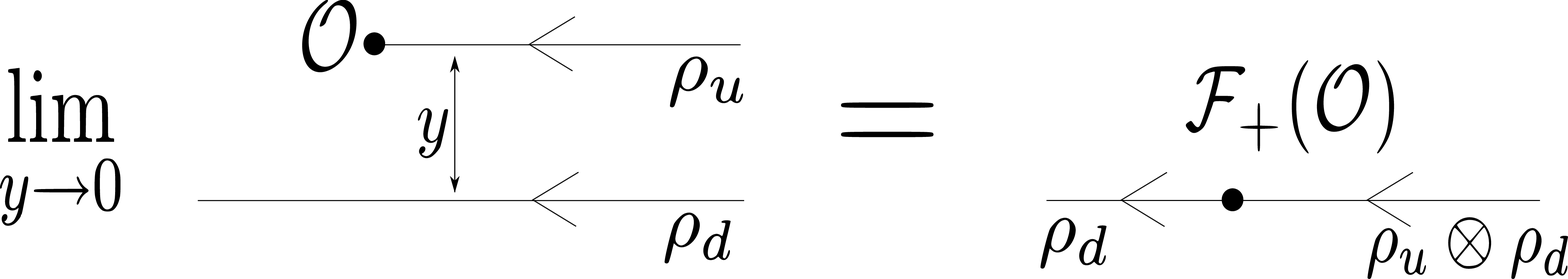}}
\vspace{0.3in}
\centerline{\includegraphics[width=3.3in]{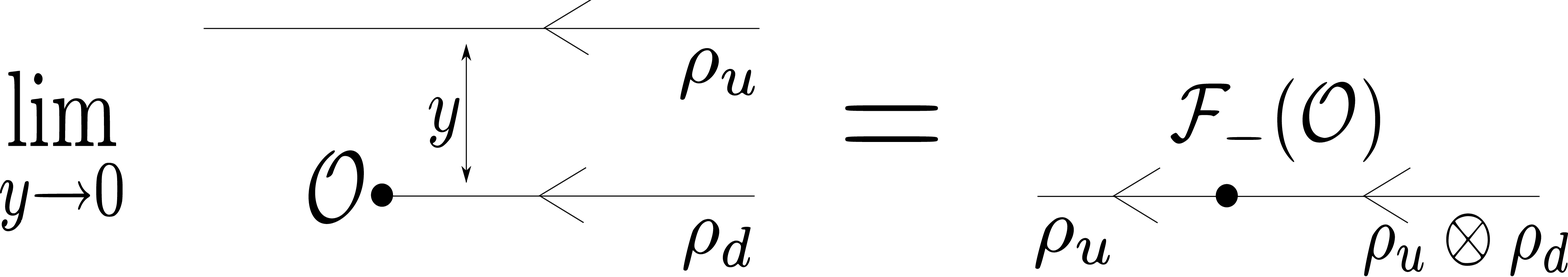}}
\vspace{0.3in}
\centerline{\includegraphics[width=3.3in]{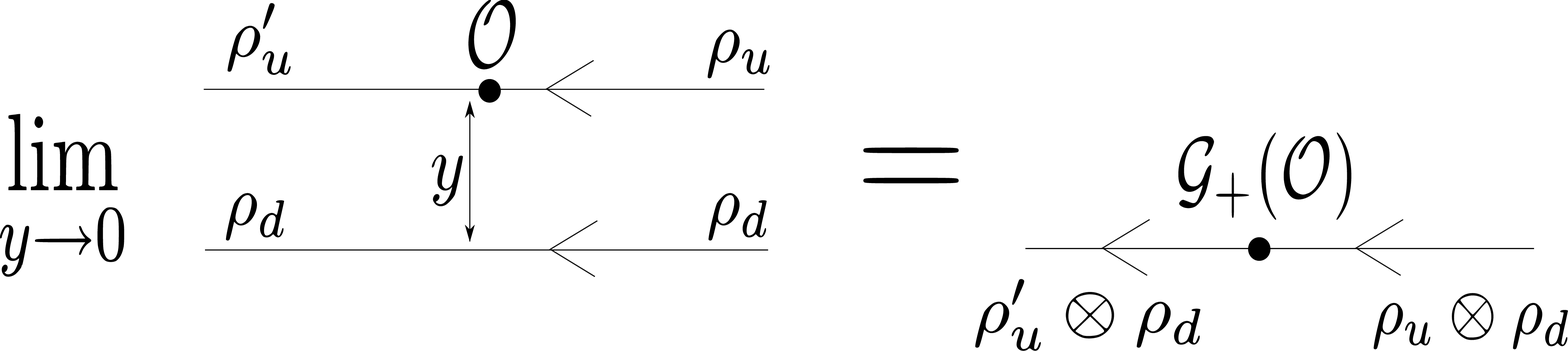}}
\caption{\small Fusion operations $\mathcal{F}_+$,  $\mathcal{F}_-$ and $\mathcal{G}_+ $\label{fig:Fusions}}
\end{figure}

We want to study the objects of this type in the situation when two contours
come close to each other.
For example, consider a Wilson line in the representation $\rho_u$ with
some operator ${\cal O}$ inserted at the endpoint. 
Let us take another Wilson line, an infinite one, carrying the representation $\rho_d$, and
put the  Wilson line with the representation $\rho_u$ on top of the the one carrying $\rho_d$. 
In the limit when the separation 
goes to zero we should  have a Wilson line carrying $\rho_u\otimes \rho_d$ at $-\infty$
and $\rho_d$ at $+\infty$.

This defines maps ${\cal F}_{\pm}$, see Figure \ref{fig:Fusions}.
If ${\cal O}$ is inserted inside the contour (rather than at the endpoint) we get
${\cal G}_{\pm}$. To summarize:
\begin{eqnarray}
{\cal F}_+ \; &:& \; {\bf H}(\rho_u^*)\to
{\bf H}(\rho_u^*\otimes \rho_d^*\otimes \rho_d)
\\
{\cal F}_- \; &:& \; {\bf H}(\rho_d^*)\to
{\bf H}(\rho_u^*\otimes \rho_d^*\otimes \rho_u)
\\
{\cal G}_+ \; &:& \; {\bf H}(\rho_u^*\otimes \rho_u')\to
{\bf H}(\rho_u^*\otimes \rho_d^*\otimes \rho_u'\otimes \rho_d)
\\
{\cal G}_- \; &:& \; {\bf H}(\rho_d^*\otimes \rho_d')\to
{\bf H}(\rho_u^*\otimes \rho_d^*\otimes \rho_u\otimes \rho_d') \,.
\end{eqnarray}

\subsubsection{Split operators}
\label{sec:SplitExchange}
We also want to be able to insert two operators:
 ${\cal O}^i_{up}$ into the upper line, 
and ${\cal O}^j_{dn}$ into the lower line, 
such that they are not separately gauge invariant, but 
$\sum_i {\cal O}^i_{up} {\cal O}^i_{dn}$ is gauge invariant. For example,
for a gauge invariant operator ${\cal O}$
we can insert ${\cal C}^{\mu\nu}
t^2_{\mu}\otimes \{t^2_{\nu},{\cal O}\}$ where 
${\cal C}_{\mu\nu}={\cal C}_{\mu\nu} (x_{up},x_{dn}, \vartheta_{up},\vartheta_{dn})$ 
is some kind of a parallel transport. This will be gauge
invariant. We will use a thin vertical line to denote such a ``split operator''

\vspace{10pt}

\hbox to \linewidth{\hfill \includegraphics[width=1.2in]{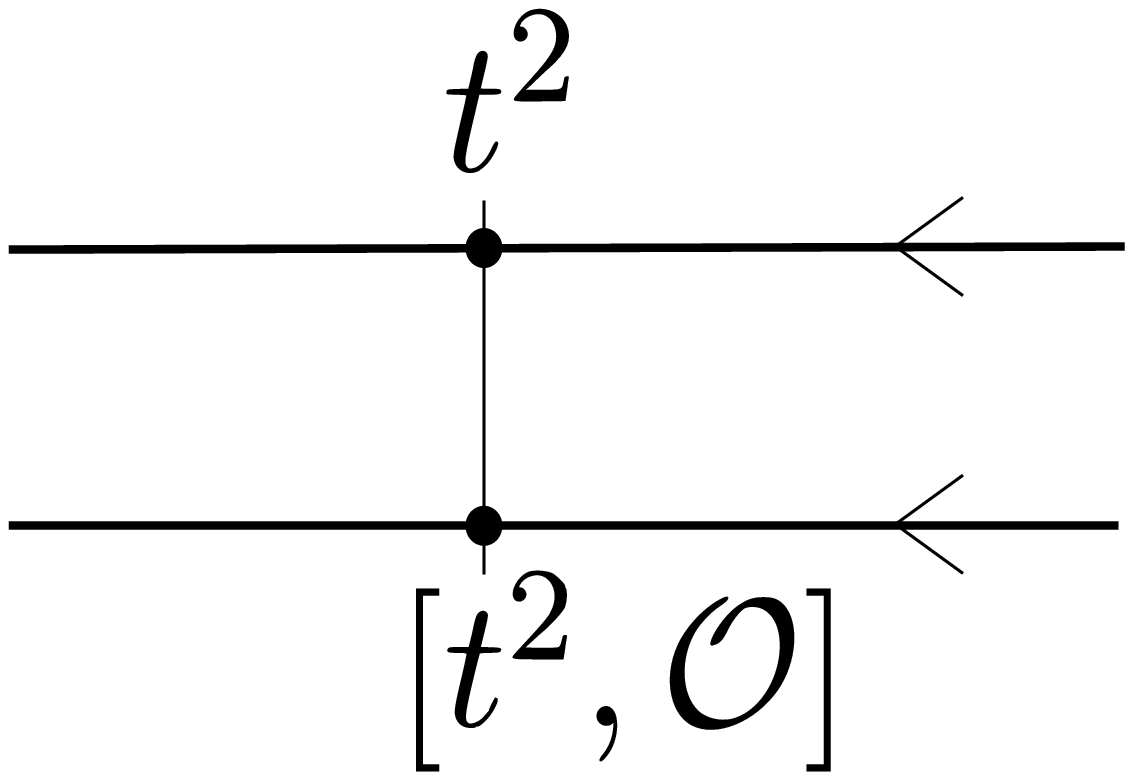} \hfill}

\noindent
In the tensor product notations, for example when we write ${\cal C}^{\mu\nu}
t^2_{\mu}\otimes \{t^2_{\nu},{\cal O}\}$, we assume that the first 
tensor generator in the tensor product (in this case $t^2_{\mu}$) acts
on the upper Wilson line, and the second (in this case $\{t^2_{\nu},{\cal O}\}$)
on the lower line.
We will need such operators in 
the limit where the upper contour approaches the lower contour. 
Strictly speaking the split operator will depend on which parallel transport
is used even in the limit of coinciding contours, by the mechanism similar
to what we described in Section \ref{sec:TheProductOfTwoT}. 
We will not discuss this dependence in this paper, because it is not important
at the tree level.

\noindent
The {\em exchange map} ${\cal R}$ acts as follows:
\begin{equation}
{\cal R}\;:\; 
{\bf H}_{split}
(\rho_1^{out}\otimes (\rho_1^{in})^* , \rho_2^{out}\otimes (\rho_2^{in})^*)
\to
{\bf H}_{split}
(\rho_2^{out}\otimes (\rho_2^{in})^* , \rho_1^{out}\otimes (\rho_1^{in})^*) \,.
\end{equation}
The pictorial representation of $\cal R$ is:

\vspace{7pt}

\hbox to \linewidth{\hfill \includegraphics[width=3.6in]{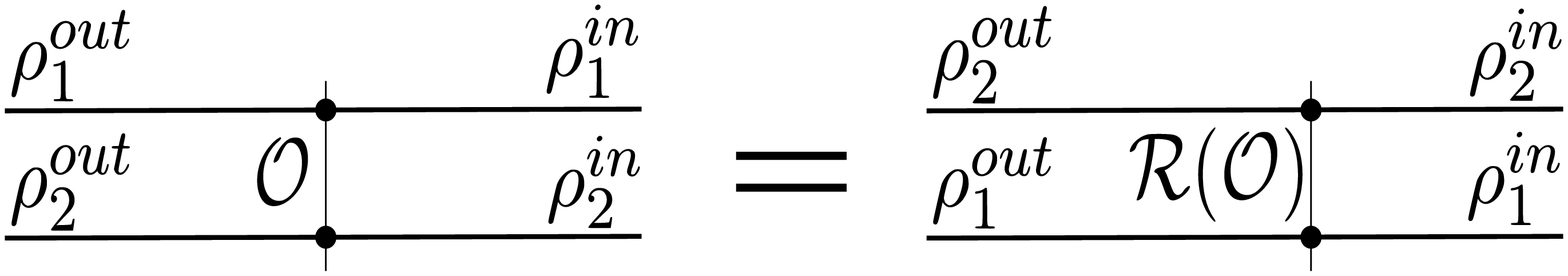} \hfill}

\subsubsection{Switch operators}
\label{sec:ExplicitExpressionsForGandR}
Given $\rho$ a representation of $\mathfrak{psu(2,2|4)}$ we denote the
evaluation representation $\rho^z$. Consider $\rho_u=\rho^{z_u^{in}}$,
$\rho'_u=\rho^{z_u^{out}}$ and $\rho_d=\rho^{z_d}$, where 
$z_u^{in}$, $z_u^{out}$ and $z_d$ are three different complex numbers.
Take ${\cal O}=1$. This is gauge invariant because
$\rho^{z_u^{in}}$ and $\rho^{z_u^{out}}$ are equivalent as representations
of the gauge group $\mathfrak{g}_0$. We can think of such ${\cal O}$ as
``the operator changing the spectral parameter'', or the ``switch operator'' 

\vspace{10pt}
\noindent
For abbreviation we write
$\rho^{in}_u=\rho^{z_u^{in}}$ and $\rho^{out}_u=\rho^{z_u^{out}}$.
 Let us first consider the operation $\mathcal{G}_+$ 
in Figure \ref{fig:Fusions}, with $\mathcal{O}=1$.  
In Section \ref{sec:StructureOfG} we will show that 
${\cal G}_+({\bf 1})$ is given (at the tree level) by this formula:
\begin{eqnarray}\label{FormulaForGPlus}
{\cal G}_+(1) & = & 
{\bf 1} + {\pi i\over 2}\left[ (r+s)|_{\rho_u^{in}\otimes {\rho_d}} -
(r+s)|_{\rho_u^{out}\otimes {\rho_d}} \right] + \ldots
\end{eqnarray}
Here the $r$ matrix appears from the diagrams involving
the interaction of currents in the bulk of the contours. 
It comes from the deformed coproduct, see Eq. (\ref{Xdefinition}).
The matrix $s$ comes from the diagrams which are localized
near the insertion of ${\cal O}$. These are the additional
diagrams existing because we inserted the impurities.

The corresponding exchange relation is:
\vspace{7pt}

\hbox to \linewidth{\hfill%
\includegraphics[width=2.9in]{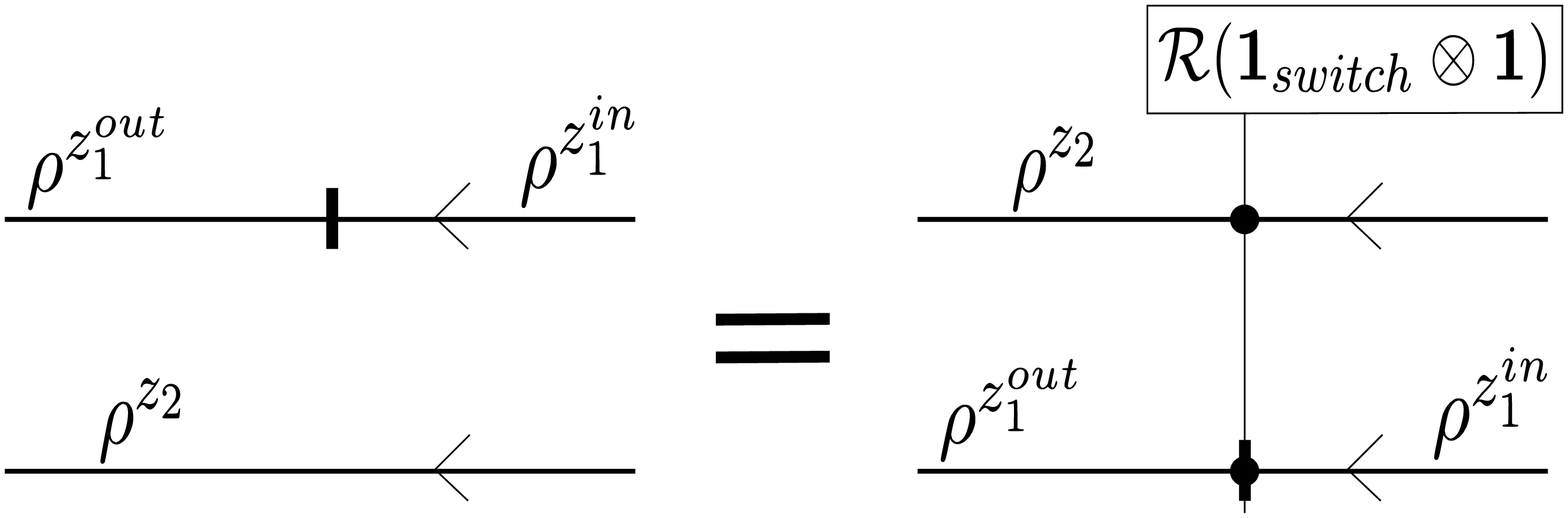} \hfill}

\noindent
where
\begin{eqnarray}
{\cal R}({\bf 1}_{switch}\otimes {\bf 1}) & = &
1 + \pi i \; r_+(z_{up}^{in},z_{dn})
  - \pi i \; r_+ (z_{up}^{out},z_{dn}) +\ldots
\label{ExchangeUp}
\\
r_+ & = & r+s \,.
\nonumber
\end{eqnarray}
Similarly, if we lift the switched contour from the lower position to the upper
position, we should insert ${\cal R}({\bf 1} \otimes {\bf 1}_{switch})$:
\vspace{7pt}

\hbox to \linewidth{\hfill%
\includegraphics[width=2.9in]{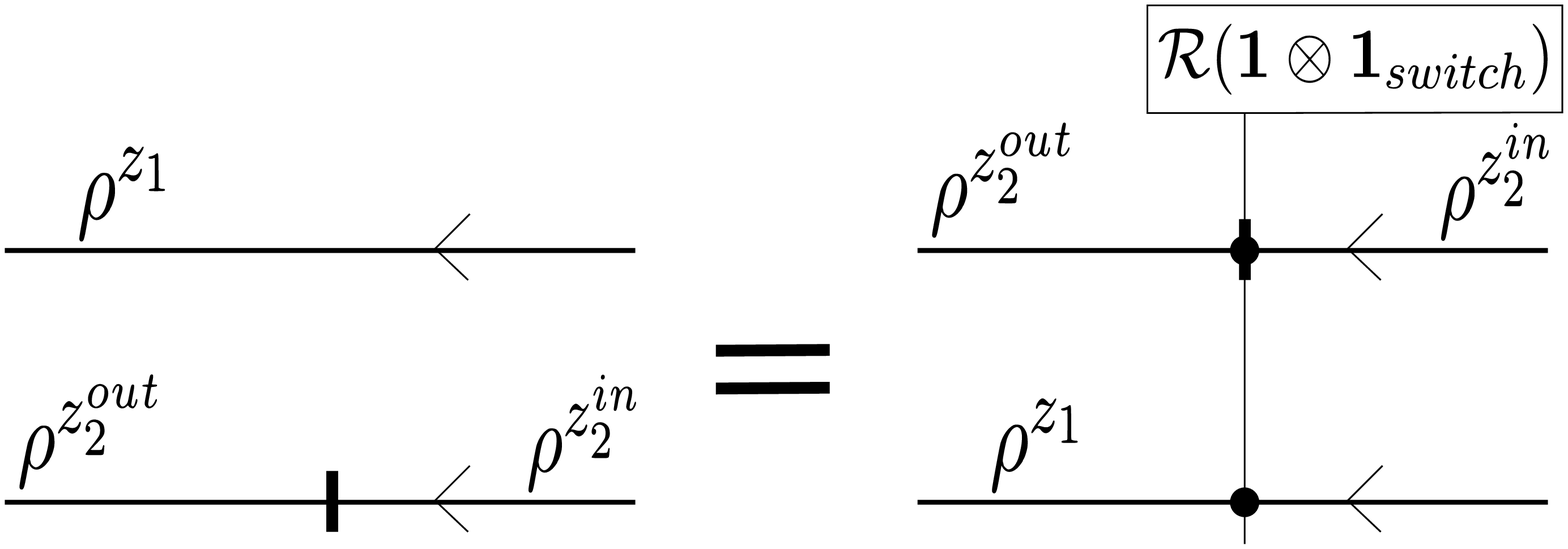} \hfill}

\begin{eqnarray}
{\cal R}({\bf 1}\otimes {\bf 1}_{switch}) & = &
1 + \pi i \; r_-(z_{up}^{in},z_{dn}) 
  - \pi i \; r_- (z_{up}^{out},z_{dn}) +\ldots
\label{ExchangeDn}
\\
r_- & = & r-s \,.
\nonumber
\end{eqnarray}
It is useful to write down explicit formulas for $r_{\pm}=r\pm s$ following from
(\ref{IntroClassicalR}) and (\ref{IntroClassicalS}):
\begin{eqnarray}
\left.{r+s\over 2}\right|_{\rho_u\otimes\rho_d} 
& = & {1\over z_u^4-z_d^4} \left[ (z_d^2-z_d^{-2})^2
(z_uz_d^3 t^1\otimes t^3 + z_u^2z_d^2 t^2\otimes t^2 + z_u^3z_d t^3\otimes t^1)
+ \right.
\nonumber \\
&&  \phantom{2\over z_u^4-z_d^4} \left. +
z_u^2z_d^2(z_u^2-z_u^{-2})(z_d^2-z_d^{-2})t^0\otimes t^0\right]   \,,  \label{rPluss}
\\[5pt]
\left.{r-s\over 2}\right|_{\rho_u\otimes\rho_d} 
& = & {1\over z_u^4-z_d^4} \left[ (z_u^2-z_u^{-2})^2
(z_uz_d^3 t^1\otimes t^3 + z_u^2z_d^2 t^2\otimes t^2 + z_u^3z_d t^3\otimes t^1)
+ \right.
\nonumber \\
&&  \phantom{2\over z_u^4-z_d^4} \left. +
z_u^2z_d^2(z_u^2-z_u^{-2})(z_d^2-z_d^{-2})t^0\otimes t^0\right]   \,.
\label{rMinuss}
\end{eqnarray}
We will use the notation 
\begin{eqnarray}
R_+ & = & {\cal R}({\bf 1}_{switch}\otimes {\bf 1})
\\
R_- & = & {\cal R}({\bf 1}\otimes {\bf 1}_{switch}) \,.
\end{eqnarray}


\subsubsection{Intersecting Wilson lines}
In this paper we mostly consider exchange and fusion as relations in the 
algebra generated by transfer matrices with insertions. It is also possible
to think of these operations as defining vertices connecting several Wilson
lines in different representations. For example the fusion can be thought of 
as a triple vertex:
\vspace{7pt}

\hbox to \linewidth{\hfill \includegraphics[width=1.4in]{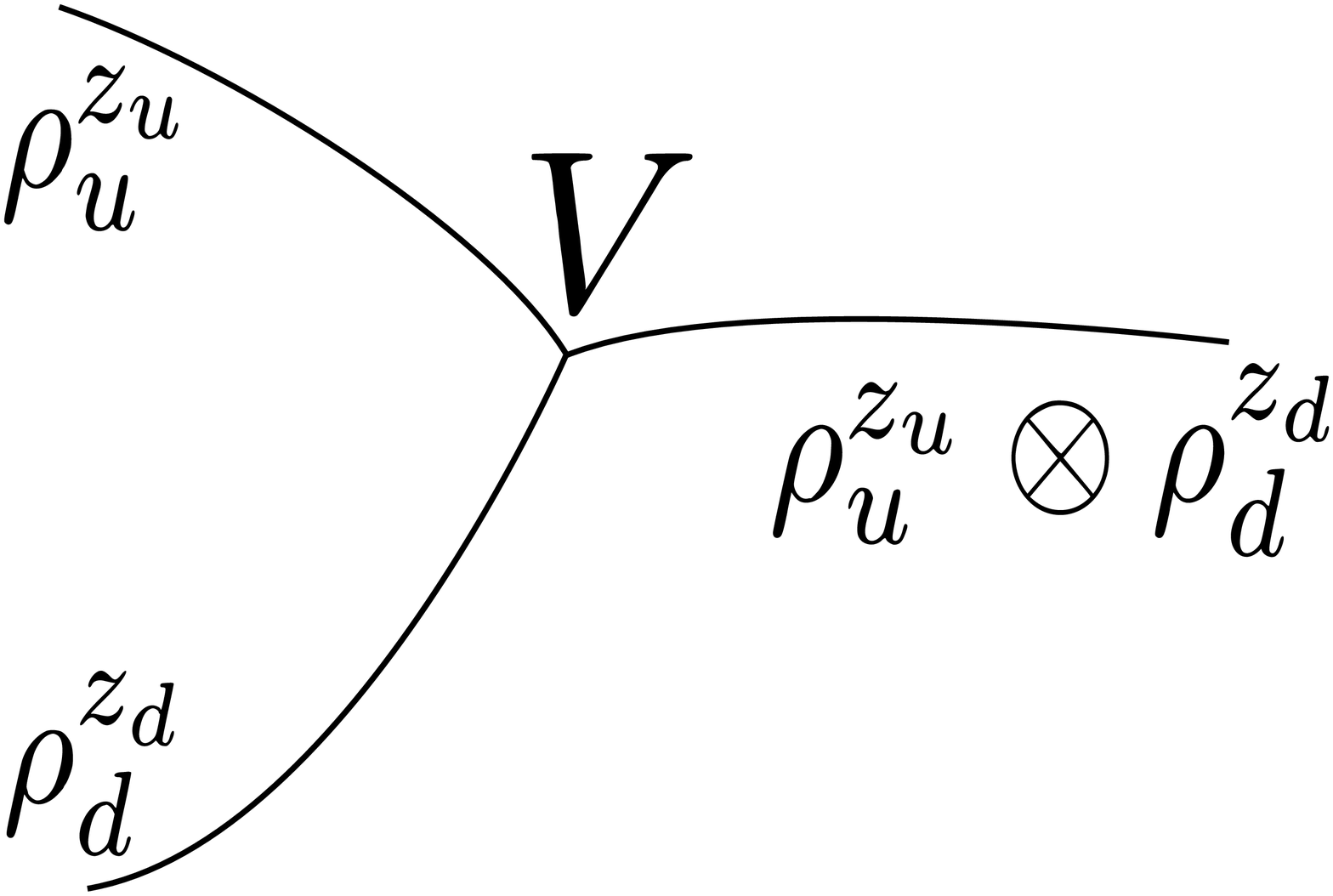} \hfill}

\noindent
Such vertices will become important if we want to consider networks
of Wilson lines.
We want to define this triple vertex so that the diagram is indepependent
of the position of the vertex, just as it is independent of the
shape of the contours. At the tree level we suggest the following prescription:
\vspace{7pt}

\hbox to \linewidth{\hfill \includegraphics[width=3.0in]{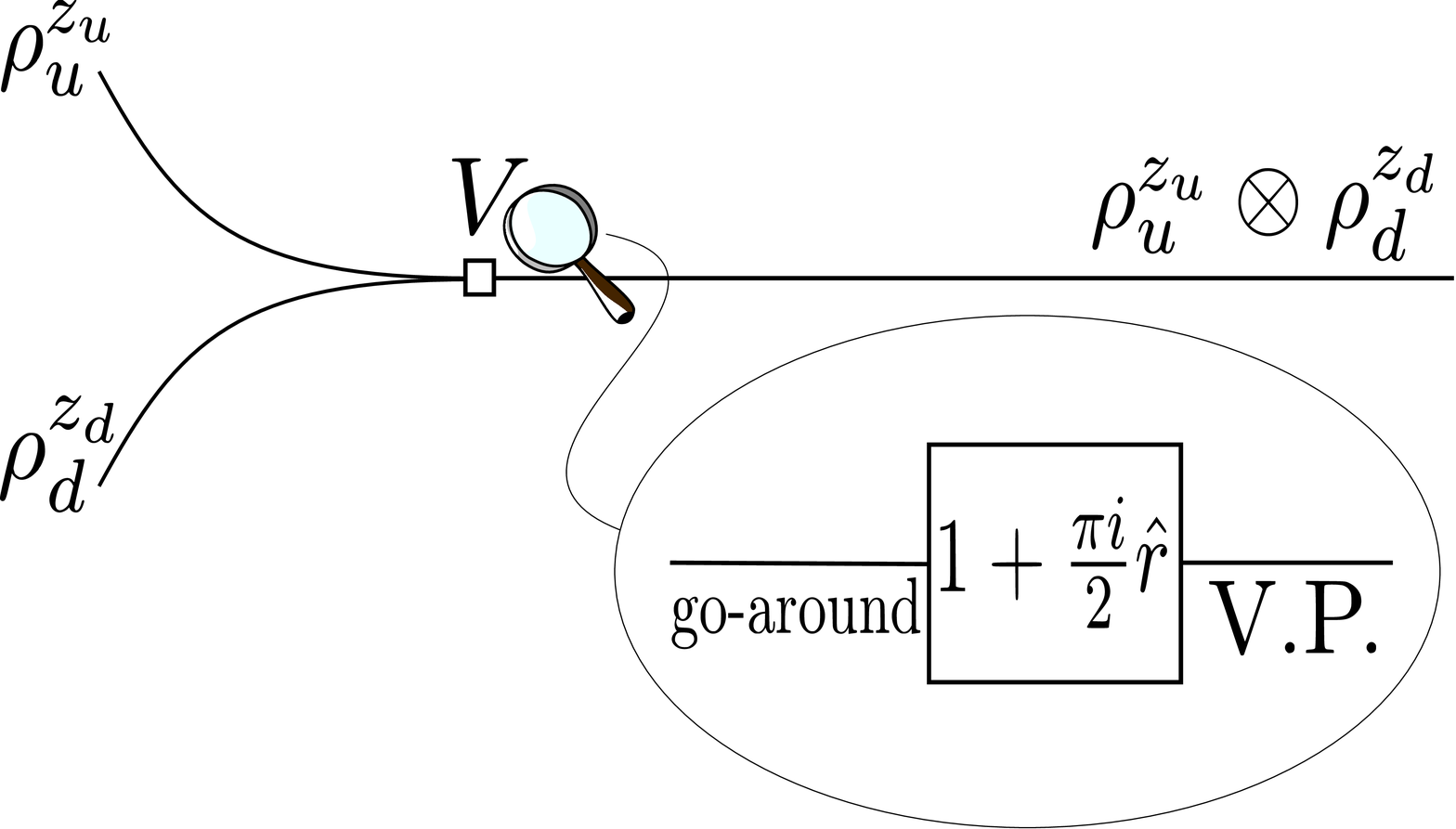} \hfill}

\noindent
The subscripts ``go-around'' and ``V.P.'' require explanation. 
They indicate different prescriptions for dealing with the collisions
of the currents coupled to $t\otimes 1$ with the currents 
coupled to $1\otimes t$. Suppose that we consider the integral
$\int dw \;J_a \;t^a\otimes 1$ and the integration contour
has to pass through several insertions of $J_b \; 1\otimes t^b$.
The prescription is such that to the right of the point $V$ 
we treat the collision as the principal value integral, while
to the left of $V$ the contour for $\int dw (J^a t_a)\otimes 1$ 
it goes around the singularity in the upper half-plane:
\vspace{7pt}

\hbox to \linewidth{\hfill \includegraphics[width=3.0in]{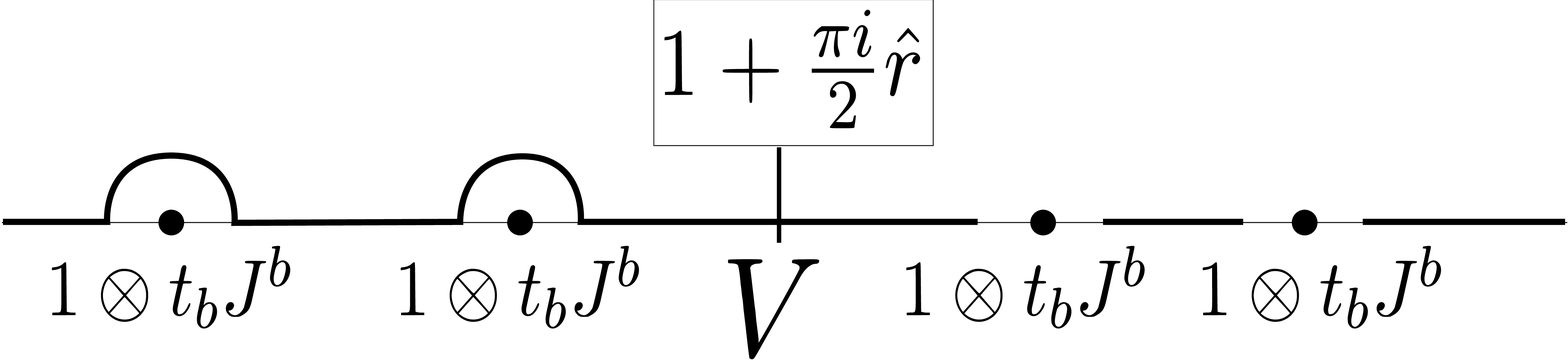} \hfill}

\noindent
The insertion of $1+{\hat{r}\over 2}$ is necessary to have
independence of the position of the vertex $V$.
Notice that in defining the worldsheet fusion we use
$r$ rather than $r+s$ or $r-s$. This is different from the 
formula (\ref{FormulaForGPlus}) for ${\cal G}_+$ which uses $r+s$.

\subsection{Outline of the calculation}

\subsubsection{Use of flat space limit}
\label{sec:UndeformableT}
We will use the near flat space expansion of $T[C+y]T[C]$, see 
Section \ref{sec:NearFlatSpace}. For our calculation it is important
that the transfer matrix is undeformable. The definition given by 
Eqs. (\ref{DefTransferMatrix}), (\ref{APlusZ}) and (\ref{AMinusZ}) cannot
be modified in any essential way. More precisely, we will use the
following statement. Suppose that there is another definition of the contour
independent Wilson line of the form 
\begin{equation}
T^{new}=P\exp\left( -\int_C I^a e_a \right) \,,
\end{equation}
where the new currents $I$ have ghost number zero and 
coincide with $J$ at the lowest order in the
near flat space expansion. In other words:
\[
I_{0\pm}=0+\ldots\;,\;\;
I_{1\pm} = -{1\over R}\partial_{\pm}\vartheta_R+\ldots \;,\;\;
I_{2\pm} = -{1\over R}\partial_{\pm}x+\ldots \;,\;\;
I_{3\pm} = -{1\over R}\partial_{\pm}\vartheta_L+\ldots
\]
where dots denote the terms of the order $1\over R^2$ or higher. Let us also
require that $T^{new}$ is invariant (up to conjugation) under the global
symmetries including the shifts (\ref{GlobalShift}). Then
\begin{equation}\label{Undeformable}
(T^{new})_A^B=\exp(\varphi(A)) T \exp(-\varphi(B)) \,,
\end{equation}
 where $\varphi(w,\bar{w})$ is a power series
in $x$ and $\vartheta$ with zero constant term. 
Eq. (\ref{Undeformable}) says that the transfer
matrix is an undeformable object.

\subsubsection{Derivation of $\hat{r}$}
 We will start in 
Section \ref{sec:CalculationOfDelta} by calculating the couplings
of $d_{\pm}x$ and $d_{\pm}\vartheta$. These are the standard couplings of the form
$R^{-1}d_{\pm}x^{\mu}(t^2_{\mu}\otimes 1 + 1\otimes t^2_{\mu})$ plus corrections
proportional to $R^{-3}d_{\pm}x$ arising as in Section \ref{sec:TheProductOfTwoT}. 
These couplings are defined up to total derivatives, {\it i.e.} up to the couplings
of $dx$. In particular, a different prescription for the order of integrations
would add a total derivative coupling. It will turn out that with one particular
choice of the total derivative terms the coupling is of the form
\begin{equation}\label{LinearCouplings}
\exp\left({\pi i\over 2}r\right) \left[ dx^{\mu} (t^2_{\mu}\otimes 1 + 1\otimes t^2_{\mu})
+ d\theta^{\alpha}_L(t^3_{\alpha}\otimes 1 + 1\otimes t^3_{\alpha})
+ d\theta^{\dot{\alpha}}_R(t^1_{\dot{\alpha}}\otimes 1 + 1 \otimes t^1_{\dot{\alpha}})
\right] \exp\left(-{\pi i\over 2}r\right) \,.
\end{equation}
where $r$ is the c-number matrix defined in Eq. (\ref{IntroClassicalR}).
These total derivative terms are important, because they
correspond to the field dependence of $\hat{r}$ in (\ref{IntroRHat}).
The same prescription for the total derivatives gives 
the right couplings
for $[x,d_{\pm}x]$ and $[\vartheta,d_{\pm}\vartheta]$ 
(Sections \ref{sec:AsymmetryXdX}, \ref{sec:AsymmetricCouplingsZuZd} and
\ref{sec:AsymmetryThetadTheta}).
The best way to fix the total derivatives in our approach is
by looking at the effects of the global shift symmetry (\ref{GlobalShift})
near the boundary, as we do in Section \ref{sec:BoundaryAndShift} 
deriving (\ref{XDependentTerms}).

According to Section \ref{sec:UndeformableT} Eq. (\ref{LinearCouplings}) implies
that: 
\begin{equation}
\lim_{y\to 0}T_{\rho_2}[C+y]T_{\rho_1}[C] = 
\exp(\varphi(A)) \exp\left({\pi i\over 2}r\right) 
T_{\rho_1\otimes \rho_2}[C]
\exp\left(-{\pi i\over 2}r\right) \exp(-\varphi(A)) \,.
\end{equation}
The right hand side is $e^{{\pi i\over 2}\hat{r}(A)}
T_{\rho_1\otimes \rho_2}[C]e^{-{\pi i\over 2}\hat{r}(B)}$, the difference
between $r$ and $\hat{r}$ is due to the field dependent gauge
transformation with the parameter $\varphi$.

\subsubsection{Boundary effects and the matrix $s$}
We then proceed to the study of the boundary effects and derive
the exchange relations for the simplest gauge invariant insertion
--- the switch operator, see Eqs. (\ref{ExchangeUp}) and
(\ref{ExchangeDn}). The matrix $s$ given by Eq. (\ref{IntroClassicalS})
arises from the diagrams localized on the insertion of the switch operator.

\subsubsection{Dynamical {\it vs.} c-number}
The $r$ and $s$ matrices appearing in the description of the exchange
relations are generally speaking field dependent, and in our
approach they are power series in $x$ and $\vartheta$. 
These series depend on which insertions we exchange, although the
leading c-number term in $\hat{r}$ given by (\ref{IntroClassicalR}) 
should be universal. For the exchange of the switch operator
we claim that $r$ and $s$ entering Eqs. (\ref{FormulaForGPlus}),
(\ref{ExchangeUp}) and (\ref{ExchangeDn}) are exactly c-number
matrices given by (\ref{rPluss}) and (\ref{rMinuss}). In other words,
all the field dependent terms cancel out. The argument based on the invariance
under the global shift symmetry is given in Section \ref{sec:StructureOfG}.

\subsubsection{BRST transformation}
The action of $Q$ on the switch operator is the insertion of 
$ (-)^F\left( {1\over z^{out}} - {1\over z^{in}} \right) \lambda $.
The consistency of this action with the exchange relation 
is verified in Section \ref{sec:BRST}.

\section{Short distance singularities in the product of currents}
\label{sec:ShortDistance}

\subsection{Notations for generators and tensor product}
\label{sec:Notations}

Recall that the notations for generators of $L \mathfrak{psu} (2,2|4)$ is
\begin{equation}
e^{-3}_{\alpha}=z^{-3}t^3_{\alpha}, \qquad
e^{-2}_{\mu}=z^{-2}t^2_{\mu}, \qquad
e^1_{\alpha}=zt^3_{\alpha}  \,.
\end{equation}
The collective notations for the generators of $\mathfrak{psu}(2,2|4)$ are:
\begin{equation}\label{CollectiveNotations}
t^i_a\qquad i\in \mathbb{Z}_4 \,, \quad a\in\{\dot{\alpha},\mu,\alpha,[\rho\sigma]\} \,.
\end{equation}
The coproduct for superalgebra involves the operator $(-1)^F$, which has
the property $(-1)^Ft_{\alpha}^3=-t_{\alpha}^3(-1)^F$,
see (\ref{Coproduct}). 
The origin of $(-)^F$ can be understood from this example: 
\begin{eqnarray}
e^{\psi_1(t\otimes 1)} e^{\psi_2(t'\otimes 1)} e^{\psi_3(t''\otimes 1)}\;\;
e^{\psi_1(1\otimes t)} e^{\psi_2(1\otimes t')} e^{\psi_3(1\otimes t'')}\;\;|0>\otimes|0> 
=
\\
=
e^{\psi_1(t\otimes 1 + (-)^F\otimes t)}
e^{\psi_2(t'\otimes 1 + (-)^F\otimes t')}
e^{\psi_3(t''\otimes 1 + (-)^F\otimes t'')}|0 \rangle\otimes |0\rangle \,,
\end{eqnarray}
where $\psi_{1,2,3}$ are three Grassman variables and $t,t',t''$ three generators
of some algebra, acting on the representation generated by a vector $|0\rangle $,
where $(-)^F|0\rangle =|0\rangle $, $(-)^Ft|0\rangle =-t|0\rangle $, $(-)^Ft't|0\rangle =t't|0\rangle $ {\it etc.}

When we write the tensor products we will omit $(-)^F$ for the purpose of
abbreviation. For example:
\begin{eqnarray}
1\otimes t^3_{\alpha} & \mapsto & (-)^F\otimes t^3_{\alpha}
\\
t^3_{\alpha}\otimes 1& \mapsto & t^3_{\alpha} \otimes 1
\\
1\otimes 1\otimes t^3_{\alpha} & \mapsto & (-)^F\otimes (-)^F\otimes t^3_{\alpha}
\\
1\otimes t^3_{\alpha}\otimes 1& \mapsto & (-)^F\otimes t^3_{\alpha} \otimes 1
\\
t^3_{\alpha}\otimes 1\otimes 1& \mapsto & t^3_{\alpha}\otimes 1\otimes 1
\\
t^3_{\alpha}\otimes t^3_{\beta} & \mapsto & t^3_{\alpha}(-)^F\otimes t^3_{\beta}
\end{eqnarray}
Generally speaking 
$1\otimes 1\otimes\ldots\otimes 1\otimes t^j_a \otimes 1 \otimes \ldots \otimes 1$
means:
\begin{equation}
(-)^{jF}\otimes (-)^{jF}\otimes\ldots\otimes (-)^{jF}\otimes  
t^j_a \otimes 1 \otimes \ldots \otimes 1 \,.
\end{equation}
With these notations we have:
\begin{equation}
(t^3_{\alpha}\otimes 1) (1\otimes t^3_{\beta}) =
- (1\otimes t^3_{\beta}) (t^3_{\alpha}\otimes 1)
= t^3_{\alpha}\otimes t^3_{\beta} \,.
\end{equation}
We also use the following abbreviations:
\begin{eqnarray}
&& e^{-1}_{\alpha}\otimes e^2_{\mu} = 
(z^{-1}t^3_{\alpha})\otimes (z^2 t^2_{\mu}) =
z_u^{-1}z_d^2 \; t^3_{\alpha}\otimes t^2_{\mu}
\\[5pt]
&& e^{-1}_{\alpha}\wedge e^2_{\mu} = 
{1\over 2} (e^{-1}_{\alpha}\otimes e^2_{\mu}
-e^2_{\mu} \otimes e^{-1}_{\alpha})
\\[5pt]
&& e^{-1}_{\alpha}\wedge e^1_{\dot{\beta}} =
{1\over 2} (e^{-1}_{\alpha}\otimes e^1_{\dot{\beta}} +
 e^1_{\dot{\beta}} \otimes e^{-1}_{\alpha}) \,.
\end{eqnarray}
When we write Casimir-like combinations of generators, we
often omit the Lie algebra index:
\begin{eqnarray}
t^1\otimes t^3 & = & C^{\dot{\alpha}\alpha} t^1_{\dot{\alpha}}\otimes t^3_{\alpha}
\nonumber
\\
t^3\otimes t^1 & = & C^{\alpha\dot{\alpha}} t^3_{\alpha}\otimes t^1_{\dot{\alpha}}
\nonumber
\\
t^2\otimes t^2 & = & C^{\mu\nu} t^2_{\mu}\otimes t^2_{\nu}
\nonumber
\\
t^0\otimes t^0 & = & C^{[\mu\nu][\rho\sigma]} t^0_{[\mu\nu]}\otimes t^0_{[\rho\sigma]} \,.
\end{eqnarray}
We will also use this notation:
\begin{equation}\label{NotationTTT}
t^i\otimes t^j\otimes t^k = f_{a'b'c'} C^{a'a} C^{b'b} C^{c'c}\;
t^i_a\otimes t^j_b\otimes t^k_c \,,
\end{equation}
where
\begin{equation}
f_{abc}=\fddu{a}{b}{c'}C_{c'c}=\mbox{Str}([t_a,t_b]t_c) \,.
\end{equation}
For example:
\begin{equation}
t^3\otimes t^1\otimes t^0 = f_{\dot{\alpha}\beta[\mu\nu]}
C^{\dot{\alpha}\alpha} C^{\beta\dot{\beta}} C^{[\mu\nu][\rho\sigma]}
t^3_{\alpha} \otimes t^1_{\dot{\beta}} \otimes t^0_{[\rho\sigma]}
\end{equation}
Using these notations we can write, for example:
\begin{equation}
[ t^i\otimes t^{4-i}\otimes {\bf 1} \; , \;
  t^j\otimes {\bf 1}\otimes t^{4-j} ] =
(-)^{i+j+ij} t^{(i+j)mod \;4}\otimes t^{4-i}\otimes t^{4-j} \,.
\end{equation}


\subsection{Short distance singularities using tensor product notations}
Short distance singularities in the products of currents were calculated
in  \cite{Puletti:2006vb,Mikhailov:2007mr}.
Here is the table in the ``tensor product'' notations:
\begin{eqnarray}
J_{1+}\otimes J_{2+} &=& - {1\over w_u-w_d}
t^1\otimes \{ t^3,\partial_+\vartheta_L\}
\nonumber
\\
J_{3+}\otimes J_{2+} &=& - {2\over w_u-w_d}
t^3\otimes \{t^1,\partial_+\vartheta_R\} -
{\overline{w}_u-\overline{w}_d\over (w_u-w_d)^2}
t^3\otimes \{t^1,\partial_-\vartheta_R\}
\nonumber
\\
J_{1+}\otimes J_{1+} &=& - {1\over w_u-w_d}
t^1\otimes [ t^3,\partial_+ x]
\nonumber
\\
J_{3+}\otimes J_{3+} &=& - {2\over w_u-w_d}
t^3\otimes [t^1,\partial_+ x] -
{\overline{w}_u-\overline{w}_d\over (w_u-w_d)^2}
t^3\otimes [t^1,\partial_- x]
\nonumber
\\
J_{0+}\otimes J_{1+} &=& -{1/2 \over w_u-w_d}
t^0\otimes [t^0, \partial_+\vartheta_R] 
-{1/2 \over (w_u-w_d)^2} 
t^0\otimes [t^0, \vartheta_R]
\nonumber
\\
J_{0+}\otimes J_{3+} &=& -{1/2\over w_u-w_d}
t^0\otimes [t^0, \partial_+\vartheta_L]
-{1/2 \over (w_u-w_d)^2} 
t^0\otimes [t^0, \vartheta_L]
\nonumber
\\
J_{1-}\otimes J_{2+} &=& -{1\over w_u-w_d} 
t^1\otimes \{t^3,\partial_-\vartheta_L\} 
\nonumber
\\
J_{1+}\otimes J_{2-} &=& -{1\over w_u-w_d}
t^1\otimes \{t^3,\partial_-\vartheta_L\}
\nonumber
\\
J_{3-}\otimes J_{2+} &=& -{1\over \overline{w}_u-\overline{w}_d}
t^3\otimes \{t^1,\partial_+\vartheta_R\}
\nonumber
\\
J_{3+}\otimes J_{2-} &=& -{1\over \overline{w}_u-\overline{w}_d}
t^3\otimes \{t^1,\partial_+\vartheta_R\}
\nonumber
\\
J_{1+}\otimes J_{1-} &=& -{1\over w_u-w_d} 
t^1\otimes \{t^3,\partial_-x\}
\nonumber
\\
J_{3+}\otimes J_{3-} &=& -{1\over \overline{w}_u-\overline{w}_d}
t^3\otimes \{t^1,\partial_+x\} \,.
\nonumber
\end{eqnarray}
Such ``tensor product notations'' are very useful and widely used in 
expressing the commutation relations of transfer matrices.
We will list the same formulas in more standard index notations
in appendix \ref{sec:IndexOPE}.


\section{Calculation of $\Delta$}
\label{sec:CalculationOfDelta}
In this section we will give the details of the calculation which was outlined 
in Section \ref{sec:TheProductOfTwoT}. 

\subsection{The order of integrations}
\label{sec:OrderOfIntegrations}
As we discussed in \cite{Mikhailov:2007mr} the intermediate calculations depend
on the choice of the order of integrations. We will use the 
symmetric prescription. This means that if we have a multiple
integral, we will average over all possible orders of integration.
For example in this picture:

\vspace{7pt}

\hbox to \linewidth{\hfill %
\includegraphics[width=2.5in]{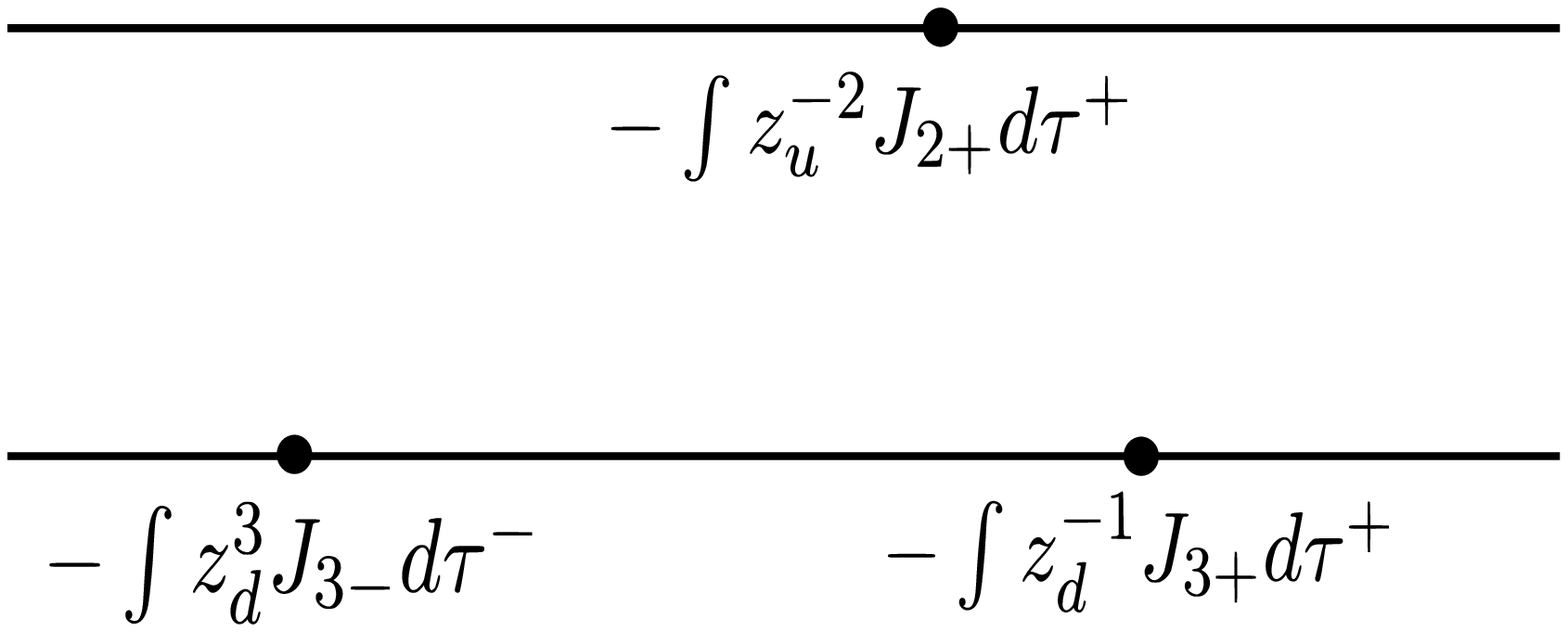} \hfill}

\noindent
we have three integrations, and therefore we average over 6 possible
ways of taking the integrals. Another prescription would give the same
answer (because after regularization the multiple integral is convergent,
and does not depend on the order of integrations), but will lead to
a different distribution of the divergences between the bulk and the boundary.


\subsection{Contribution of triple collisions to $\Delta$}
\label{sec:TripleCollisions}

Triple collisions contribute to the comultiplication because
of the double pole. Let us for example consider this triple
collision:

\vspace{7pt}

\hbox to \linewidth{\hfill %
\includegraphics[width=2.1in]{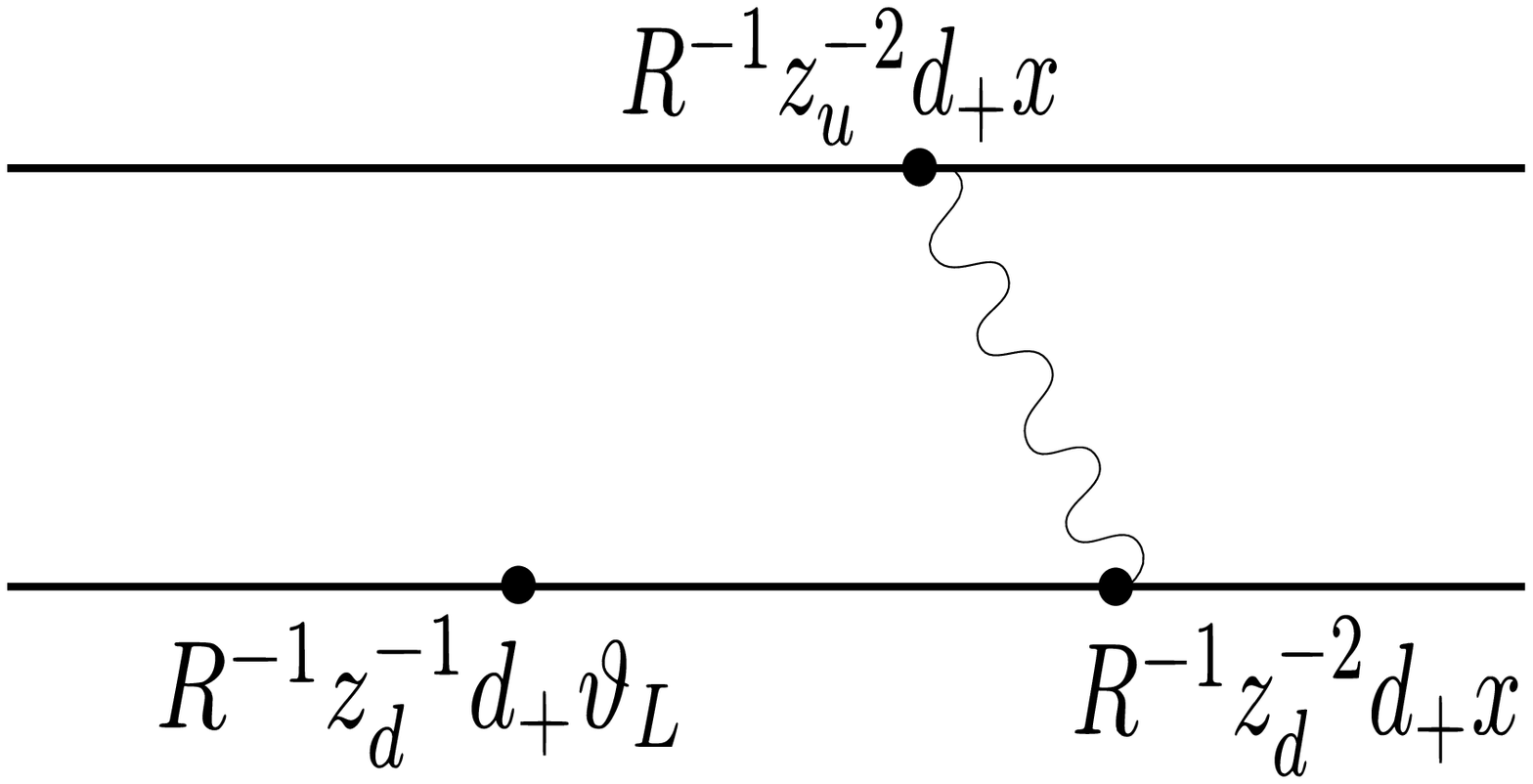} \hfill}

\noindent
Of course this is not really a collision, since only the lower
two points collide. But we still call it a ``triple collision''
This has to be compared to:

\vspace{7pt}

\hbox to \linewidth{\hfill %
\includegraphics[width=2.8in]{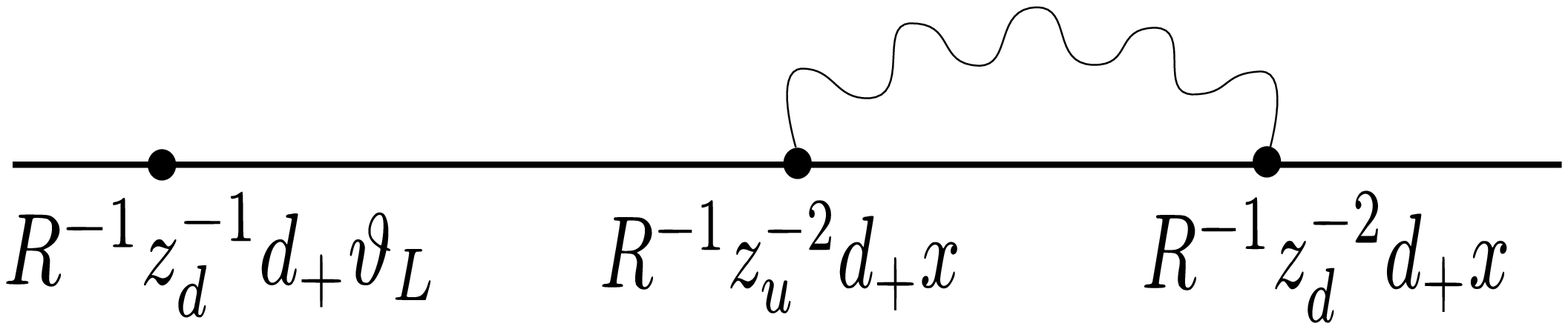} \hfill}

\noindent
where the integrals are understood in the sense of taking the principal 
value. We have to average over two ways of integrating: (1) first integrating over the position
of the $z_u^{-2}d_+x$ on the upper contour, and then $z_d^{-2}d_+x$ on the lower
contour and (2) first integrating over the position of $z_d^{-2}d_+x$ and then
integrating over the position of $z_u^{-2}d_+x$. 
The first way of doing integrations does not contribute to $\Delta$, and the second
does. Indeed, the contraction $\langle d_+ x d_+ x \rangle$ gives
$-{1\over (w_u-w_d)^2} z_u^{-2} z_d^{-2}\; t^2\otimes t^2$, and after we integrate 
over $w_d$ we get:

\vspace{7pt}

\hbox to \linewidth{\hfill %
\includegraphics[width=2.1in]{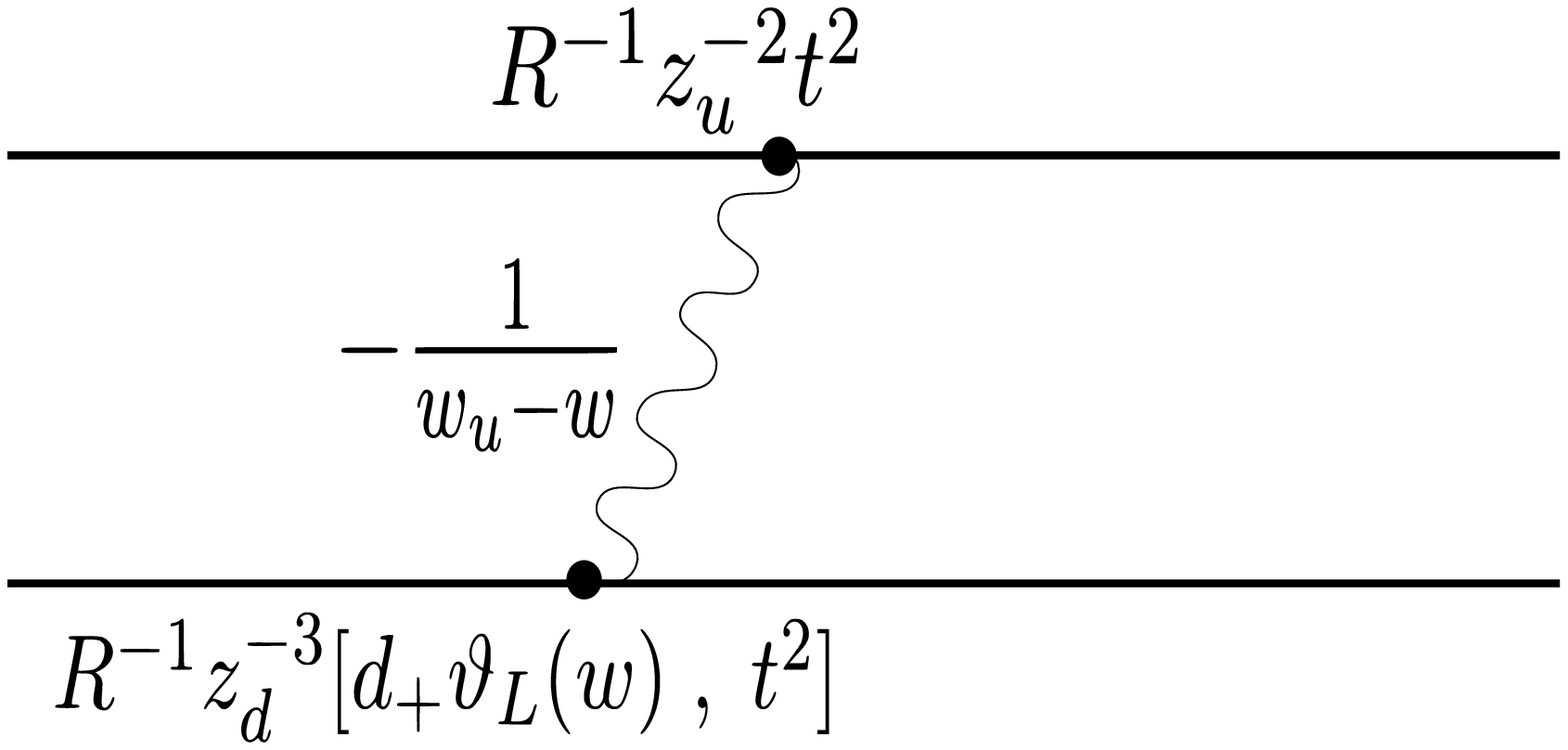} \hfill}

\noindent
Then integration over $w_u$ gives the imaginary contribution 
$\int \left(-{dw_u\over w_u-w}\right) =-\pi i$:

\vspace{7pt}

\hbox to \linewidth{\hfill %
\includegraphics[width=2.1in]{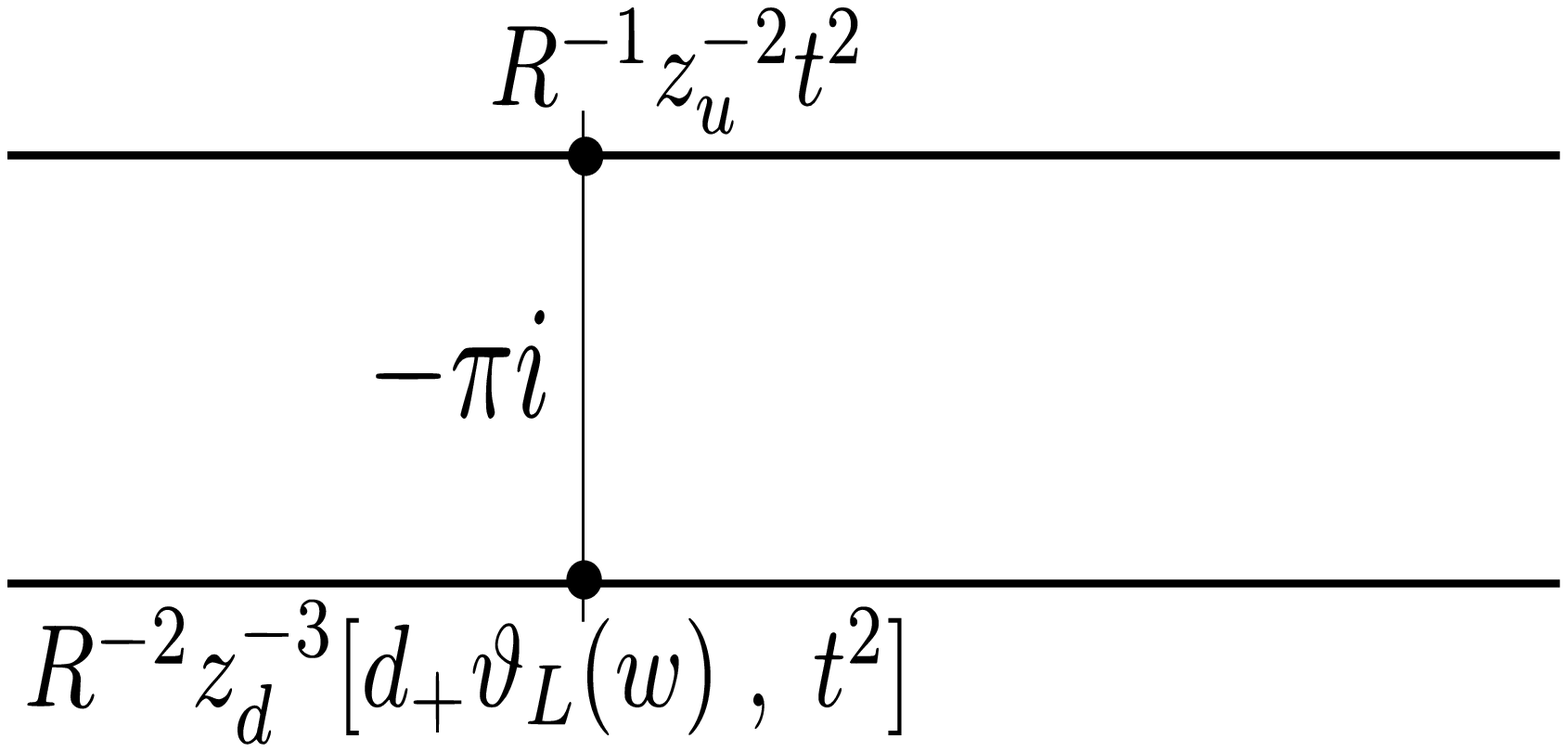} \hfill}

\noindent
The contribution from the contractions $\langle d_+\vartheta_L d_+\vartheta_R\rangle$
is similar, and the result for the contribution of triple collisions to $\Delta$ is:
\begin{equation}\label{GeneralTripleCollision}
\Delta^{triple}(e_a^m)=
\pi i{1\over 2}[C_+-C_- , 1\otimes e_a^m-e_a^m\otimes 1]   \,,
\end{equation}
where $1/2$ is because we average over two different orders of integration,
and $C_{\pm}$ is defined as
\begin{eqnarray}
C_+ & = & (z^{-1}t^3)\otimes (z^{-3}t^1) 
        + (z^{-2}t^2)\otimes (z^{-2}t^2)
        + (z^{-3}t^1)\otimes (z^{-1}t^3)
\label{CplusDef}
\\[5pt]
C_- & = & (z^3 t^3)\otimes (z t^1) 
        + (z^2 t^2)\otimes (z^2 t^2)
        + (z   t^1)\otimes (z^3 t^3) \,.
\label{CminusDef}
\end{eqnarray}
The expression (\ref{GeneralTripleCollision}) for $\Delta^{trpl}$
 should be added to $\Delta^{dbl}$ which is generated by the double collisions.
We will now calculate $\Delta^{dbl}$ and $\Delta'=\Delta^{dbl}+\Delta^{trpl}$.


\subsection{Coupling of $dx$}
\label{sec:CouplingOfX}

We have just calculated the contribution of triple collisions;
now we will discuss the contribution of double collisions and the issue
of total derivatives.

\noindent
\underline{Effect of double collisions}
\begin{eqnarray}
\mbox{Collision } &\hspace{10pt}& \mbox{contributes }\pi i \mbox{ times:}
\nonumber
\\[5pt]
 J_{1+}J_{1+} : 
&&
-z_u^{-3}z_d^{-3} \; t^1\wedge [t^3,d_+x] +
\nonumber
\\
 J_{1-}J_{1-} :
&&
+2z_u z_d \; t^1\wedge [t^3,d_- x] 
  + z_u z_d \; t^1\wedge [t^3,d_+ x] +
\nonumber
\\
 J_{3+}J_{3-} :
&&
+2z_u^{-1} z_d^3 \; t^3\wedge [t^1,d_+ x] +
\nonumber
\\
 J_{3-}J_{3-} :
&&
+z_u^3 z_d^3 \; t^3 \wedge [t^1, d_-x] -
\nonumber
\\
 J_{3+}J_{3+} :
&&
-2z_u^{-1} z_d^{-1} \; t^3 \wedge [t^1,d_+x] 
   - z_u^{-1} z_d^{-1} \; t^3 \wedge [t^1,d_-x] -
\nonumber
\\
 J_{1-}J_{1+} :
&&
-2z_u z_d^{-3} t^1\wedge [t^3,d_-x] +
\nonumber
\\
 J_{0\pm}J_{2\pm'} :
&&
+{3\over 2} (z_d^2-z_d^{-2}) [dx,t^2]\wedge t^2 \,.
\end{eqnarray}
In the calculation of the contribution of $J_{0\pm}J_{2\pm'}$
we take an average of first taking an integral over the position
of $J_{0\pm}$ and then taking an integral over the position of
$J_{2\pm'}$.  To summarize:
\begin{eqnarray}
{1\over \pi i}\Delta^{dbl}(dx) & = &
\spacemaker{+} (-z_u^{-3}z_d^{-3} + z_u z_d) t^1\wedge [t^3,d_+x] +
\nonumber 
\\
&& +(z_u^{-1}z_d^3 + z_u^3 z_d^{-1} -2 z_u^{-1}z_d^{-1})
t^3\wedge [t^1,d_+ x] +
\nonumber
\\
&& +(-z_uz_d^{-3} -z_u^{-3}z_d +2z_uz_d) 
t^1\wedge [t^3,d_- x] +
\nonumber
\\
&& +(z_u^3z_d^3-z_u^{-1}z_d^{-1}) t^3\wedge [t^1,d_-x] +
\nonumber
\\
&& + {3\over 2} (z_u^{2}-z_u^{-2}) t^2\wedge [t^2,dx] \,.
\end{eqnarray}
\underline{Effect of triple collisions:}
\begin{eqnarray}
{1\over \pi i}\Delta^{trpl}(dx) & = &
[C^+-C^-, 1\wedge (z^{-2}d_+x + z^2 d_-x)] =
\nonumber
\\
&& = (z_u^{-3}z_d^{-3} -z_u z_d) t^1\wedge [t^3,d_+ x] 
   + (z_u^{-1} z_d^{-5} -z_u^3 z_d^{-1}) t^3\wedge [t^1,d_+ x] +
\nonumber
\\
&& + (z_u^{-2} z_d^{-4}- z_u^2) t^2\wedge [t^2,d_+ x] +
\nonumber
\\
&& + (z_u^{-3} z_d - z_u z_d^5) t^1\wedge [t^3,d_- x] 
   + (z_u^{-1}z_d^{-1} - z_u^3z_d^3) t^3\wedge [t^1,d_-x] +
\nonumber
\\
&& + (z_u^{-2}-z_u^2z_d^4) t^2\wedge [t^2,d_-x]\,.
\nonumber
\end{eqnarray} 
This leads to the following expression for the total $\Delta'$:
\begin{eqnarray}
{1\over \pi i}\Delta' (dx) & = &\spacemaker{+}
{1\over 2} ((z_u^2-z_u^{-2})^2+(z_d^2-z_d^{-2})^2)
 z_u^{-1}z_d^{-1} \; t^3\wedge [t^1,d_+ x] -
\nonumber
\\
&& 
-{1\over 2} ((z_u^2-z_u^{-2})^2+(z_d^2-z_d^{-2})^2)
z_uz_d \; t^1\wedge [t^3,d_- x]
\nonumber
\\
&&
+ (z_u^{-2}z_d^{-4}-z_u^2) \;
 t^2\wedge [t^2,d_+x] +
\nonumber
\\
&&
+ (z_u^{-2}-z_u^2z_d^4) \; 
t^2\wedge [t^2,d_-x]+
\nonumber
\\
&& + {3\over 2} (z_u^{2}-z_u^{-2}) t^2\wedge [t^2,dx] \,.
\label{Coproduct_dx_UptoTotalDerivatives}
\end{eqnarray}
The calculations of this section can only fix the coupling of $d_{\pm}x$ up to
total derivatives, {\it i.e.} terms proportional to $dx=d_+x+d_-x$. Only
the terms proportional to $*dx=d_+x - d_-x$ are fixed. To fix the terms proportional
to $dx$, we have to either study the couplings of $xdx$ or look at what
happens at the endpoint of the contour. We will discuss this in 
Sections \ref{sec:FixingTotalDerivatives} and \ref{sec:BoundaryEffects}. 
The result it that the following \underline{additional coupling:}
\begin{equation}\label{AdditionalCouplingDX}
{1\over 2} (z_u^{2}-z_u^{-2}) t^2\wedge [t^2,dx] \,,
\end{equation}
\underline{should be added} to (\ref{Coproduct_dx_UptoTotalDerivatives}).


\subsection{Coupling of $d\vartheta_L$}
\label{sec:CouplingOfTheta}

Similar to the $dx$ terms, we can discuss the $d\vartheta$ coproduct.

\noindent
\underline{Effect of double collisions.}
Here is the table:
\begin{eqnarray}
\mbox{Collision } && \mbox{contributes $\pi i$ times}
\nonumber
\\[5pt]
J_{1+}J_{2+} 
&&
- 2 z_u^{-3} z_d^{-2} \; t^1\wedge \{ t^3,d_+ \vartheta_L\} +
\nonumber
\\
J_{1-}J_{2-}
&&
+ 2 z_u z_d^2 \; t^1\wedge \{ t^3, d_+ \vartheta_L\} +
  4 z_u z_d^2 \; t^1\wedge \{ t^3, d_- \vartheta_L\} -
\nonumber
\\
J_{1-}J_{2+}
&&
- 2 z_u z_d^{-2} \; t^1\wedge \{ t^3, d_- \vartheta_L\} -
\nonumber
\\
J_{1+}J_{2-}
&&
- 2 z_u^{-3} z_d^2 \; t^1\wedge \{ t^3, d_- \vartheta_L\} +
\nonumber
\\
J_0J_3
&&
+ {3\over 2} ((z^3-z^{-1})t^3)\wedge \{t^1,d\vartheta_L\}
\nonumber
\end{eqnarray}

\vspace{10pt}

\noindent
\underline{Contribution of  triple collisions}
\begin{eqnarray}
{1\over \pi i} \Delta^{trpl}(d\vartheta_L) & = &
[C^+-C^-, 1\wedge (z^{-1}d_+\vartheta_L + z^3 d_-\vartheta_L)] =
\nonumber
\\
&& 
= z_u^{-3} z_d^{-2}(1-z_u^4z_d^4) \; t^1 \wedge \{ t^3 , d_+\vartheta_L \}
+ z_u^{-2} z_d^{-3}(1-z_u^4z_d^4) \; t^2 \wedge \{ t^2 , d_+\vartheta_L \} +
\nonumber 
\\
&&
+ z_u^{-1} z_d^{-4}(1-z_u^4z_d^4)      \; t^3 \wedge \{ t^1 , d_+\vartheta_L \} +
\nonumber
\\
&&
+ z_u^{-3} z_d^2(1-z_u^4z_d^4) \; t^1 \wedge \{ t^3 , d_-\vartheta_L \}
+ z_u^{-2} z_d(1-z_u^4z_d^4) \; t^2 \wedge \{ t^2 , d_-\vartheta_L \} +
\nonumber 
\\
&&
+ z_u^{-1} (1-z_u^4z_d^4)      \; t^3 \wedge \{ t^1 , d_-\vartheta_L \} 
\nonumber
\\[5pt]
& = &
(z_u^{-3} z_d^{-2} + z_u^{-2} z_d^{-3}) (1-z_u^4z_d^4) 
\; t^1 \wedge \{ t^3 , d_+\vartheta_L \} +
\nonumber 
\\
&&
+ z_u^{-1} z_d^{-4}(1-z_u^4z_d^4)      \; t^3 \wedge \{ t^1 , d_+\vartheta_L \} +
\nonumber
\\
&&
+ (z_u^{-3} z_d^2 + z_u^{-2} z_d) (1-z_u^4z_d^4)
\; t^1 \wedge \{ t^3 , d_-\vartheta_L \} +
\nonumber
\\
&&
+ z_u^{-1} (1-z_u^4z_d^4)      \; t^3 \wedge \{ t^1 , d_-\vartheta_L \} \,.
\nonumber
\end{eqnarray}
Just as in case of the couplings of $dx$, we observe that only the couplings
proportional to $d_+x - d_-x$ are fixed by the calculation in this section.
In fact the analysis of Section \ref{sec:FixingTotalDerivatives} will show that
we have to \underline{add the following total derivative coupling}:
\begin{equation}\label{AdditionalCouplingDTheta}
 (1/2) ((z^3-z^{-1})t^3)\wedge \{t^1,d\vartheta_L\} \,.
\end{equation}
Adding this to $\Delta^{dbl} + \Delta^{trpl}$ we get:
\begin{eqnarray}
{1\over \pi i}\Delta'(d\vartheta_L) & = & 
-z_uz_d^2 [(z_d^2-z_d^{-2})^2 + (z_u^2-z_u^{-2})^2]  \;
t^1\wedge \{t^3,d_-\vartheta_L\} +
\nonumber
\\
&& +(2z_u^3-z_u^{-1}-z_d^4z_u^3)t^3\wedge \{t^1,d_-\vartheta_L\}-
\nonumber
\\
&& -(2z_u^{-1}-z_u^3-z_u^{-1}z_d^{-4})t^3\wedge \{ t^1,d_+\vartheta_L\} \,.
\label{DeltaDTheta}
\end{eqnarray}

\subsection{The structure of $\Delta$}
\label{sec:StructureOfDelta}

At the first order of perturbation theory $\Delta = \Delta_0 + \Delta'$ 
where $\Delta_0 (t) = t\otimes 1 + 1 \otimes t $ is the trivial coproduct. 
It follows from Sections \ref{sec:CouplingOfX} and \ref{sec:CouplingOfTheta} 
that $\Delta'$ is given by the following formula:
\begin{equation}\label{CoproductAsConjugation}
\Delta'={\pi i\over 2}\left[r,\Delta^0\right]  \,,
\end{equation}
where
\begin{equation}
r \; = \; {\Phi (z_u, z_d)\over z_u^4 - z_d^4}
(z_uz_d^3 t^1 \otimes t^3 + z_u^3z_d t^3 \otimes t^1 + z_u^2z_d^2 t^2 \otimes t^2) 
+ 2 {\Psi(z_u, z_d) \over z_u^4 - z_d^4} t^0 \otimes t^0 \,.
\label{ClassicalR}
\end{equation}
We used the notations:
\begin{eqnarray}
\Phi(z_u, z_d) & = & (z_u^2 -z_u^{-2})^2 + (z_d^2 -z_d^{-2})^2 
\nonumber
\\
\Psi(z_u, z_d) & = & 1+ z_u^4 z_d^4 - z_u^4- z_d^4 \,.
\nonumber
\end{eqnarray}
The following identities are useful in deriving (\ref{CoproductAsConjugation}).
\begin{equation}
\begin{aligned}
{ [z_uz_d^3 t^1\otimes t^3 + z_u^2z_d^2 t^2\otimes t^2 + z_u^3z_d t^3\otimes t^1
   \,, \,     (z_u^{-1}t^3_{\alpha})\otimes 1 + 1\otimes (z_d^{-1}t^3_{\alpha})] }
& = 2 z_u^3 t^3\bullet \{ t^1,t^3_{\alpha}\} 
 \cr
  [t^0\otimes t^0, (z_u^{-1}t^3_{\alpha})\otimes 1 + 1\otimes (z_d^{-1}t^3_{\alpha})]
&     =-2 (z^{-1}t^3)\bullet \{t^1,t^3_{\alpha}\} 
\cr
 [z_uz_d^3 t^1\otimes t^3 + z_u^2z_d^2 t^2\otimes t^2 + z_u^3z_d t^3\otimes t^1
   \, , \,
   (z_u^{-2} t^2_{\mu})\otimes 1 + 1\otimes (z_d^{-2} t^2_{\mu})  ]  
& = 2z_u^{-1} z_d^3 [t^1,t^2_{\mu}]\bullet t^3 \cr
&   + 2z_d^2 [t^2,t^2_{\mu}]\bullet t^2 \,. 
\label{SymmetricTensorProduct}
\end{aligned}
\end{equation}
Here $\bullet$ denotes the symmetric tensor product; it is the opposite
of $\wedge$. The minus sign in the last line of (\ref{SymmetricTensorProduct})
is because $C^{\dot{\alpha}\beta}=-C^{\beta\dot{\alpha}}$. So in particular 
$2 z_u^3 t^3\bullet \{ t^1,t^3_{\alpha}\} =
        (z_u^3 t^3)\otimes \{ t^1,t^3_{\alpha}\} - 
         \{ t^1,t^3_{\alpha}\}\otimes (z_d^3 t^3) 
$.

\section{Generalized gauge transformations}
\label{sec:FixingTotalDerivatives}

\subsection{Dress code}
The coupling of fields to the generators of the algebra is strictly speaking
not defined unambiguously, because of the possibility of a 
``generalized gauge transformation''
\begin{equation}\label{ChangeOfDressing}
J\mapsto f(d+J)f^{-1}\,,
\end{equation}
where $f$ is a group-valued function of fields, depending on the spectral parameter $z$.
A ``proper'' gauge transformation would not depend on $z$ 
and would belong to the Lie group of $\mathfrak{g}_0$, while $f$
in (\ref{ChangeOfDressing})  belongs to the Lie group of $\mathfrak{g}$ and
does depend on $z$.
Therefore it would perhaps be appropriate to
call (\ref{ChangeOfDressing}) ``generalized gauge transformation'' 
or maybe ``change of dressing''  
If there is some insertion $A$ into the contour, then we should also
transform $A\mapsto fAf^{-1}$.

One of the reasons to discuss the transformations (\ref{ChangeOfDressing})
is that different prescriptions for the order of integrations are related
to each other by such a ``change of dressing'' A similar story for log divergences
was discussed in \cite{Mikhailov:2007mr}. Different choices of the order of
integration lead to different distribution of the log divergences between
the bulk and the boundary.

We agreed in Section \ref{sec:OrderOfIntegrations} 
to use the ``symmetric prescription'' for the order of integrations. 
It turns out that with this prescription 
$\lim_{y\to 0}T_{\rho_2}[C+y]T_{\rho_1}[C]$ comes out
in the ``wrong dressing'' in the sense that the limit cannot be immediately
presented in the form
\begin{equation}\label{DressCode}
P\exp \left( -\int J^a \Delta(t_a) \right) \,.
\end{equation}
In particular $x^{\mu}\partial_+x^{\nu}$ couples to a different algebraic expression
than $x^{\mu}\partial_-x^{\nu}$, while in (\ref{DressCode}) they should both
couple to $\Delta(t^0_{[\mu\nu]})$.
However, it turns out that it is possible to satisfy the ``dress code'' (\ref{DressCode})
by the change of dressing of the type (\ref{ChangeOfDressing}).

We will now {\em stick to the symmetric prescription} for the order of integrations
and study the asymmetry between the couplings of $xd_+x$ and $xd_-x$,
and the asymmetry between the couplings of $\vartheta d_+\vartheta$ and
$\vartheta d_-\vartheta$.
Then we will determine the generalized gauge transformation needed to
satisfy (\ref{DressCode}), and this will fix the total derivative couplings
discussed in Section \ref{sec:CouplingOfX}. It turns out that in the symmetric
prescription we will have
to do the generalized gauge transformation (\ref{ChangeOfDressing})
with the parameter:
\begin{equation}\label{GeneralizedGaugeParameter}
f=1-{\pi i\over 2}\left(\; (z^{-2}-z^2)t^2)\wedge [t^2,dx]
   +((z^{-1}-z^3)t^3)\wedge \{t^1,d\vartheta_L\}
   +((z^{-3}-z)t^1)  \wedge \{t^3,d\vartheta_R\} \;\right) + \ldots
\end{equation}
In the next Sections \ref{sec:AsymmetryXdX} and \ref{sec:AsymmetryThetadTheta}
we will show that the gauge transformation with this parameter indeed removes
the asymmetry.
In Section \ref{sec:BoundaryAndShift} we will derive (\ref{GeneralizedGaugeParameter})
using the invariance under the shift symmetries.

\subsection{Asymmetry between the coupling of $xd_+x$ and $xd_-x$}
\label{sec:AsymmetryXdX}

\subsubsection{Coupling proportional to $z_u^{-4} xdx$ }
The most obvious asymmetry is that there is a term with $z_u^{-4} xd_+x$ but no
term with $z_u^{-4} xd_-x$. The term with $z_u^{-4} xd_+x$ comes from this 
collision:

\vspace{10pt}

\hbox to \linewidth{%
\hfill \upupdown{$z^{-2}d_+x$}{$z^{-2}d_+x$}{${1\over 2}[x,d_+ x]$}{50} \hfill}

\noindent 
The result is: 
\begin{equation}\label{UnwantedCouplingZ-4xdx}
\pi i
\left[
(z^{-2}d_+x)\otimes 1 \; , \;
{1\over 4}(z^{-2}t^2)\otimes [x,t^2]
\right] \,.
\end{equation}
This is unwanted, so we want to do the generalized gauge transformation with the
parameter
\begin{equation}\label{ChangeOfDressingParameter}
-{\pi i\over 2} (z^{-2}t^2)\wedge [t^2,x]\,.
\end{equation}
which removes this coupling and adds instead a total derivative coupling
to $dx$:
\begin{equation}\label{CouplingAddedByChangeOfDressing}
-{\pi i\over 2} (z^{-2}t^2)\wedge [t^2,dx]\,.
\end{equation}
We will now argue that the change of dressing with the parameter 
(\ref{ChangeOfDressingParameter}) also removes the asymmetry between
the coupling of $xd_+x$ and $xd_-x$. 

Also the coefficient of $z_u^{-2}z_d^{-2} \; xd_+x$ is different
from the coefficient of $z_u^{-2}z_d^{-2} \; xd_-x$. Let us explain this.

\subsubsection{Asymmetric couplings of the form $z_u^{-2}z_d^{-2} \; xdx$ }
\label{sec:AsymmetricCouplingsZuZd}
There is a contribution from a double collision, and from a triple collision. 
The double collision is:

\hbox to \linewidth{\hfill%
\updown{$z^{-2}J_{2+}$}{$z^{-2}J_{2+}$}{10}{40}\hfill}

\vspace{10pt}
\noindent and we have to take into account  the interaction vertex in the action:
\begin{equation}
-S\mapsto {1\over 6\pi} \mbox{str} [x,\partial_+x][x,\partial_-x] \,.
\end{equation}
The calculation is in Section \ref{sec:XdXinJ2J2}, and 
the result is:
\begin{equation}\label{ContributionFrom_J2_J2}
{1\over 2} \pi i \; C^{\mu\nu} (z^{-2}[t^{2}_{\mu},[x,d_-x]])
\wedge (z^{-2} t^{2}_{\nu}) \,.
\end{equation}
There is also a triple collision:

\hbox to \linewidth{\hfill%
\upupdown{$z^{-2}d_+x$}{$(1/2) [x,d_+x]$}{$z^{-2}d_+x$}{50} \hfill}

\noindent It contributes:
\begin{equation}\label{ContributionFrom_dx_xdx_atop_dx}
{1\over 4} \pi i (z^{-2}[[d_+x,x],t^2]\wedge (z^{-2}t^2) \,.
\end{equation}

\vspace{10pt}
\noindent
The sum of equations (\ref{ContributionFrom_J2_J2}) and
(\ref{ContributionFrom_dx_xdx_atop_dx}) amounts to the following
asymmetry of the form $z_u^{-2}z_d^{-2}$:
\begin{equation}\label{AsymmetryZ-2Z-2}
-{1\over 4}\pi i (z^{-2}[[d_+x,x],t^2]) \wedge (z^{-2}t^2) \,.
\end{equation}
We see that (\ref{UnwantedCouplingZ-4xdx})+(\ref{AsymmetryZ-2Z-2}) is:
\begin{equation}
\left[ \;
(z^{-2} d_+x)\otimes 1 + 1\otimes (z^{-2}d_+x)\;\; ,\; 
{1\over 2}\pi i \; (z^{-2} t^2)\wedge [x,t^2] \; \right] \,.
\end{equation}
This is undone with the generalized gauge transformation with the
parameter ${1\over 2}\pi i \; (z^{-2} t^2)\wedge [x,t^2]$,
which adds an additional total derivative coupling:
\begin{equation}
{1\over 2}\pi i \; (z^{-2} t^2)\wedge [dx,t^2] \,.
\end{equation}
This is the ``additional coupling'' of Eq. (\ref{AdditionalCouplingDX}).

\vspace{15pt}

\subsection{Asymmetry in the couplings of $\vartheta  d\vartheta$}
\label{sec:AsymmetryThetadTheta}

The situations with the couplings of $\vartheta d\vartheta$ is similar.
There are asymmetric couplings of the form $z_u^{-4} \vartheta_L d_+\vartheta_R$
which are removed by the generalized gauge transformation. This generalized
gauge transformation should also remove the asymmetry in the couplings
of $z_u^{-2}z_d^{-2}\vartheta d_+\vartheta$ and 
$z_u^{-2}z_d^{-2}\vartheta d_- \vartheta$, but we did not 
check this. 

Terms of the form  $z_u^{-4} \vartheta_L d_+\vartheta_R$
come from $z^{-3} d_+\vartheta_R \leftrightarrow z^{-1} d_+\vartheta_L\atop
{1\over 2}[\vartheta_L,d\vartheta_R] + {1\over 2}[\vartheta_R,d\vartheta_L]$.
They are similar to (\ref{UnwantedCouplingZ-4xdx}):
\begin{equation}
\pi i
\left[
(z^{-3}d_+\vartheta_R )\otimes 1 \; , \;
\left(-{1\over 4}\right)(z^{-1}t^3)\otimes \{t^1,\vartheta_L\}
\right] \,.
\end{equation}
This should be removed with the generalized gauge transformation
which simultaneously introduces the total derivative coupling:
\begin{equation}
- {\pi i\over 2}(z^{-1}t^3)\wedge \{t^1,d\vartheta_L\} \,.
\end{equation}
This is the ``additional coupling'' of (\ref{AdditionalCouplingDTheta}).

\section{Boundary effects}
\label{sec:BoundaryEffects}

\subsection{The structure of ${\cal G}_{\pm}$}
\label{sec:StructureOfG}
\subsubsection{Introducing the matrix $s$}
Here we will derive Eq. (\ref{FormulaForGPlus}) in 
Section \ref{sec:ExplicitExpressionsForGandR}. We inserted the switch
operator on the upper line, which turns $z_u^{in}$ into $z_u^{out}$.
Naively Eq. (\ref{CoproductAsConjugation}) implies that:

\vspace{7pt}

\hbox to \linewidth{\hfill %
\includegraphics[width=3.5in]{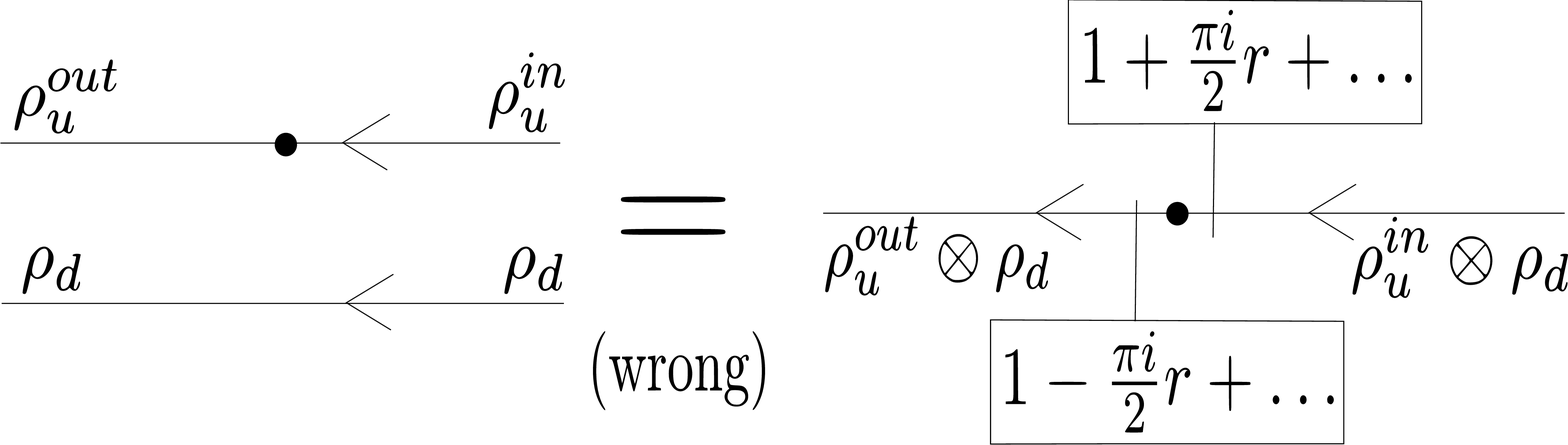} \hfill}

\noindent
But this is wrong because there is an additional boundary contribution
related to the second order poles in the short distance singularities of
the products of currents. Notice that these second order poles
correspond to the $\delta'$ terms in the approach of 
\cite{Maillet:1985ec,Maillet:1985ek,Maillet:1985fn} 
(see Appendix \ref{VeryBriefSummaryOfMaillet}). At the first order in the
$x$-expansion the contributing diagram is this one:

\vspace{7pt}

\hbox to \linewidth{\hfill %
\includegraphics[width=2.4in]{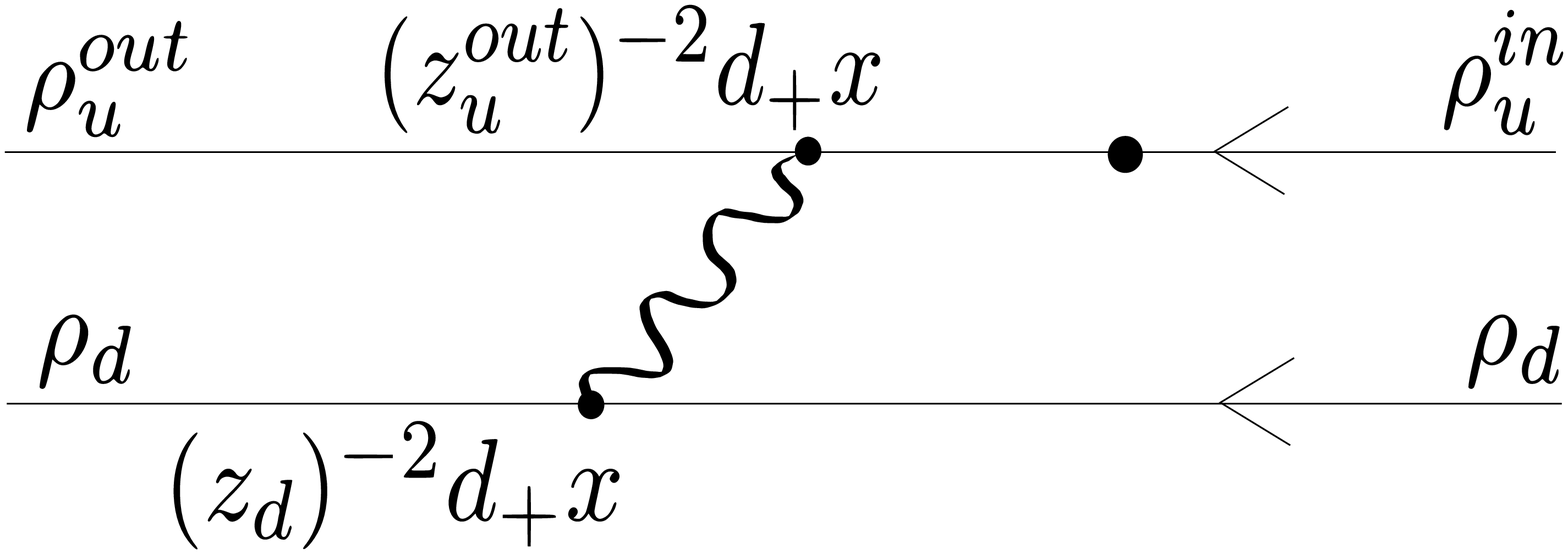} \hfill}

\noindent
and similar ones. This turns $1\pm {\pi i\over 2}r+\ldots$ into $1\pm {\pi i\over 2}(r+s)+\ldots$
where 
\begin{equation}
s=C_+-C_- \,,
\end{equation}
and $C_{\pm}$ are given by (\ref{CplusDef}) and (\ref{CminusDef}).
Therefore ${\cal G}_+$ of the switch operator is the following split operator:
\begin{equation}\label{GSplitOperator}
{\cal G}_+(1_{switch})=1+{\pi i\over 2}(r(z_u^{in},z_d)+s(z_u^{in},z_d)) - 
  {\pi i\over 2}(r(z_u^{out},z_d)+s(z_u^{out},z_d))+\ldots
\end{equation}
\subsubsection{Cancellation of field dependent terms}
Dots in (\ref{GSplitOperator}) denote the contribution of the higher orders of the
string worldsheet perturbation theory. Those are the terms of the order
$\hbar^2$ and higher. The terms with ${1\over 2}(r+s)$ are of the order $\hbar$.
Remember that we are also expanding in powers of elementary fields.
It turns out that all the terms of the order $\hbar$ ({\it i.e.} tree level) 
in ${\cal G}_+(1_{switch})$
are c-number terms written in (\ref{GSplitOperator}), there are no corrections
of the higher powers in $x$ and $\vartheta$. 
This is because such corrections would
contradict the invariance with respect to the global shifts (\ref{GlobalShift}).
Indeed, suppose that ${\cal R}(1_{switch}\otimes 1)$ 
contained $x$ and $\vartheta$. For example,
suppose that there was a term linear in $x$, something like $x\;t\otimes t$.
Then the variation under the global shift (\ref{GlobalShift}) will be proportional
to $\xi\; t\otimes t$ and there is nothing to cancel it\footnote{
If we inserted some operator ${\cal O}$ which is not gauge invariant,
for example ${\cal O}=t^2_{\mu}$, the variation under the global shift 
will give $[t^2_{\mu},[\xi,x]]$. This is linear in $x$, but $x$ will contract with
$d_+x$ in $\int (z^{-2}J_{2+}d\tau^+ + z^2J_{2-}d\tau^-)$ resulting in the 
$x$-independent expression
of the form $z^{-2}t^2\otimes [t^2,\xi]$, which will cancel the $\xi$-variation of
the field dependent terms(\ref{XDependentTerms}).}. 
This implies that ${\cal R}(1_{switch}\otimes 1)$ is a c-number insertion, 
{\it i.e.} no
field-dependent corrections to (\ref{ExchangeUp}), (\ref{ExchangeDn}),
(\ref{rPluss}), (\ref{rMinuss}).

\subsection{Boundary effects and the global symmetry}
\label{sec:BoundaryAndShift}

We explained in Section 3 of \cite{Mikhailov:2007mr} that the global shifts
act on the ``capital'' currents by the gauge transformations (normal gauge
transformation, not generalized):
\begin{eqnarray}
S_{\xi}.J & = & -dh h^{-1} + hJh^{-1}
\nonumber
\\
h & = & 1-{1\over 2} R^{-2} [x,\xi]+\ldots \,.
\end{eqnarray}
Suppose that the outer contour is open-ended, then this is not invariant
under the global shifts:

\vspace{7pt}

\hbox to \linewidth{\hfill %
\includegraphics[width=2.1in]{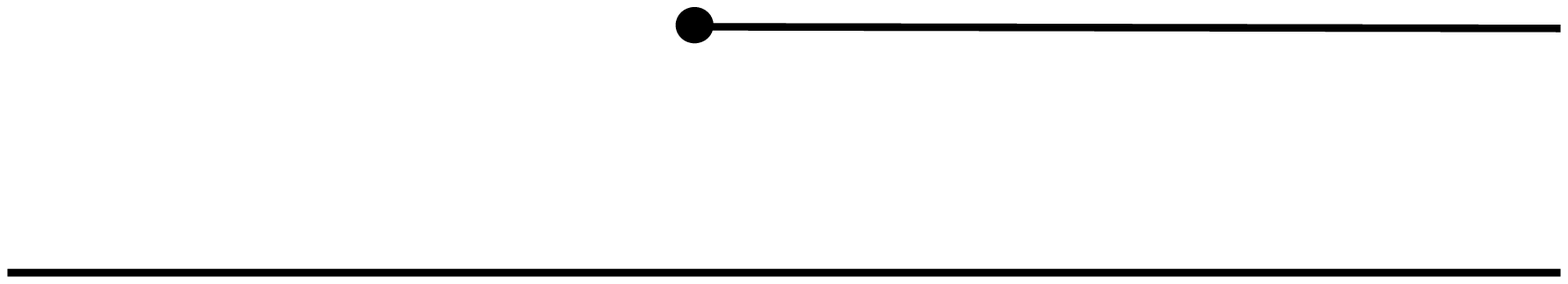} \hfill}

\noindent
The infinitesimal shift of this is equal to:

\vspace{7pt}

\hbox to \linewidth{\hfill %
\includegraphics[width=2.1in]{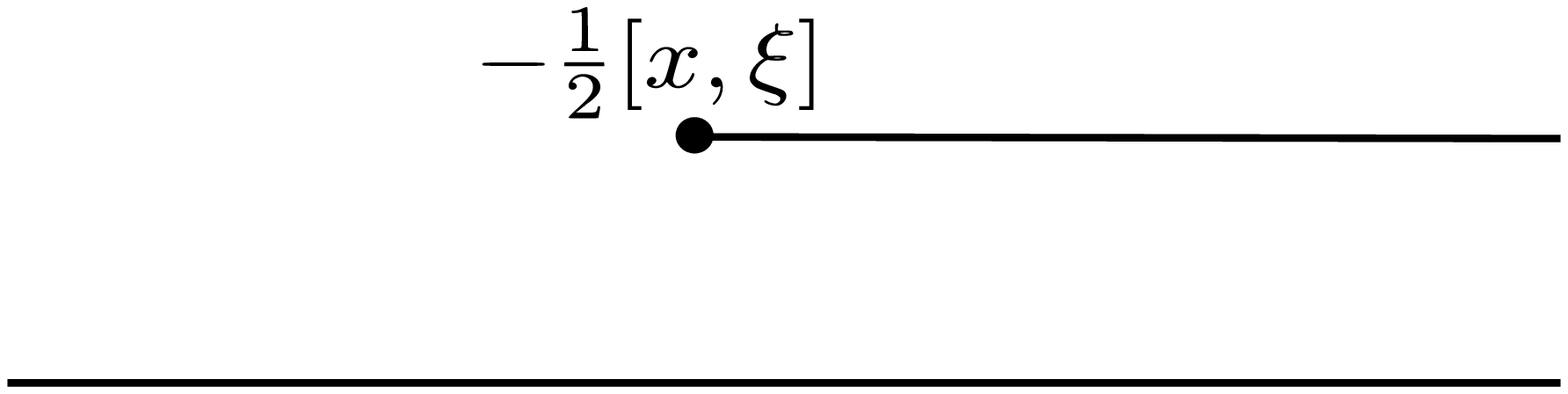} \hfill}

\noindent
Therefore because of this contraction:

\vspace{7pt}

\hbox to \linewidth{\hfill %
\includegraphics[width=2.1in]{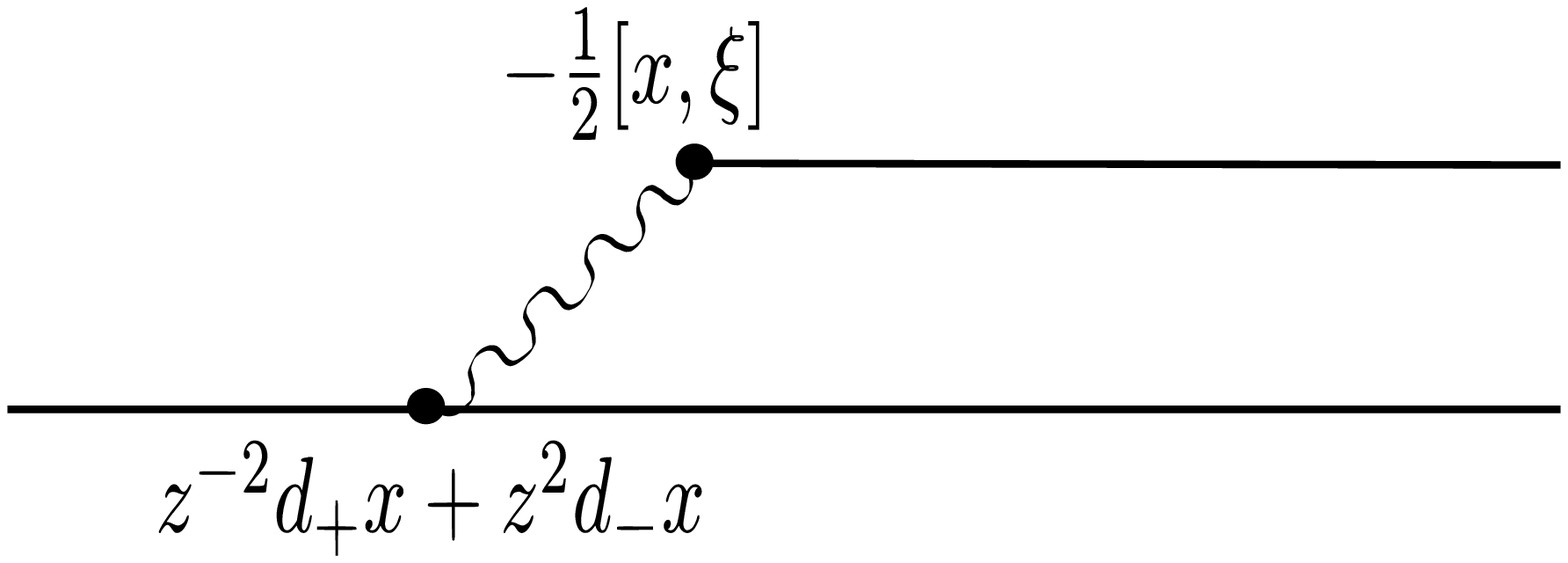} \hfill}

\noindent
We have the imaginary contribution:

\vspace{7pt}

\hbox to \linewidth{\hfill %
\includegraphics[width=2.1in]{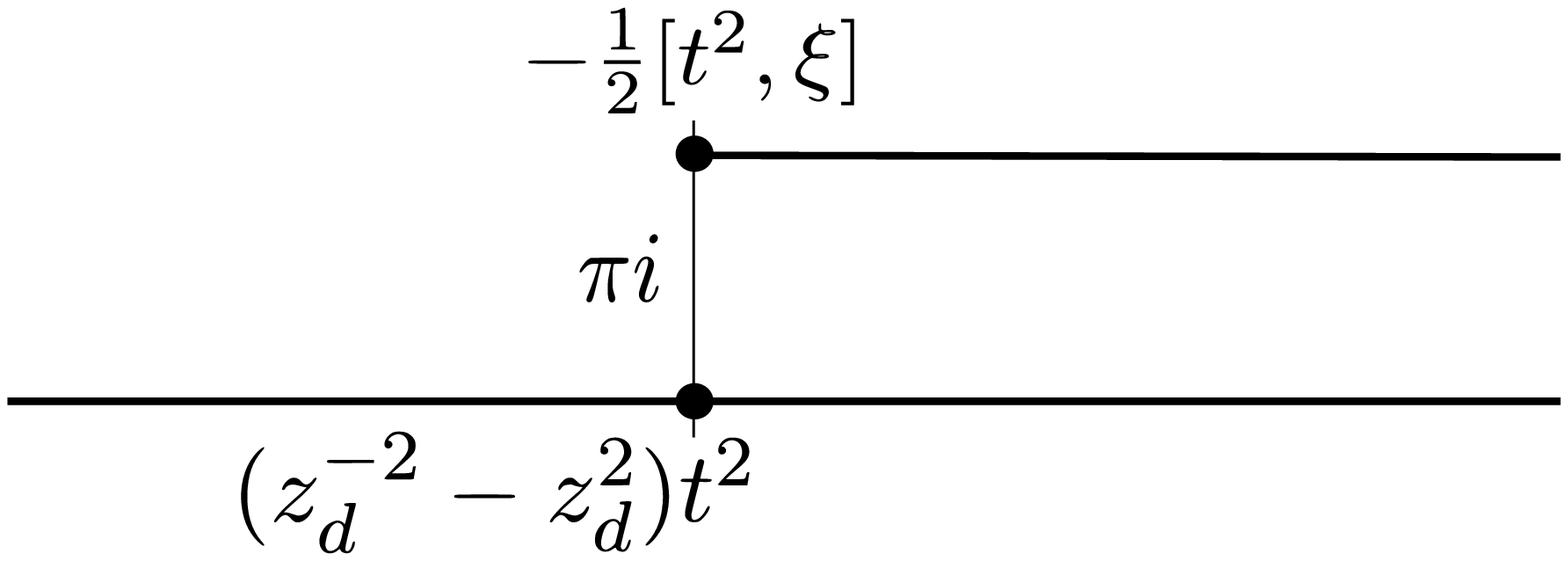} \hfill}

\noindent
Using the terminology from Section \ref{sec:WhatToExpect} we should say
that ${\cal F}_+(1)$ is such that:
\begin{equation}
S_{\xi}{\cal F}_+(1)=-\pi i {1\over 2} [t^2,\xi]\otimes (z_d^{-2}-z_d^2)t^2 \,.
\end{equation}
There are similar considerations for the super-shifts.
Therefore:
\begin{eqnarray}
{\cal F}_+(1) & = & \mbox{const} +
  \pi i {1\over 2} [x,t^2]\otimes (z_d^{-2}-z_d^2)t^2 +
\nonumber \\
&&\phantom{\mbox{const}}
+ \pi i {1\over 2} \{\vartheta_L,t^1\}\otimes (z_d^{-1}-z_d^3)t^3 +
\nonumber \\
&&\phantom{\mbox{const}}
+ \pi i {1\over 2} \{\vartheta_R,t^3\}\otimes (z_d^{-3}-z_d^1)t^1+\ldots \,.
\label{FPlusOfOne}
\end{eqnarray}
The relation between this formula and the generalized gauge transformation
with the parameter (\ref{ChangeOfDressingParameter}) is the following.
Part of (\ref{FPlusOfOne}) comes from (\ref{ChangeOfDressingParameter}),
and another part from the following diagrams:

\vspace{7pt}

\hbox to \linewidth{\hfill %
\includegraphics[width=3.2in]{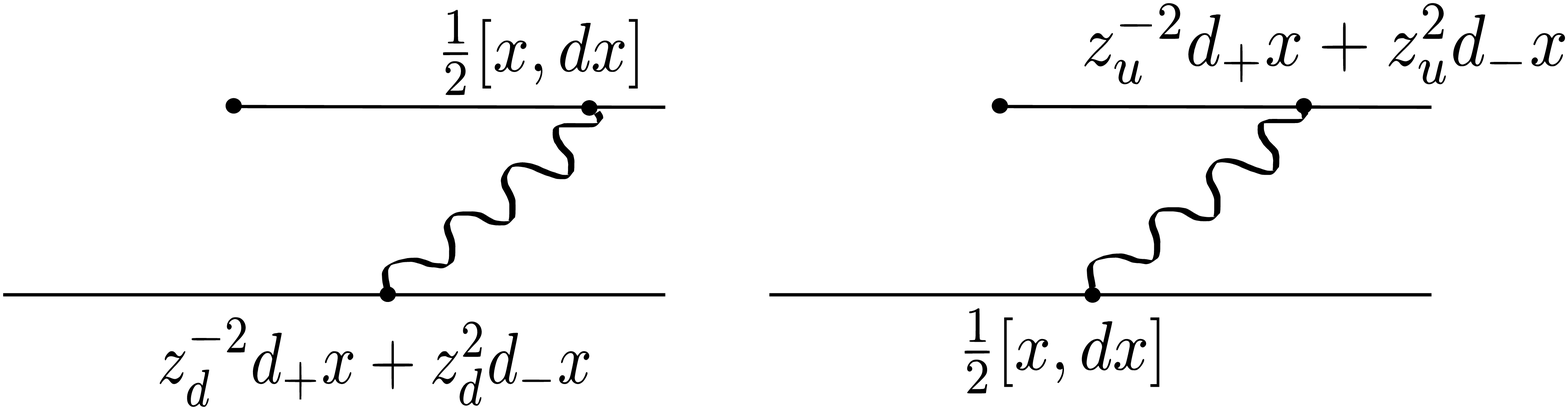} \hfill}

\noindent
These two diagrams contribute:
\begin{equation}\label{ContributionOfTwoDiagramms}
 \pi i {1\over 4} [x,t^2]\otimes (z_d^{-2}-z_d^2)t^2 
+\pi i {1\over 4} (z_u^{-2}-z_u^2)t^2\otimes [x,t^2] \,.
\end{equation}
And the generalized dressing transformation with the
parameter (\ref{ChangeOfDressingParameter}) gives the
boundary term $\pi i {1\over 4} [x,t^2]\otimes (z_d^{-2}-z_d^2)t^2 
-\pi i {1\over 4} (z_u^{-2}-z_u^2)t^2\otimes [x,t^2]$ which
in combination with (\ref{ContributionOfTwoDiagramms}) gives:
\begin{equation}
\pi i {1\over 2} [x,t^2]\otimes (z_d^{-2}-z_d^2)t^2 \,,
\end{equation}
which is in agreement with (\ref{FPlusOfOne}). Similar diagrams with
fermions give terms with $\vartheta$ in (\ref{FPlusOfOne}).

Notice that the constant terms in (\ref{FPlusOfOne}) are essentially
the same as in Section \ref{sec:StructureOfG}:
\begin{equation}
{\cal F}_+(1)=1+{\pi i\over 2}(r(z_u,z_d)+s(z_u,z_d))+\ldots
\end{equation}
The difference between ${\cal F}_+(1)$ and ${\cal G}_+(1_{switch})$ is that
${\cal G}_+(1_{switch})$ is a c-number while ${\cal F}_+(1)$ is field-dependent.
That is because $1_{switch}$ is invariant under the gauge transformations,
because $\rho_u^{z^{out}}$ and $\rho_u^{z^{in}}$ are the same as representations
of the finite dimensional $\mathfrak{g}_0\subset L\mathfrak{psu(2,2|4)}$.


\section{BRST transformations}
\label{sec:BRST}
Here we discuss  the action of $Q_{BRST}$ on the switch operators
end verify that it commutes with ${\cal G}_+$. There are two BRST currents,
holomorphic $Q$ and antiholomorphic $\overline{Q}$. They are both nilpotent
and $\{Q,\overline{Q}\}=0$. The
total BRST operator is their sum:
\begin{equation}
Q_{BRST}=Q+\overline{Q}
\end{equation}
Here we will consider the holomorphic BRST operator $Q$. 
The action of $Q$ on the currents is:
\begin{equation}
[ \epsilon Q , J_{\pm}(z) ] = D_{\pm}^{(z)}(\epsilon z^{-1}\lambda) \,,
\end{equation}
(see for example Section 2 of \cite{Mikhailov:2007mr}).
In other words
\begin{equation}
[ \epsilon Q , T_A^B(z) ] = {1\over z}\epsilon \lambda(B)\; T_A^B(z)
-T_A^B(z)\; {1\over z}\epsilon \lambda(A)
\end{equation}
The switch operator
turns $z^{in}$ into $z^{out}$. We have:
\begin{equation}
Q.{\bf 1}_{switch} = \left({1\over z^{out}} - {1\over z^{in}}\right)\lambda \,.
\end{equation}
According to (\ref{FormulaForGPlus}) the fusion of the switch operator
on the upper contour is the split operator ${\pi i\over 2}(-r_+^{out}+r_+^{in})$.
Therefore:
\begin{eqnarray}
Q{\cal G}_+ {\bf 1}_{switch}& = & {\pi i\over 2}\left[\;\;
(z_{out}^{-1}(\lambda\otimes 1)+z^{-1}(1\otimes \lambda))
\;(-r_+^{out}+r_+^{in}) -
\right.
\\
&& \hspace{18pt}\left.- (-r_+^{out}+r_+^{in}) \;
(z_{in}^{-1}(\lambda\otimes 1)+z^{-1}(1\otimes \lambda))
\;\;\right]
\end{eqnarray}
Now we have to caclulate ${\cal G}_+ Q {\bf 1}_{switch}$. 
The action of ${\cal G}_+$ on 
$\left({1\over z^{out}} - {1\over z^{in}}\right)\lambda$ is essentially the same 
as the action on the switch:
\[
{\pi i\over 2}\left[\;\;
-r_+^{out} (z_{out}^{-1}(\lambda\otimes 1) - z_{in}^{-1}(\lambda\otimes 1))
+ (z_{out}^{-1}(\lambda\otimes 1) - z_{in}^{-1}(\lambda\otimes 1)) r_+^{in}
\;\;\right]
\]
plus the contribution of this diagram:

\vspace{7pt}

\hbox to \linewidth{\hfill %
\includegraphics[width=2.3in]{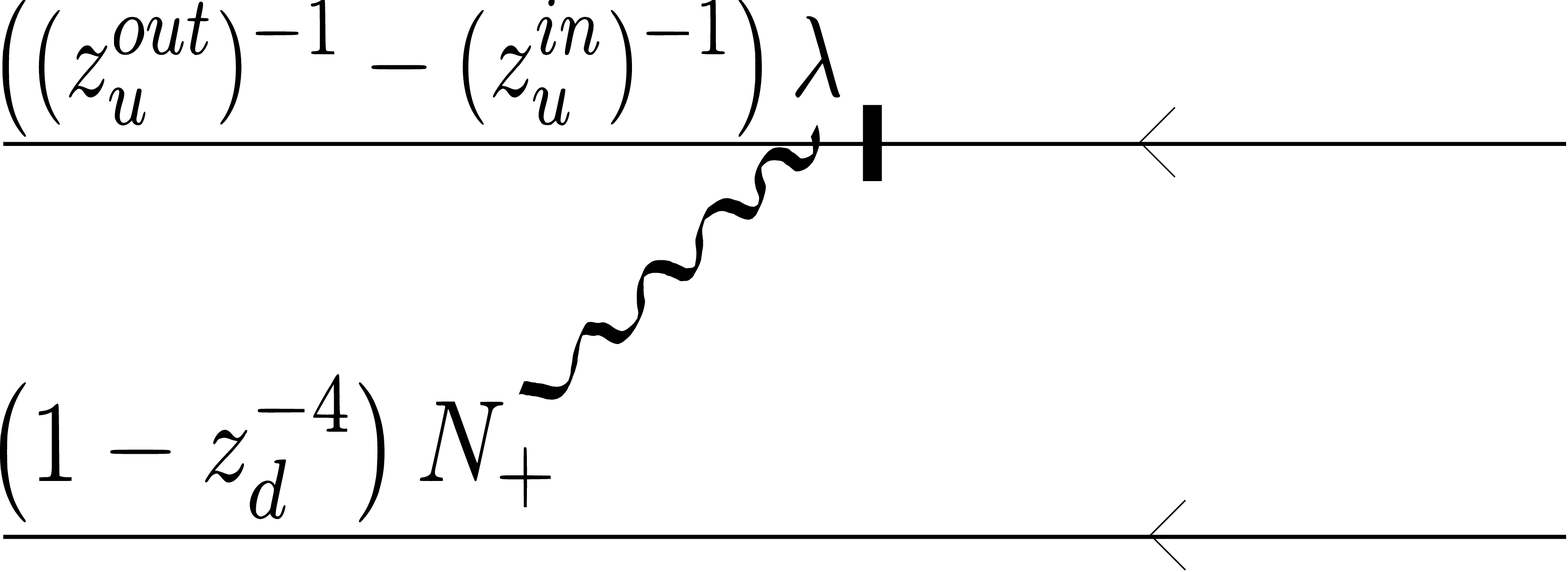} \hfill}

\noindent
This diagram contributes:
\begin{equation}
-\pi i \; \left((z_u^{out})^{-1} - (z_u^{in})^{-1}\right)
(1-z_d^{-4})\; t^3\otimes \{t^1, \lambda\} \,,
\end{equation}
where we have used the short distance singularity:
\begin{equation}
\lambda(w_u)\otimes (-\{w_+,\lambda\}(w_d)) 
= -{1\over w_u-w_d} t^3\otimes \{t^1,\lambda\}
\end{equation}
Therefore the condition $[Q,{\cal G}_+]=0$ can be written as follows:
\begin{eqnarray}
0 & = &
\left[ {r_+(z_u^{out},z_d)\over 2}\; , \; 
{1\over z_u^{out}}\lambda\otimes 1 + {1\over z_d} 1\otimes\lambda \right]
-
\left[ {r_+(z_u^{in} ,z_d)\over 2}\; , \; 
{1\over z_u^{in}}\lambda\otimes 1 + {1\over z_d} 1\otimes\lambda \right]
+
\nonumber \\
&+&
\left({1\over z_u^{out}} - {1\over z_u^{in}}\right) \left(1-{1\over z_d^4}\right)\; 
t^3\otimes \{t^1, \lambda\} \,.
\end{eqnarray}
This can be verified using the identity:
\begin{equation}\label{QR}
\left[ (z_u)^{-1}\lambda\otimes 1 + (z_d)^{-1}1\otimes\lambda 
\;,\;  
{r_+(z_u,z_d) \over 2} \right] = 
\left( 1 - {1\over z_d^4}\right)
(z_u^{-1} t^3\otimes \{t^1,\lambda\} - z_d^3 \{\lambda,t^1\}\otimes t^3) \,.
\end{equation} 
Notice that this identity can be used to derive equation (\ref{rPluss}) for $r_+$. 
Also notice that the second term on the right hand side 
$-\left( 1 - {1\over z_d^4}\right) z_d^3 \{\lambda,t^1\}\otimes t^3$ 
is a gauge transformation and does not contribute because the switch operator
is gauge invariant. 

As another example let us consider the upper Wilson line terminating on $\tilde{\lambda}$.  
In this case the second term on the right hand side of (\ref{QR})
does contribute:
\begin{equation}\label{SecondTermQR}
-\left( 1 - {1\over z_d^4}\right) z_d^3 [\{\lambda,t^1\},\tilde{\lambda}]\otimes t^3
\end{equation}
But this cancels with the contribution of the contraction of $\tilde{\lambda}$ with
the integral $\int d\tau^-(1-z_d^4) N_-$ arizing in the expansion of the lower Wilson
line: 

\vspace{7pt}

\hbox to \linewidth{\hfill %
\includegraphics[width=2.3in]{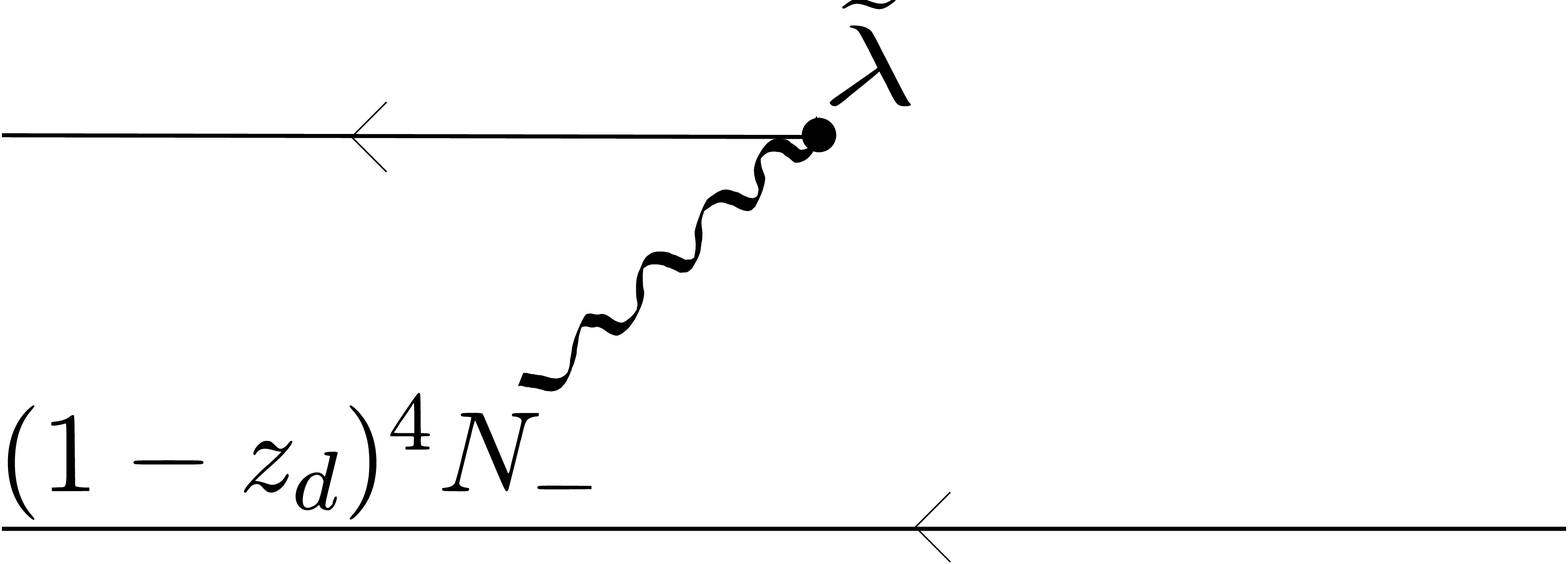} \hfill}

\noindent
Indeed, this contraction gives
\begin{equation}\label{ContractionLambdaTildeN}
t^1\otimes (1-z_d^4)\{t^3,\tilde{\lambda}\} 
\end{equation}
and therefore there is an additional contribution to $[Q,{\cal G}]$:
\begin{equation}\label{AdditionalContributionToQG}
-t^1\otimes \left(1 - z_d^4\right) {1\over z_d}[\lambda,\{t^3,\tilde{\lambda}\}]
\end{equation}
Notice that (\ref{SecondTermQR}) cancels with (\ref{AdditionalContributionToQG})
because 
$[\{\lambda,t^1\},\tilde{\lambda}]\otimes t^3  
- t^1\otimes [\lambda,\{t^3,\tilde{\lambda}\}] = 0$

\section{Generalized YBE}
\label{sec:GCYBE}
The $r$-matrix (\ref{ClassicalR}) does not satisfy the classical Yang-Baxter
equation in its usual form, but the deviation from zero is
a polynomial in $z_1,z_2,z_3,z_1^{-1},z_2^{-1},z_3^{-1}$.
Using the notation of Eq. (\ref{NotationTTT}):
\begin{eqnarray}\label{cYBE}
&& [r_{12}, r_{13}] + [r_{12}, r_{23}] + [r_{13}, r_{23}] =
\\[5pt]
&&=\phantom{=}
t^0\otimes t^2\otimes t^2 
\left(z_2^2 z_3^2 z_1^4-\frac{z_3^2}{z_2^2}-\frac{z_2^2}{z_3^2}+\frac{1}{z_2^2 \
z_3^2 z_1^4}\right)+
\nonumber\\
&& \phantom{=}
+t^3\otimes t^3\otimes t^2 
\left(-z_1^3 z_3^2 z_2^3+\frac{4}{z_1 z_3^2 \
z_2}-\frac{1}{z_1^5 z_3^2 z_2}-\frac{1}{z_1 z_3^6 z_2}-\frac{1}{z_1 z_3^2 \
z_2^5}\right)+
\nonumber\\
&& \phantom{=}
+t^0\otimes t^1\otimes t^3 \left(-z_2 z_3^3 \
z_1^4+\frac{z_3^3}{z_2^3}+\frac{z_2}{z_3}-\frac{1}{z_2^3 z_3 z_1^4}\right)+
\nonumber \\
&&\phantom{=}
+t^1\otimes t^1\otimes t^2
\left(-z_1 z_2 z_3^6-z_1 z_2^5 z_3^2-z_1^5 z_2 z_3^2+4 z_1 z_2 z_3^2
-\frac{1}{z_1^3 z_2^3 z_3^2}\right) +
\nonumber \\
&&\phantom{=} +\mbox{permutations} \,.
\nonumber
\end{eqnarray}
We will now explain why (\ref{cYBE}) is not zero and what replaces the classical
Yang-Baxter equation. We will also derive a set of generalized YBE which we conjecture 
to be relevant in the quantum theory. 

\begin{figure}[th]
\centerline{\includegraphics[width=14cm]{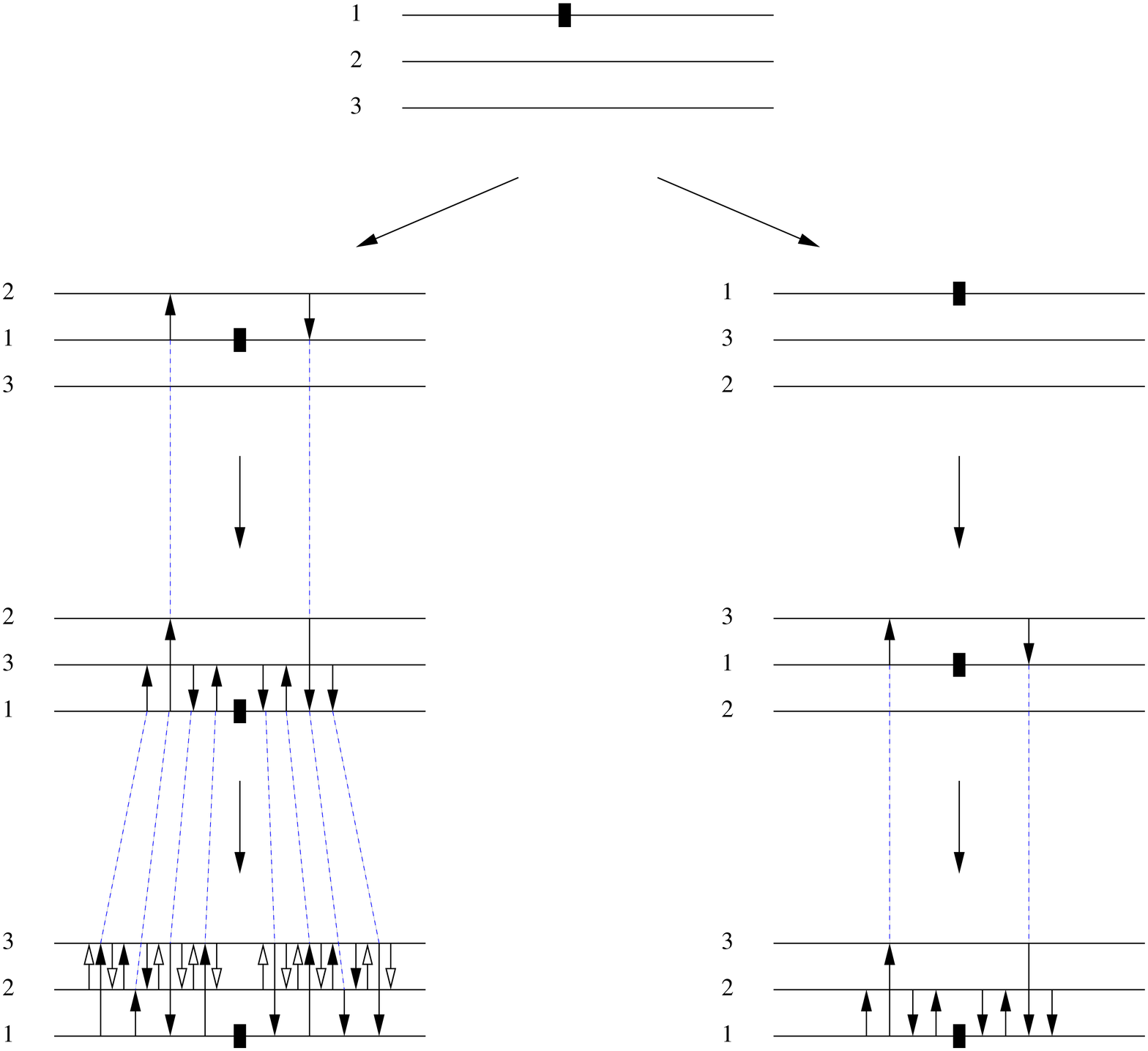}}
\caption{\label{FiguregYBE1}\small 
Generalized YBE 1.}
\end{figure}

The consistency conditions follow from considering the different ways of exchanging
the product of three Wilson lines with insertions. We first consider the case
of gauge invariant insertions; in this case the R-matrices are c-numbers.
Then we will consider the case of non-gauge-invariant insertions, namely
loose endpoints. In this case the R-matrices are field-dependent, and the generalized
Yang-Baxter equations are of the dynamical type.


\subsection{Generalized quantum YBE}

\begin{figure}[th]
\centerline{\includegraphics[width=14cm]{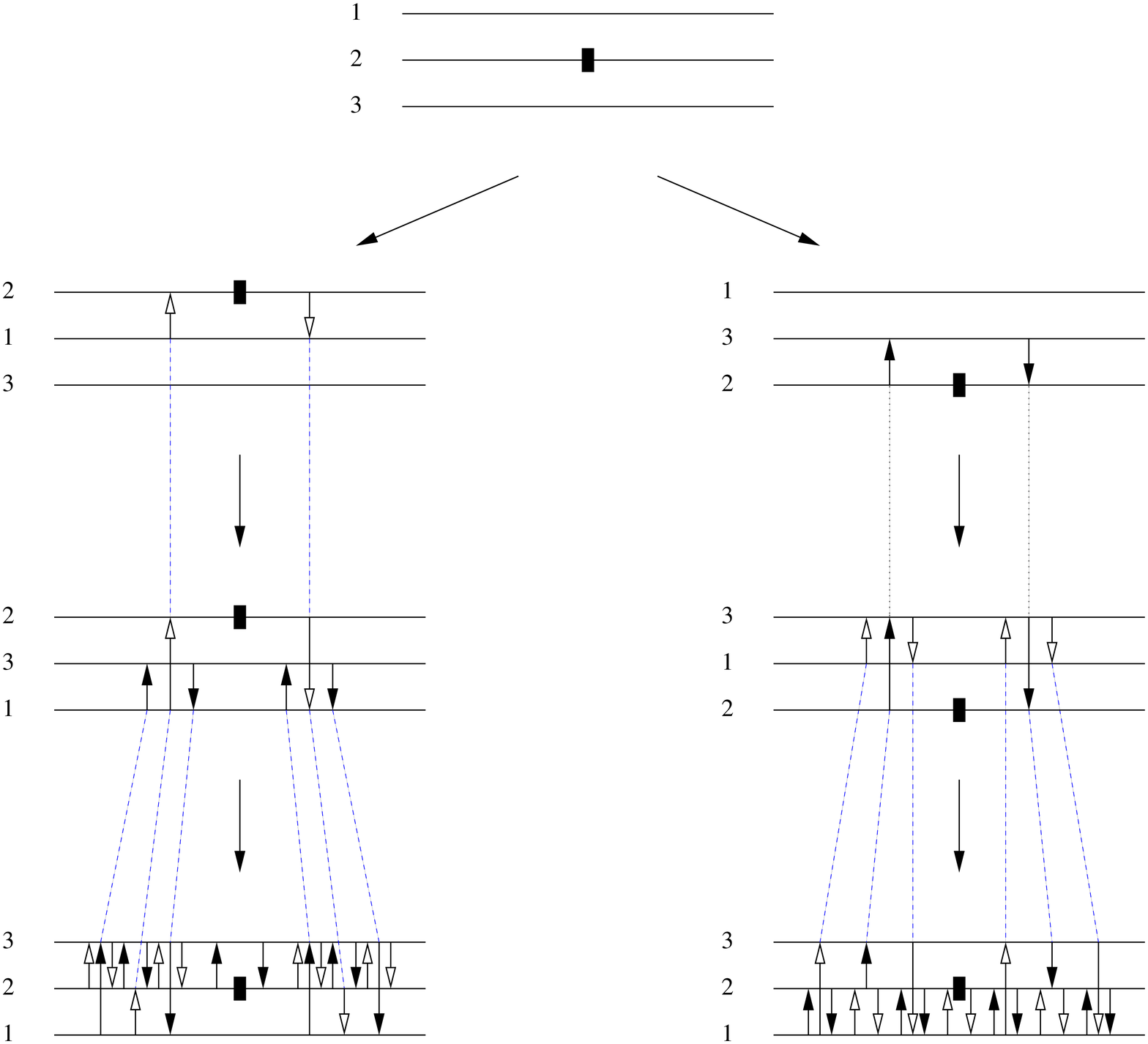}}
\caption{\label{FiguregYBE2}\small 
Generalized YBE 2.}
\end{figure}

To understand the quantum consistency conditions for the $R$ matrices let
us put the Wilson line with the spectral parameter switch on top of two
other Wilson lines, the other two Wilson lines  having no operator insertions. 
The equations of this section will not change if we put a constant
gauge invariant operator at the point on the upper contour where
we switch the spectral parameter (instead of just $1$). 
For example, $C^{\mu\nu}t^2_{\mu}t^2_{\nu}$
is a constant gauge invariant operator. It is gauge invariant because commutes
with $\mathfrak{g}_{\bar{0}}$.

The generalized quantum Yang-Baxter equations (qYBE) are obtained from the exchanges 
illustrated in figure \ref{FiguregYBE1}.
The notations are: \includegraphics[width=5pt]{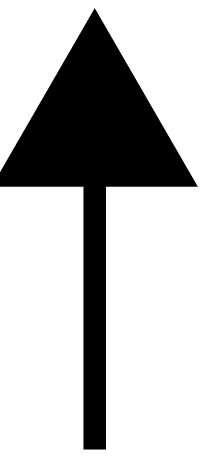}$\; = R_+$, 
\includegraphics[width=5pt]{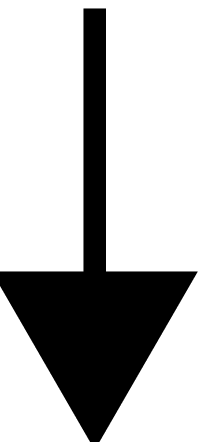}$\; =R_+^{-1}$, 
\includegraphics[width=5pt]{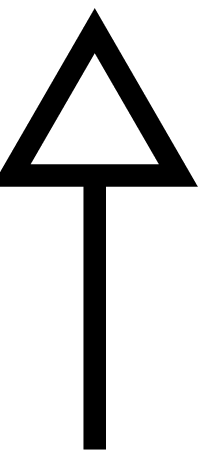}$\; =R_-$,
\includegraphics[width=5pt]{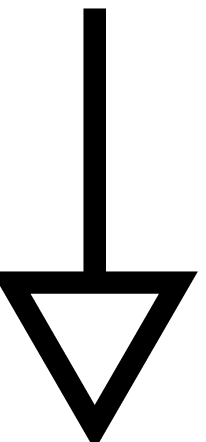}$\; =R_-^{-1}$.
The insertion of the spectral parameter changing operator is marked by a black bar.

Equating LHS and RHS in figure \ref{FiguregYBE1} yields
\begin{equation}
R_{23,-} R_{13,+} R_{23,-}^{-1}		R_{23,+} R_{12,+} R_{23,+}^{-1}
R_{23,-} R_{13,+}^{-1} R_{23,-}^{-1}	R_{23,-} R_{13,+} R_{23,-}^{-1} 
\cong
R_{12,+} R_{13,+} R_{12,+}^{-1} R_{12,+} \,.
\end{equation}
After cancellations of $RR^{-1}$:
\begin{equation}\label{generalizedYBE}
R_{23,-} R_{13,+} R_{23,-}^{-1}		R_{23,+} R_{12,+} R_{23,+}^{-1}
\cong R_{12,+} R_{13,+}   \,.
\end{equation}
Here the sign $\cong$ means that the ratio of the left hand side and the right
hand side commutes with ${\cal O}$:
\begin{eqnarray}\label{RRRRRRRR}
R_{13,+}^{-1} R_{12,+}^{-1} R_{23,-} R_{13,+} R_{23,-}^{-1} 
R_{23,+} R_{12,+} R_{23,+}^{-1} & = & T_{123}
\\
T_{123} {\cal O}_1 T_{123}^{-1} & = & {\cal O}_1 \,.
\label{GaugeInvarianceInYB}
\end{eqnarray}
At the first order of perturbation theory the left hand side of (\ref{RRRRRRRR}) is,
{\it cf.} (\ref{rPluss}) and (\ref{rMinuss}),
\begin{equation}\label{LHS}
[r_{23,-},r_{13,+}] + [r_{13,+},r_{12,+}] + [r_{23,+},r_{12,+}]\;= 
\end{equation}
And the right hand side of (\ref{RRRRRRRR}) is: 
\begin{eqnarray}
& = & - 4\; t^0 \otimes (z_2^2-z_2^{-2})t^2\otimes (z_3^2-z_3^{-2})[t^0,t^2] -
\nonumber
\\
&& - 4\; t^0\otimes (z_2-z_2^{-3})t^1 \otimes (z_3^3-z_3^{-1})[t^0,t^3] -
\label{RHS}
\\
&& - 4\; t^0\otimes (z_2^3-z_2^{-1})t^3 \otimes (z_3-z_3^{-3})[t^0,t^1]
\nonumber
\end{eqnarray}
(the indices of $t^0$ contract with the indices of another $t^0$, so
$t^0\otimes t^0$ stands for 
$C^{[\mu_1\nu_1][\mu_2\nu_2]}\;t^0_{[\mu_1\nu_1]}\otimes t^0_{[\mu_2\nu_2]}$;
similarly the indices of $t^2$ contract with the indices of another $t^2$,
and $t^1$ with $t^3$.)
Our ${\cal O}$ is just the spectral parameter switch, it is a constant
gauge invariant operator. In particular, $[t^0,{\cal O}]=0$, 
{\it cf.} (\ref{GaugeInvarianceInYB}). In other words,
eq. (\ref{LHS},\ref{RHS}) is the generalized classical Yang-Baxter equation
{\em modulo gauge transformation}.

Similarly, putting the switch operator on the lower contour (see Figure \ref{FiguregYBE1})
we get the following consistency condition:
\begin{equation}
R_{23,+}R_{13,-}R_{23,+}^{-1} R_{23,-} R_{12,-} R_{23,-}^{-1} =
R_{12,-}R_{13,-} \,.
\end{equation}

Finally we turn to the exchange of $R_+$ and $R_-$, which is derived  in Figure \ref{FiguregYBE2}. Equating the LHS and RHS of this graph, we obtain
\begin{equation}
\begin{aligned}
  & (R_{23-} R_{13+} R_{23-}^{-1} ) 
      (R_{23+}   R_{12-}   R_{23+}^{-1})
       (R_{23-}   R_{13-}^{-1}     R_{23-}^{-1})  R_{23+}  \cr
= &   (R_{12+} R_{13-} R_{12+}^{-1} ) 
      (R_{12-}   R_{23+}   R_{12-}^{-1})
       (R_{12+}   R_{13-}^{-1}    R_{12+}^{-1})  R_{12-}  \,.
\end{aligned}
\end{equation}
Note that we equate of course only one side of the insertion at a time. Note also, that for $R_+ =R_-$ this returns to a standard YBE. Another way of writing it:
\begin{equation}
\begin{aligned}
&R_{12-}  R_{23+}^{-1}  \cr
&=(   \hbox{ad}_{R_{12+}} (R_{13-})
    \hbox{ad}_{R_{12-}}  (R_{23+}) 
            \hbox{ad}_{R_{12+}} (R_{13-}^{-1}) 
     )^{-1}
( \hbox{ad}_{R_{23-}} (R_{13+}) 
    \hbox{ad}_{R_{23+}} (R_{12-}) 
     \hbox{ad}_{R_{23-}} (R_{13-}^{-1})
     ) \,.
\end{aligned}
\end{equation}
In \cite{Freidel:1991jx, Freidel:1991jv} another generalization 
of quantum YBE was proposed as 
the quantum version of a more restricted set of classical YBE.
The main difference to the equations here is that 
the ones in \cite{Freidel:1991jx, Freidel:1991jv} impose the standard qYBE on $R$ 
(and thus the standard YBE on the classical $r$-matrix) and supplement these by equations
 of the type $RSS = SSR$. However the main problem with this approach is that the case of 
 principal chiral models and strings on $AdS_5 \times S^5$ do not fall in the class of 
 models where $r$ satifies the YBE separately from $s$.


\subsection{Some speculations on charges}
\label{sec:Charges}

Strictly speaking our derivation of equations like (\ref{generalizedYBE})
only applies to the terms quadratic in $r$ ({\it i.e.} tree level).
Although the derivation outlined in Figures \ref{FiguregYBE1} and \ref{FiguregYBE2}
seems to apply also at the level of higher loops, in fact there might be
subtleties associated to overlapping diagramms involving all three lines.

Nevertheless, let us for a moment take the proposed generalized qYBE (\ref{generalizedYBE})
seriously and see how it could  
be put to use in order to construct a quadratic algebra of $RTT$ type. 
The relation (\ref{generalizedYBE}) can be thought of in the following way. 
Let us begin with the standard YBE, which reads
\begin{equation}
R_{12} R_{13} R_{23} = R_{23} R_{13} R_{12} \,.
\end{equation}
This can formally be thought of as ``$R_{12}$ and $R_{13}$ commute up to conjugation by $R_{23}$'', or explicitly
\begin{equation}
R_{12} R_{13} = \left( R_{23} R_{13}  R_{23}^{-1} \right)  \left( R_{23} R_{12}  R_{23}^{-1} \right) 
\end{equation}
The relation (\ref{generalizedYBE}) generalizes this version of the YBE naturally, in that
\begin{equation}\label{YBEasCommute}
 R_{12,+} R_{13,+} =  \left( R_{23,-} R_{13,+} R_{23,-}^{-1}\right)		
  				   \left( R_{23,+} R_{12,+} R_{23,+}^{-1}\right) \,.
\end{equation}
If we interpret this as $RTT$ relations, we obtain
\begin{equation}
T_{12,+} T_{13,+} =  \left( R_{23,-} T_{13,+} R_{23,-}^{-1}\right)		
  				   \left( R_{23,+} T_{12,+} R_{23,+}^{-1}\right) \,.
\end{equation}
Naively one might then conclude that this equation is in fact of the type that has been discussed in \cite{Freidel:1991jx}, equation (14)
\begin{equation}
A_{12} T_{1} B_{12} T_2 = T_2 C_{12} T_1 D_{12} \,,
\end{equation}
where $T_{1} = T_{12,+}$ and $T_2 = T_{13,+}$ and 
\begin{equation}
A_{12} = R_{23,-}^{-1} \,,\quad
B_{12} = {\bf 1} \,, \quad 
C_{12} = R_{23,-}^{-1} R_{23,+} \,,\quad
D_{12} = R_{23,+}^{-1} \,.
\end{equation}
However, in \cite{Freidel:1991jx} it is required that the matrices $A,B,C,D$ satisfy a set of equations, in particular $A$ and $D$ have to separately satisfy the standard YBE, as well as equations of the type $ACC = CCA$ and $DCC = CCD$ as well as $[A_{12}, C_{13} ] =0$ and $[D_{12}, C_{32}] =0$ have to hold (note that it is pointed out in \cite{Freidel:1991jx} that these are only sufficient conditions). We do not require these equations, but only seem to be imposing the equation (\ref{generalizedYBE}). This is in fact a much weaker equation, but has the vital advantage that it gives as a classical limit an the algebra of $r-s$-matrices as we require it.

In view of the algebra (\ref{YBEasCommute}) the standard argument of construction of commuting charges does not go through, namely $[ \hbox{tr}_2 (T_{12,+}) , \hbox{tr}_3 (T_{13,+} ) ]$ is not obviously vanishing, as the conjugation in this case is by $R_{23, +}$ and $R_{23,-}$ respectively, which do not agree in the present case.
At this point a construction that appears in \cite{Freidel:1991jx} is useful, despite the fact that their transfer matrix algebra is different from ours. 
First let us simplify notation and suppress the physical space index of the $T$-matrices, so we consider the exchange relation of $T_{2}$ and $T_3$. $T_i$ is an element of End$(\rho_a) \equiv \rho_a \otimes \rho_a^\ast$ (at least for finite dimensional representations). Thus we can label them by $T_{(a ,\bar{a})}$, where $\bar{a}$ denotes the dual representation. The generalized RTT relations then become
\begin{equation}\label{InhomogeneousYBE}
T_{(2,\bar{2})} T_{(3,\bar{3})}  = R_{23-} T_{(3, \bar{3})} R_{2 \bar{3}-}^{-1} R_{23+} T_{(2, \bar{2})} R_{\bar{2} 3+}^{-1} 
						=: {\cal R}_{ (3,\bar{3})(2,\bar{2})} T_{(3, \bar{3})} T_{(2, \bar{2})}
\,,
\end{equation} 
where $R_{ab}$ acts on the $\rho_a$ part of $T_{(a, \bar{a})}$ etc.
and we defined
\begin{equation}
{\cal R}_{ (3,\bar{3})(2,\bar{2})} = R_{23-}  R_{2 \bar{3}-}^{-1} R_{23+} R_{\bar{2} 3+}^{-1}  \,.
\end{equation}
We require that ${\cal R}$ satisfies the YBE, in order for the exchange algebra of 
$T_{(a, \bar{a})}$ to be consistent. At this point the deviation from the construction 
in \cite{Freidel:1991jx} is necessary. 
Our $R_{ab\pm}$ matrices obey the generalized YBE (\ref{generalizedYBE})  
and the complete set of consistency conditions on $R_{\pm}$ should imply YBE for ${\cal R}$. 
This requires in particular additional relations for $R_{12-}$ and $R_{21-}$. 
Once the YBE for $\mathcal{R}$ are established, we define the dual RTT algebra as
\begin{equation}\label{DualRTT}
 \hat{T}_{(2,\bar{2})} \hat{T}_{(3,\bar{3})}   {\cal R}_{ (3,\bar{3})(2,\bar{2})} 
						= \hat{T}_{(3, \bar{3})} \hat{T}_{(2, \bar{2})}
\,,
\end{equation}
Consider a matrix representation (scalar matrix) of  (\ref{DualRTT}) given by $\hat{\tau}_{(2, \bar{2})}$ and 
$\hat{\tau}_{(3, \bar{3})}$.
There is a natural inner product between the representations and their duals, in particular 
$\hat{\tau}_{a \bar{a}} . T_{a \bar{a}}$. Thus acting with $\hat{\tau}_{2 \bar{2}} \hat{\tau}_{3 \bar{3}}$ on the generalized YBE in the form (\ref{InhomogeneousYBE}) we obtain
\begin{equation}
(\hat{\tau}_{2 \bar{2}} . T_{2 \bar{2}}) (\hat{\tau}_{3 \bar{3}} . T_{3 \bar{3}})
= (\hat{\tau}_{3 \bar{3}} . T_{3 \bar{3}}) (\hat{\tau}_{2 \bar{2}} . T_{2 \bar{2}}) \,,
\end{equation}  
and thus
\begin{equation}
[(\hat{\tau}_{2 \bar{2}} . T_{2 \bar{2}}),  (\hat{\tau}_{3 \bar{3}} . T_{3 \bar{3}})] =0 \,.
\end{equation}
This allows for construction of a family of infinite commuting charges by expanding these expressions in powers of the spectral parameter.


\subsection{Contours with loose endpoints}
The consistency condition for the exchange of contours with endpoints is more complicated.
Again, we can compare $(123)\to (213)\to (231)\to (321)$ to
$(123)\to (132)\to (312)\to (321)$. When we exchange $(123)\to (213)$ we get
the insertion of the split operator ${\cal F}_-^{-1}({\cal F}_+(1))$:

\vspace{7pt}

\hbox to \linewidth{\hfill %
\includegraphics[width=2.0in]{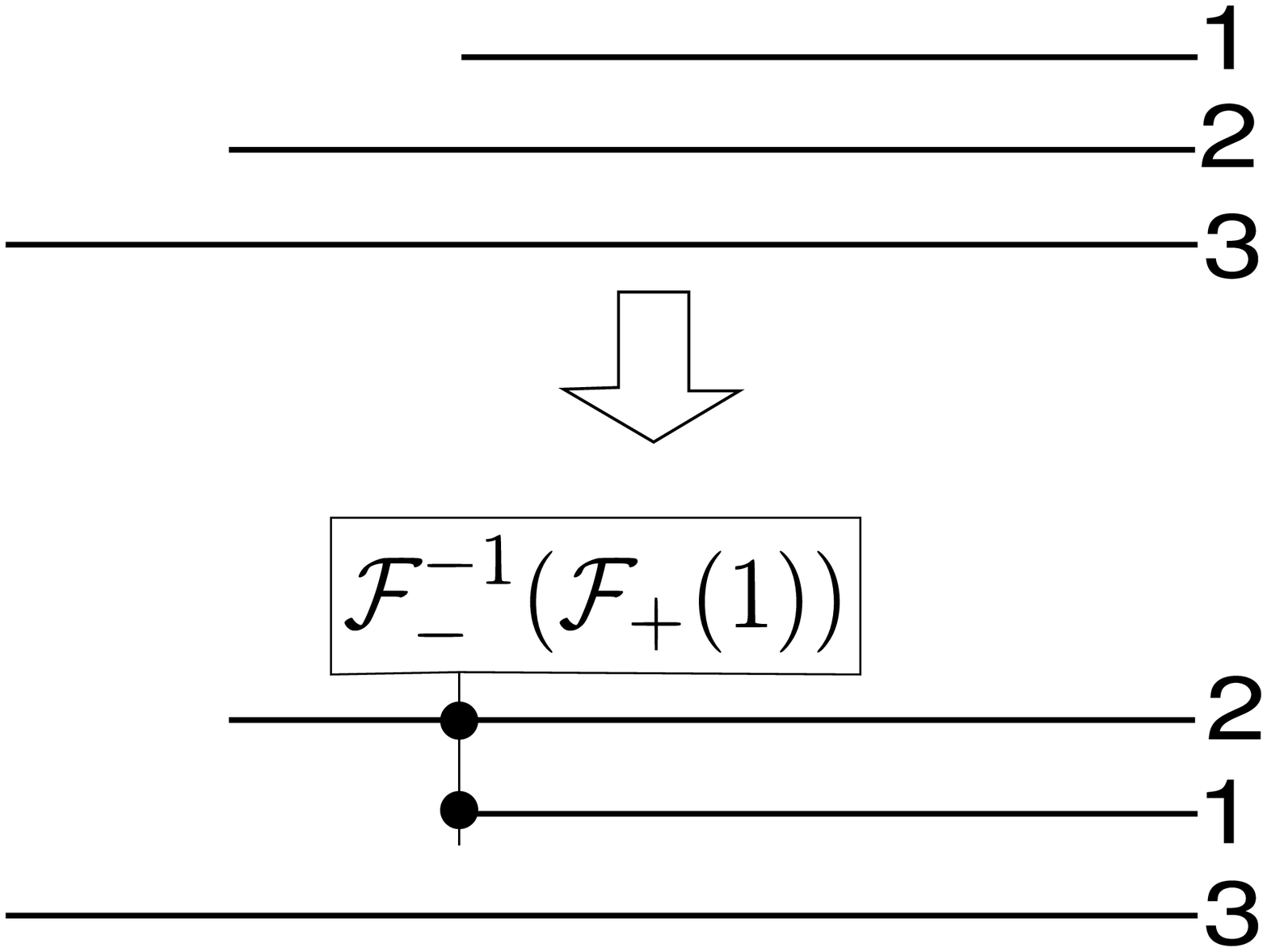} \hfill}

\noindent
At the first order of perturbation theory:
\begin{equation}
 {\cal F}_-^{-1}({\cal F}_+(1)) =  \mathfrak{1} + r + s  + q \,,
\end{equation}
where $q$ are {\em field-dependent} (= dynamical) terms. Indeed, the main difference
between the exchange of the switch operator and the exchange of the
endpoint is that the endpoint is not gauge invariant and therefore
the exchange matrix is field dependent. The expansion of $q$ in powers
of $x$ and $\vartheta$ starts with:
\begin{equation}\label{tMatrix}
q   =  {1\over 2} \left([x,t^2]\otimes t^2 + \{\vartheta^3,t^1\}\otimes t^3 + 
\{\vartheta^1,t^3\}\otimes t^1 \right) + \ldots 
\end{equation}
Then, when we exchange $(213)\to (231)\to (321)$ we get additional contributions
coming from the contraction of $q_{12}$ with the currents integrated over line 3,
for example:

\begin{equation}\label{QContraction}
\hbox to 2.2in{\hfill %
\includegraphics[width=2.0in]{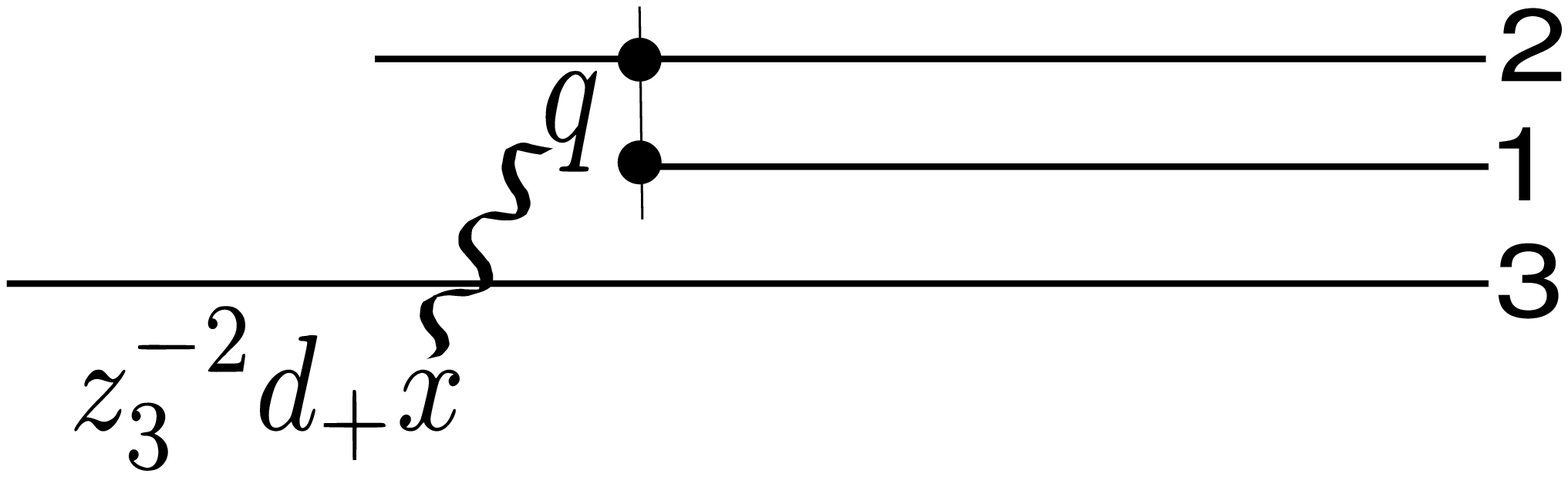} \hfill}
\end{equation}
On the other hand, if we look at the field independent (leading) terms,
we will get an equation identical to (\ref{RRRRRRRR}), but now $T_{123}$
does not act as the identity on the endpoint, because the endpoint is not
gauge invariant. But in fact the $T$ on the right hand side of (\ref{RRRRRRRR})
cancels with the terms arising from the contractions (\ref{QContraction}).

To summarize, we have the following two types of consistency conditions:
\begin{enumerate}

\item Consistency conditions for the exchange of gauge invariant operators.
In this case the right hand side of (\ref{RRRRRRRR}) does not spoil the 
consistency because of equation (\ref{GaugeInvarianceInYB}), which expresses
the gauge invariance of the inserted operators.

\item Consistency conditions for the insertions which are not gauge
invariant. In this case the right hand side of (\ref{RRRRRRRR}) cancels
against the terms arising from the diagrams like (\ref{QContraction}).

\end{enumerate}


\section{Conclusions and Discussion}

We have setup a formalism in which to compute the product of two transfer matrices, using the operator algebra of the currents. In particular, to leading order in the expansion around flat space-time, a structure reminiscent of classical $r$-matrices appears. This is however modified in that we require a system of $r$ and $s$-matrices, which satisfy a generalized classical YBE.  This is related in the approach of \cite{Maillet:1985fn} to Poisson brackets being non-ultralocal. 

We consider it a first step towards constructing the analog of a quantum $R$-matrix, 
which satisfies a generalized quantum YBE. 
The situation is different from  
\cite{Freidel:1991jx, Freidel:1991jv}, 
because the classical $r$-matrix in our case does not satisfy the standard classical YBE 
(which is one of the assumptions that goes into the construction in 
\cite{Freidel:1991jx, Freidel:1991jv}) but the combined equation for $r$ and $s$ (\ref{LHS}). 

The most promising direction to extend this work is to construct the quantum conserved charges from the $T$-matrices, as outlined in section \ref{sec:Charges}. It would also be interesting to test the generalized  quantum YBE explicitly at higher orders in the $1/R$ expansion.

It would also be interesting to understand how the $r$-$s$-matrices found here relate to the classical  $\mathfrak{su}(2|2)$
$r$-matrices found from the light-cone string theory and super-Yang Mill dual in \cite{Torrielli:2007mc, Moriyama:2007jt, Beisert:2007ty}. The connection, if it exists, would presumably  be along the line of our speculations in section \ref{sec:Charges}.


\subsection*{Acknowledgments}

We thank Jean-Michel Maillet for very interesting discussions. 
The research of AM is supported by the Sherman Fairchild 
Fellowship and in part by the RFBR Grant No. 06-02-17383 and in part by the 
Russian Grant for the support of the scientific schools NSh-8065.2006.2. 
The research of SSN is supported by a John A. McCone Postdoctoral Fellowship of Caltech. 
We thank the Isaac Newton Institute, Cambridge, for generous hospitality 
during the completion of this work. 


\newpage

\setcounter{section}{0} 

\appendix

\section{Calculation of the products of currents}
\label{sec:CalculationOfProductsOfCurrents}
Here we will describe some methods for calculating the singularities
in the product of two currents. We will only discuss two examples.
The first example is the collision $J_{3+}J_{3+}$ and the collision
$J_{1+}J_{2+}$. The second is
the singularities proportional to $xdx$ in the collision $J_{2+}J_{2+}$,
which we needed in Section \ref{sec:AsymmetricCouplingsZuZd}.

\subsection{Collisions $J_{3+}J_{3+}$ and $J_{1+}J_{2+}$.}
\noindent\underline{Collision $J_{3+}J_{3+}$}
\begin{equation}\label{CollisionJ3+J3+}
( \partial_+ \vartheta_L + [\vartheta_R , \partial_+x] )^{\alpha}(w_a)
\llra
( \partial_+ \vartheta_L + [\vartheta_R , \partial_+x] )^{\beta}(w_b) \,.
\end{equation}
The cubic vertex $([\vartheta_L, \partial_-\vartheta_L] \partial_+ x)$
does not contribute to the singularity, but the other cubic vertex does:
\begin{equation}
-S\mapsto {1\over \pi}\int d^2 v 
\;{1\over 2} \mbox{str}\; 
\left( [\vartheta_R, \partial_+\vartheta_R] \partial_- x\right) \,.
\end{equation}
After integration by parts the interaction vertex becomes:
\begin{equation}
- {1\over \pi}\int d^2 v 
\;{1\over 2} \mbox{str}\; 
\left( [\partial_-\vartheta_R, \partial_+\vartheta_R]  x
+ \mbox{str}\; [\vartheta_R, \partial_+\partial_-\vartheta_R]  x\right) \,.
\end{equation}
Integrating by parts $\partial_+$ in the second term we get:
\begin{equation}
 {1\over \pi}\int d^2 v \; \mbox{str}\; 
\left( - [\partial_-\vartheta_R, \partial_+\vartheta_R]  x
+ {1\over 2}\mbox{str}\; [\vartheta_R, \partial_-\vartheta_R] \partial_+ x\right) \,.
\end{equation}
This implies that (\ref{CollisionJ3+J3+}) gives the same singularity as 
the following collision {\em in the free theory:}
\begin{equation}
\left( \partial_+ \vartheta_L - [\partial_+\vartheta_R,x] 
 + {1\over 2} [\vartheta_R , \partial_+x] \right)^{\alpha}(w_a) \llra
\left( \partial_+ \vartheta_L - [\partial_+\vartheta_R,x] 
 + {1\over 2} [\vartheta_R , \partial_+x] \right)^{\beta}(w_b) \,.
\end{equation}
The singularity is:
\begin{eqnarray}
&& {1\over (w_a-w_b)^2} ([t^1,x(w_a)]\otimes t^3 + t^3\otimes [t^1,x(w_b)])+
\nonumber \\
&& +{1\over 2} {1\over w_a-w_b}([t^1,\partial_+ x(w_a)]\otimes t^3 -
t^3\otimes [t^1,\partial_+ x(w_b)]) \,.
\end{eqnarray}
This is equal to:
\begin{equation}
{2\over w_a-w_b} [t^1,\partial_+x]\otimes t^3 +
{\overline{w}_a - \overline{w}_b \over (w_a - w_b)^2}
[t^1,\partial_-x]\otimes t^3 \,.
\end{equation}

\underline{Collision $J_{1+}J_{2+}$}
\begin{equation}
(\partial_+\vartheta_R+[\vartheta_L,\partial_+x])
\llra
(\partial_+x + 1/2 [\vartheta_L, \partial_+\vartheta_L]+\ldots) \,.
\end{equation}
We have to take into account the interaction vertex in the action:
\begin{equation}
-S\mapsto {1\over \pi}\int d^2 v 
{1\over 2} \mbox{str}\; 
\left( [\vartheta_L, \partial_-\vartheta_L] \partial_+ x\right) \,.
\end{equation}
It is convenient to denote the contracted fields by using prime. 
For example, this notation:
\begin{equation}\label{FirstContraction}
{1\over 2} \mbox{str}\; 
\left( [\vartheta'_L, \partial_-\vartheta_L] \partial_+ x' \right) \,.
\end{equation}
means that ${\vartheta_L}$ is contracted with the $\partial_+ \vartheta_R$
in $J_{1+}$, and ${\partial_+ x}$ with $\partial_+x$ 
in $J_{2+}$. Therefore $\partial_-\vartheta_L$ remains uncontracted.
There is another possible contraction:
\begin{equation}\label{SecondContraction}
{1\over 2} \mbox{str}\; 
\left( [\vartheta_L, \partial_-\vartheta'_L] \partial_+ x' \right) \,.
\end{equation}
In the interaction vertex (\ref{FirstContraction})
 let us integrate by parts $\partial_-$.
We will get:
\begin{equation}
-{1\over 2} \mbox{str}
\left( [\partial_- \vartheta'_L, \vartheta_L] \partial_+ x' \right)
-{1\over 2} \mbox{str}
\left( [ \vartheta'_L, \vartheta_L] \partial_-\partial_+ x' \right) \,.
\end{equation}
In the second expression let us integrate by parts $\partial_+$.
The result is:
\begin{equation}
-{1\over 2} \mbox{str}
\left( [\partial_- \vartheta'_L , \vartheta_L] \partial_+ x' \right)
+{1\over 2} \mbox{str}
\left( [\partial_+ \vartheta'_L , \vartheta_L] \partial_- x' \right)
+{1\over 2} \mbox{str}
\left( [\vartheta'_L , \partial_+ \vartheta_L] \partial_- x' \right) \,.
\end{equation}
The first term coincides with (\ref{SecondContraction}), and
together with (\ref{SecondContraction}) gives:
\begin{equation}
-\mbox{str}
\left( [\partial_- \vartheta'_L , \vartheta_L] \partial_+ x' \right) \,.
\end{equation}
This is easy to contract, and precisely cancels the ``direct
hit'' $[\vartheta_L,\partial_+x]\llra \partial_+ x$.
The second and thrid terms combine with the ``direct hit''
\begin{equation}\partial_+\vartheta_R \llra 1/2[\vartheta_L,\partial_+\vartheta_L]\end{equation}
to give the same contribution as the collision
\begin{equation}
\partial_+\vartheta_R' \llra [\vartheta_L' , \partial_+\vartheta_L] \,,
\end{equation}
which gives:
\begin{equation}
{\partial_+\vartheta_L^{\alpha}\over w_a-w_b}
\fduu{\alpha}{\dot{\alpha}}{\mu} \,.
\end{equation}

\subsection{Terms $xdx$ in the collision $J_{2+}J_{2+}$}
\label{sec:XdXinJ2J2}
Consider this collision:

\hbox to \linewidth{\hfill%
\updown{$z^{-2}J_{2+}$}{$z^{-2}J_{2+}$}{10}{40}\hfill}

\vspace{10pt}

\noindent More explicitly, we are looking at:
\begin{equation}
\left(\partial_+ x +{1\over 6}[x,[x,\partial_+x]]\right)(w_u)
\stackrel{\otimes}{\leftrightarrow}
\left(\partial_+ x +{1\over 6}[x,[x,\partial_+x]]\right)(w_d) \,.
\end{equation}
Couplings to $xdx$ receive contributions from the quartic interaction vertex:
\begin{equation}
-S\mapsto {1\over 6\pi} \mbox{str} [x,\partial_+x][x,\partial_-x] \,.
\end{equation}
We denote the contracted fields $x'(w_u)$ and $x''(w_d)$.
When $\partial_-x$ in the interaction vertex gets contracted
with  $\partial_+x$ in one of the $J_{2+}$, this contribution
cancels the ``direct hit'' $\partial_+x \llra {1\over 6}[x,[x,\partial_+x]]$.
Let us study the diagrams in which $\partial_-x$ in the interaction
vertex remains uncontracted. There are the following possibilities:
\begin{eqnarray}
& {1\over 6\pi}\int d^2 v\; \mbox{str} & 
( 2 [x',\partial_+ x''][x,\partial_- x] +
\\
&&
+   [x,\partial_+ x'][x'',\partial_- x] +
\\
&&
+   [x,\partial_+ x''][x',\partial_- x] ) \,.
\end{eqnarray}
Here prime and double prime mark the contracted elementary fields; for
example in the first term $2 [x',\partial_+ x''][x,\partial_- x]$ the elementary
field 
$x'$ contracts with $\partial_+x(w_u)$  in $z_u^{-2}J_{2+}(w_u)$ 
and $\partial_+ x''$ contracts with $\partial_+x(w_d)$ in $z_d^{-2}J_{2+}(w_d)$;
while $[x,\partial_- x]$ remains uncontracted.
This gives:
\begin{eqnarray}
& {1\over 6\pi}\int d^2 v \; \mbox{str} &
\left( { (-2)\over (v-w_u)(v-w_d)^2 }
C^{\mu\nu} [t^2_{\mu},[x,\partial_-x]] \otimes t_{\nu}^2 -
\right.
\nonumber
\\
&& - { (-1) \over (v-w_u)^2(v-w_d) }
   C^{\mu\nu} [x, [t^2_{\mu},\partial_-x]]\otimes t_{\nu}^2 -
\nonumber
\\
&& \left.
   - { (-1) \over (v-w_u)(v-w_d)^2 }
   C^{\mu\nu} t^2_{\mu}\otimes [x,[t^2_{\nu},\partial_- x]]
\right) =
\nonumber
\\
=& -{1\over 2\pi}\int d^2 v \; \mbox{str} &
{1\over (v-w_u)(v-w_d)^2 }
C^{\mu\nu} [t^2_{\mu},[x,\partial_-x]] \otimes t_{\nu}^2 
=
\\
&& = -{1\over 2} {\overline{w}_d-\overline{w}_u\over (w_d - w_u)^2}
C^{\mu\nu} [t^2_{\mu},[x,\partial_-x]] \otimes t_{\nu}^2 \,.
\end{eqnarray}
(A simple way to get the singularity of this integral is by considering
${\partial\over\partial\overline{w}_u}$.)
This contributes to the current-generator coupling:
\begin{equation}\label{ContributionFrom_J2_J2_Appendix}
{1\over 2} \pi i \; C^{\mu\nu} (z^{-2}[t^{2}_{\mu},[x,\partial_-x]])
\otimes (z^{-2} t^{2}_{\nu})
\end{equation}
Taking into account that 
$C^{\mu\nu}[t^2_{\mu},t^0]\otimes t^2_{\nu}=
-C^{\mu\nu}t^2_{\mu}\otimes [t^2_{\nu},t^0]$
we can rewrite (\ref{ContributionFrom_J2_J2_Appendix}) using the 
$\wedge$-product:
\[
{1\over 2} \pi i \; C^{\mu\nu} (z^{-2}[t^{2}_{\mu},[x,\partial_-x]])
\wedge(z^{-2} t^{2}_{\nu}) \,.
\]

\subsection{Short distance singularities using index notations}
\label{sec:IndexOPE}
In the main text we gave the expressions for the short distance
singularities in the tensor product notations.
Here we list the singularities using more ``conservative'' index notations:
\begin{eqnarray}
J_{1-}^{\dot{\alpha}}(w_1)J_{2+}^{\mu}(w_2) 
& = & {1\over R^3} 
{\partial_-\vartheta_L^{\gamma}
\over w_1 - w_2}
\fduu{\gamma}{\dot{\alpha}}{\mu} 
\\
J_{1+}^{\dot{\alpha}}(w_1)J_{2-}^{\mu}(w_2)
& = & 
{1\over R^3}
{\partial_-\vartheta^{\gamma}_L\over w_1 -w_2} 
\fduu{\gamma}{\dot{\alpha}}{\mu} 
\\
J_{3-}^{\alpha}(w_1)J_{2+}^{\mu}(w_2) 
& = & 
{1\over R^3}
{\partial_+\vartheta^{\dot{\gamma}}_R\over \bar{w}_1 - \bar{w}_2} 
\fduu{\dot{\gamma}}{\dot{\alpha}}{\mu} 
\\
J_{3+}^{\alpha}(w_1)J_{2-}^{\mu}(w_2)
& = &   {1\over R^3} 
{\partial_+\vartheta^{\dot{\gamma}}_R\over \bar{w}_1 -\bar{w}_2}
\fduu{\dot{\gamma}}{\alpha}{\mu} 
\\
J_{1+}^{\dot{\alpha}}(w) J_{1-}^{\dot{\beta}}(0)
& = & -{1\over R^3}
{\partial_- x^{\mu}\over w_a-w_b} 
\fduu{\mu}{\dot{\alpha}}{\dot{\beta}}
\\
J_{3+}^{\alpha}(w) J_{3-}^{\beta}(0)
& = & -{1\over R^3}
{\partial_+ x^{\mu}\over \bar{w}_a -\bar{w}_b} \fduu{\mu}{\alpha}{\beta}
\end{eqnarray}

\begin{eqnarray}
J_{1+}^{\dot{\alpha}}(w_1)J_{2+}^{\mu}(w_2) 
& = &
 {1\over R^3} { \partial_+ \vartheta_L^{\gamma} \over w_1 - w_2 }
 \fduu{\gamma}{\dot{\alpha}}{\mu} 
+ O\left({1\over R^4}\right)
\\
J_{3+}^{\alpha}(w_3)J_{2+}^{\mu}(w_2)
& = &
{2\over R^3} {  \partial_+\vartheta_R^{\dot{\beta}} \over w_3 - w_2 }
\fduu{\dot{\beta}}{\alpha}{\mu} 
+  {1\over R^3} {\bar{w}_3 -\bar{w}_2 \over (w_3 - w_2)^2 }  
\partial_- \vartheta_R^{\dot{\gamma} }
 \fduu{\dot{\gamma}}{\alpha}{\mu}+ O\left({1\over R^4}\right)
\label{OPEJ1PlusJ2Plus}
\\
J_{1+}^{\dot{\alpha}}(w_a) J_{1+}^{\dot{\beta}}(w_b) 
& = &
-{1\over R^3} {\partial_+ x^{\mu}\over w_a - w_b }
 \fduu{\mu}{\dot{\alpha}}{\dot{\beta}}
+ O\left({1\over R^4}\right)
\label{OPEJ1PlusJ1Plus}
\\
J_{3+}^{\alpha}(w_a) J_{3+}^{\beta}(w_b) 
& = &
-{2\over R^3} {\partial_+ x^{\mu} \over w_a - w_b }
\fduu{\mu}{\alpha}{\beta}
-{1\over R^3} {\bar{w}_a -\bar{w}_b \over (w_a - w_b)^2 }\partial_- x^{\mu}
\fduu{\mu}{\alpha}{\beta}
+ O\left({1\over R^4}\right)
\label{OPEJ3PlusJ3Plus}
\\
J_{1+}^{\dot{\alpha}}(w_1) J_{3+}^{\alpha}(w_3) 
& = &
-{1\over R^2} {1\over (w_1-w_3)^2}
C^{\dot{\alpha}\alpha} 
+ O\left({1\over R^4}\right)
\\
J_{2+}^{\mu}(w_m) J_{2+}^{\nu}(w_n) 
&=&
-{1\over R^2} {1\over (w_m - w_n)^2 }
C^{\mu\nu} 
+ O\left({1\over R^4}\right)
\\
J_{0+}^{[\mu\nu]}(w_0) J_{1+}^{\dot{\alpha}}(w_1) 
& = &
- {1\over 2 R^3}  \left(
{\vartheta_R^{\dot{\beta}}(w_0) \over (w_0 - w_1)^2}
+
{\partial_+\vartheta_R^{\dot{\beta}}(w_0) \over (w_0 - w_1)}
\right)
\fduu{\dot{\beta}}{\dot{\alpha}}{[\mu\nu]}
+ O\left({1\over R^4}\right)
\\
J_{0+}^{[\mu\nu]}(w_0) J_{3+}^{\alpha}(w_3) & = &
- {1\over 2 R^3}  \left(
{\vartheta_L^{\beta}(w_0) \over (w_0 - w_3)^2}
+
{\partial_+\vartheta_L^{\beta}(w_0) \over (w_0 - w_3)}
\right)
\fduu{\beta}{\alpha}{[\mu\nu]}
+ O\left({1\over R^4}\right)
\\
J_{0+}^{[\mu\nu]}(w_0) J_{2+}^{\lambda}(w_2) & = &
- {1\over 2 R^3}  \left(
{x^{\kappa}(w_0) \over (w_0 - w_2)^2}
+
{\partial_+ x^{\kappa}(w_0) \over (w_0 - w_2)}
\right)
\fduu{\kappa}{\lambda}{[\mu\nu]}
+ O\left({1\over R^4}\right) \,.
\end{eqnarray}



\section{Very brief summary of the Maillet formalism}
\label{VeryBriefSummaryOfMaillet}

Let us briefly review the situation in Maillet et al's work and how this connects to our present analysis. 
In \cite{Maillet:1985fn} a formalism was developed which generalizes the classical YBE to incorporate the case of non-ultralocal Poisson brackets. 
Consider the algebra of $L$-matrices (spatial component of the Lax operator)
\begin{equation} \label{LLalgebra}
\begin{aligned} 
\{L (\sigma_1, z_1) , L (\sigma_2,  z_2) \}
 &=  [r(\sigma_1, z_1, z_2) , \mathfrak{1} \otimes L(\sigma_1, z_1)+ L(\sigma_1, z_1) \otimes \mathfrak{1} ] 
     \delta (\sigma_1 -\sigma_2) \cr
 &+     [s(\sigma_1, z_1, z_2) ,  \mathfrak{1} \otimes L(\sigma_1, z_1)-  L(\sigma_1, z_1) \otimes \mathfrak{1}]
     \delta (\sigma_1 -\sigma_2)  \cr
 &    - \left( s(\sigma_1, z_1, z_2) + s(\sigma_2, z_1, z_2) 
       \right) \delta' (\sigma_1 - \sigma_2) \,.
\end{aligned}
\end{equation}
The terms proportional to $\delta '$ are the so-called non-ultralocal terms. 
The algebra (\ref{LLalgebra}) is a deformation of the standard ultra-local one by terms depending on the matrix $s$, which unlike $r$ is symmetric. Jacobi-identity for $\{,\}$ yields a generalized, dyamical YBE\footnote{Note that the signs are slightly different in \cite{Maillet:1985fn} from the equations we will be using.}
\begin{equation} \label{rsYBE}
 [r_{12} - s_{12} , r_{13} + s_{13}] + [r_{12} + s_{12}, r_{23} + s_{23} ] + [r_{13} + s_{13} , r_{23} +s_{23}] 
 + H_{1,23}^{(r+s)} - H_{2,13}^{(r+s)}
 =0 \,.
\end{equation}
The dynamicity is due to the terms $H_{i, jk}$, which arise if $r+s$ is field dependent and are defined by
\begin{equation}
\{L (\sigma_1, z_1)\otimes \mathfrak{1} \otimes \mathfrak{1}, \mathfrak{1} \otimes (r+s)_{23}(\sigma_2, z_2, z_3)  \}
=  H_{1, 23}^{(r+s)} (\sigma_1, z_1,z_2, z_3 ) \delta (\sigma_1-\sigma_2) \,.
\end{equation}
In the case of $s=0$ and $r$ constant (field-independent) the relation (\ref{rsYBE}) reduces to the standard classical YBE. 
This formulation was applied to the $O(n)$ model \cite{Maillet:1985fn} and the complex Sine-Gordon model \cite{Maillet:1985ek} (where in both cases the $r-s$-matrices are dynamical), as well as the principal chiral field \cite{Maillet:1985ec}, in which case the terms $H_{i, jk}$ vanish. Note that the field-dependence of the $r-s$-matrices seems to be due to the field-dependence of the non-ultralocal term.


\newpage

\bibliographystyle{JHEP} \renewcommand{\refname}{Bibliography}
\addcontentsline{toc}{section}{Bibliography}

\providecommand{\href}[2]{#2}\begingroup\raggedright\endgroup


\end{document}